\documentclass[twoside,twocolumn,9pt]{article}
\usepackage{extsizes}
\usepackage[super,sort&compress,comma]{natbib}
\usepackage[version=3]{mhchem}
\usepackage[left=1.5cm, right=1.5cm, top=1.785cm, bottom=2.0cm]{geometry}
\usepackage{balance}
\usepackage{mathptmx}
\usepackage{sectsty}
\usepackage{graphicx}
\usepackage{lastpage}
\usepackage[format=plain,justification=justified,singlelinecheck=false,font={stretch=1.125,small,sf},labelfont=bf,labelsep=space]{caption}
\usepackage{float}
\usepackage{fancyhdr}
\usepackage{fnpos}
\usepackage[english]{babel}
\addto{\captionsenglish}{%
  
}
\usepackage{array}
\usepackage{droidsans}
\usepackage{charter}
\usepackage[T1]{fontenc}
\usepackage[usenames,dvipsnames]{xcolor}
\usepackage{setspace}
\usepackage[compact]{titlesec}
\usepackage{hyperref}

\usepackage{amsmath}
\usepackage{amsfonts}
\usepackage{amssymb}

\definecolor{cream}{RGB}{222,217,201}

\begin{document}

\pagestyle{fancy}
\thispagestyle{plain}
\fancypagestyle{plain}{
\renewcommand{\headrulewidth}{0pt}
}

\makeFNbottom
\makeatletter
\renewcommand\LARGE{\@setfontsize\LARGE{15pt}{17}}
\renewcommand\Large{\@setfontsize\Large{12pt}{14}}
\renewcommand\large{\@setfontsize\large{10pt}{12}}
\renewcommand\footnotesize{\@setfontsize\footnotesize{7pt}{10}}
\makeatother

\renewcommand{\thefootnote}{\fnsymbol{footnote}}
\renewcommand\footnoterule{\vspace*{1pt}%
\color{cream}\hrule width 3.5in height 0.4pt \color{black}\vspace*{5pt}}
\setcounter{secnumdepth}{5}

\makeatletter
\renewcommand\@biblabel[1]{#1}
\renewcommand\@makefntext[1]%
{\noindent\makebox[0pt][r]{\@thefnmark\,}#1}
\makeatother
\renewcommand{\figurename}{\small{Fig.}~}
\sectionfont{\sffamily\Large}
\subsectionfont{\normalsize}
\subsubsectionfont{\bf}
\setstretch{1.125} 
\setlength{\skip\footins}{0.8cm}
\setlength{\footnotesep}{0.25cm}
\setlength{\jot}{10pt}
\titlespacing*{\section}{0pt}{4pt}{4pt}
\titlespacing*{\subsection}{0pt}{15pt}{1pt}

\fancyfoot{}
\fancyfoot[RO]{\footnotesize{\sffamily{1--\pageref{LastPage} ~\textbar  \hspace{2pt}\thepage}}}
\fancyfoot[LE]{\footnotesize{\sffamily{\thepage~\textbar\hspace{3.45cm} 1--\pageref{LastPage}}}}
\fancyhead{}
\renewcommand{\headrulewidth}{0pt}
\renewcommand{\footrulewidth}{0pt}
\setlength{\arrayrulewidth}{1pt}
\setlength{\columnsep}{6.5mm}
\setlength\bibsep{1pt}

\makeatletter
\newlength{\figrulesep}
\setlength{\figrulesep}{0.5\textfloatsep}

\newcommand{\topfigrule}{\vspace*{-1pt}%
\noindent{\color{cream}\rule[-\figrulesep]{\columnwidth}{1.5pt}} }

\newcommand{\botfigrule}{\vspace*{-2pt}%
\noindent{\color{cream}\rule[\figrulesep]{\columnwidth}{1.5pt}} }

\newcommand{\dblfigrule}{\vspace*{-1pt}%
\noindent{\color{cream}\rule[-\figrulesep]{\textwidth}{1.5pt}} }

\makeatother

\twocolumn[
  \begin{@twocolumnfalse}
\vspace{1em}
\sffamily
\begin{tabular}{m{4.5cm} p{13.5cm} }

\includegraphics{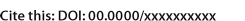} & \noindent\LARGE{\textbf{Multiscale morphology and contact mechanics of phy-sisorbed Al and Cu nanoparticles$^\dag$}} \\
\vspace{0.3cm} & \vspace{0.3cm} \\

 & \noindent\large{Mykola Prodanov,$^{\ast}$\textit{$^{a}$} Oleksii Khomenko\textit{$^{b}$}} \\

\includegraphics{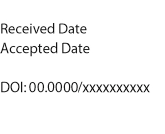} & \noindent\normalsize{

Using large-scale molecular dynamics simulations, we investigate the scaling of morphological and contact mechanics properties of Al and Cu nanoparticles (NPs) physisorbed on suspended graphene. The characteristic linear size of a NP ranges from 1 nm to 49 nm, covering a length scale of 1.5 decades. The NPs were obtained using a procedure mimicking thermal dewetting of thin films. Calculations show that NPs with a surface area-to-volume ratio above about 1.8~nm$^{-1}$, or with a linear size under 3--6~nm, behave differently from larger particles. For these smaller NPs, scaling of their total surface area and volume with the linear size can deviate from quadratic and cubic dependencies, respectively. Their mean interfacial separation and relative contact area change rapidly with size, exhibiting substantial variation. In contrast, for larger NPs, these quantities approach the thermodynamic limit. The height distributions of all particles exhibit a narrow spike and a decaying tail, both of which can be fit to Gaussians for larger NPs. In contrast, the interfacial gap distributions are close to a single Gaussian. The height power spectrum density (PSD) heatmaps of the smaller NPs are smeared and do not manifest a clear structure in contrast to the sixfold symmetry of the PSD of the larger ones. The maximum spatial frequency of the hexagonal 2D PSD roughly corresponds to the nearest-neighbor atomic distance of Al and Cu. For larger NPs with diameters of 20--25~nm, the isotropic height PSD exhibits power-law regions, which can be interpreted as self-affine roughness with Hurst exponents of 0.1--0.56. We also calculate the relative difference between the apparent contact area and the approximated area of the bottom atomic layer. Small NPs have errors above 10\%, which decrease with size. Our simulations illustrate how surface topography evolves with NP size and suggest that larger NPs can have random surface roughness. These results highlight the size-dependent morphology and contact mechanics of Al and Cu NPs, which differ qualitatively at smaller length scales.}
\\
\end{tabular}

 \end{@twocolumnfalse} \vspace{0.6cm}

  ]

\renewcommand*\rmdefault{bch}\normalfont\upshape
\rmfamily
\section*{}
\vspace{-1cm}

\footnotetext{\dag~Electronic Supplementary Information (ESI) available: videos of the time evolution of nanoparticles.}
\footnotetext{\textit{Sumy State University, 116 Kharkivska Str., UA-40007 Sumy, Ukraine.}}
\footnotetext{\textit{$^{a}$ E-mail: prodanov.my@gmail.com}}
\footnotetext{\textit{$^{b}$ E-mail: o.khomenko@mss.sumdu.edu.ua}}




%
%
%

\section{Introduction}
\label{sec:intro}

Metal nanoparticles (NPs) are known for their prominent size-dependent structural, thermodynamic, electronic, chemical, optical, tribological, and other properties, which are relevant to a wide range of practical applications~\cite{Min2008,Eker2024mol,Suhas2026,Harsha2024,Kim2023,
Badan2022nm,Khomenko2022ppm,Gawande2016,Kart2014,Medasani2009}. For example, NPs with a larger surface area-to-volume ratio ($SA/V$) exhibit better catalytic properties~\cite{Min2008,Saleh2022,Pozzi2024,Eker2024mol,Suhas2026}. Compared to bulk material samples, Al and Cu NPs have lower melting points that grow with the NP size~\cite{Yalamanchali2017,Kart2014}. Of special interest are the tribological properties of metal NPs adsorbed on various surfaces~\cite{Dietzel2008prl,Khomenko2010,Gao2022fic,
Konrad2023prb,Khomenko2025ppm,Cihan2023prl,Oo2024}. Antimony, gold, copper, and silver NPs adsorbed on highly oriented pyrolytic graphite (HOPG) and MoS$_{2}$ substrates can exhibit structural lubricity and superlubricity~\cite{Gao2022fic,Oo2024}, frictional duality~\cite{Dietzel2008prl}, and frictional anisotropy~\cite{Khomenko2013cmp,Konrad2023prb,Khomenko2025ppm} as well as a strong dependence of friction on the surface structure~\cite{Cihan2023prl} and ambient environment~\cite{Oo2024}.

In general, the NP properties can change nonlinearly with decreasing the NP linear size within the 100~nm length range, i.e., the scale, which is by convention used to separate NPs from the bulk material~\cite{Min2008,Saleh2022}. In particular, the smallest NPs, a few nm in size, also known as nanoclusters, can exhibit unique behavior. For example, the melting point of Al nanoclusters smaller than about 2~nm shows fluctuations and does not follow the thermodynamic scaling of larger ones~\cite{Yalamanchali2017}. It was also shown that the contact line length of Pt NPs increases with decreasing NP size below 10~nm, leading to higher hydrogen evolution reaction activity in electrochemical devices~\cite{Harsha2024}.

As for the tribological and contact mechanics (CM) properties of the NPs, such as the mean distance between the contacting surfaces (also called \textit{interfacial separation} or \textit{gap}~\cite{Almqvist2011jmps,Prodanov2014triblett}) $\bar{u}$, height \textit{power spectrum density (PSD)} $C_{h}(q)$, and the \textit{real contact area} $A_{real}$ of a NP, no comprehensive studies of their scaling behavior can be found in the literature. The reason is that measuring these quantities experimentally with atomic-scale precision is still challenging in practice~\cite{Rovatti2011,Rodriguez2025triblett,Chen2026ns}. High-resolution atomic force microscopes (AFMs) used to probe NPs in the experiments~\cite{Dietzel2008prl,Dietzel2010prb,Oo2024} do not allow direct measurement of the contact area. Additionally, the AFM probes have some limitations~\cite{Rovatti2011,Rodriguez2025triblett, Chen2026ns}. First, the AFM tip can change during the scanning experiment~\cite{Rodriguez2025triblett}. Second, there are difficulties controlling the tip geometry~\cite{Rovatti2011,Rodriguez2025triblett, Chen2026ns}. For example, the tip size of an AFM is typically a few nm, so the height-resolution error is of this size~\cite{Chen2026ns}. In principle, it is possible to reconstruct the height profile of a surface scan with~\AA~precision using transmission electron microscopy~\cite{Jacobs2022mrsbul}. However, this approach is still difficult to use for full 3D surface reconstruction. Therefore, reliable PSDs are usually measured to a few nm and do not resolve smaller, atomic-scale details~\cite{Persson2023triblett}. This also means that precise measurements of the morphological properties, such as the total surface area $S$ and volume $V$ of the smallest NPs of less than a few nm in size, are challenging at the moment. Note that in the tribological experiments~\cite{Dietzel2010prb} it was argued that Sb NPs on HOPG have a flat surface and that the apparent, i.e., calculated from the visible NP sizes, $A_{0}$ and the real $A_{real}$ contact areas are equal. Therefore, $A_{0}$ can be used to characterize the tribological properties of NPs. A similar approach has been used in several atomistic simulations~\cite{Khomenko2010,Khomenko2013cmp}. However, this assumption should be justified, and possible precision errors estimated.

In contrast to the contact mechanics of NPs, well-established experimental, theoretical, and numerical results provide reasonable predictions for the size scaling of the CM parameters of macroscopic randomly rough interfaces~\cite{Persson2005jpcm,Almqvist2011jmps,
Prodanov2014triblett,Persson2023triblett}. In particular, it has been shown that the contact stiffness $K$ (derived from the mean interfacial gap $\bar{u}$) exhibits finite-size scaling effects~\cite{Pastewka2013}. Namely, when the system size or the squeezing pressure is sufficiently small that a single asperity remains in contact, $K$ scales with the applied pressure as a power law with power $1/(1+H)$, where $H$ is the Hurst exponent of the self-affine surface roughness. Additionally, it was shown that numerical corrections should be considered when analyzing randomly rough surfaces in the thermodynamic, fractal, and continuum limits~\cite{Prodanov2014triblett}. Unfortunately, the aforementioned results do not extend to the atomic scale because macroscopic models rely on continuum mechanics, whose assumptions are not applicable to NPs, which require an atomistic, discrete treatment~\cite{Rodriguez2025triblett}.

Therefore, there is a need to fill the knowledge gap regarding the size scaling of morphological and CM quantities of NPs. Note that the precise, atomic-scale knowledge of the surface topography is important not only in the tribological context. At the fundamental level, surface topography has recently been recognized as an inherent parameter of a material~\cite{Jacobs2022mrsbul}. Height PSD $C_{h}(q)$ is one of the quantities that allows multiscale characterization of any surface topography and, hence, the dependent CM quantities~\cite{Persson2005jpcm,Jacobs2017stmp,Persson2023triblett}, such as $A_{real}$ and $\bar{u}$. Moreover, surface topology is relevant not only to CM and tribology but also affects related quantities, e.g. heat-/electroconductivity, and fluid leakage amount~\cite{Persson2005jpcm,Jacobs2022mrsbul,Harsha2024}, which depend on $A_{real}$ and $\bar{u}$.

Large-scale atomistic molecular dynamics (MD) simulations are the tool that overcomes the  mentioned experimental and theoretical obstacles~\cite{Frenkel2002,Rapaport2004,Griebel2007,Almqvist2011jmps}.
It provides atomic-scale spatial resolutions for NPs of different sizes and allows measuring the needed quantities with high precision. In this work, we report the results of large-scale MD simulations of the scaling of morphological and CM parameters of Al and Cu NPs physisorbed on a suspended graphene sheet. The linear size of the NPs in our model ranges from 1~nm to 49~nm, spanning 1.5 decades in length scale, which is a reasonable range for investigating size scaling. We chose Al and Cu metals for a couple of reasons. First, established methods for obtaining Al and Cu NPs with well-defined parameters exist~\cite{BenAissa2015,Jacobson2020,Chung2022}, allowing corresponding experiments to be conducted and the numerical results to be verified. Second, bulk Al and Cu are FCC metals with a similar interaction energy per atom with graphene~\cite{Vanin2010}. However, they have different lattice constants~\cite{Zhou2001}, which allows investigation of the effect of lattice constant on the NPs' parameters. Suspended graphene~\cite{Gass2008}, as the substrate material, was selected for its lattice-structure similarity to the experimentally used HOPG, as well as for its unique structural properties~\cite{Novoselov2004sc,Szroeder2019om,Sahalianov2020el,Radchenko2021}. In particular, the experiments suggest that suspended graphene is not atomically flat but can exhibit curvature in the form of waves~\cite{Gass2008,Neek-Amal2009}. Additionally, MD simulations of Ag NPs located on another 2D material, MXene, suggest that its surface can deform at the contact interface, leading to the formation of wrinkles~\cite{Borysiuk2024}. Therefore, the surface structure of suspended graphene may affect the real contact area between the NP and the surface, potentially leading to different behavior compared to the atomically flat HOPG surface.

Our calculations pursue several goals. First, to find out how the morphological parameters, such as $S$, $V$, $SA/V$ of NPs, scale with their lateral sizes $L_{x}$ and $L_{y}$. Second, to investigate the size scaling of the CM and related parameters, such as the number of atoms in the contact layer $N_{b}$, height $P(h)$ and interfacial separation $P(u)$ distributions, the mean $\bar{u}$ and most probable $u_{p}$ gaps, height $C_{h}(q)$ and gap $C_{u}(q)$ PSDs and determine whether there is a transition from the atomistic or finite-size to thermodynamic behavior, which is observed for some thermodynamic and electronic properties of NPs~\cite{Yalamanchali2017,Harsha2024} or macroscopic surfaces~\cite{Pastewka2013}. Third, to check whether the apparent contact area $A_{0}$ is a good approximation for the area of the contact atomic layer of NP, and to try using the distance criterion to interpret the contact at the atomic level, which is not well-defined from the macroscopic point of view.

\section{Model and methodology}
\subsection{Atomistic model and computational details}
\label{sec:model}

\begin{figure}[h]
\centering
  \includegraphics[height=3.2cm]{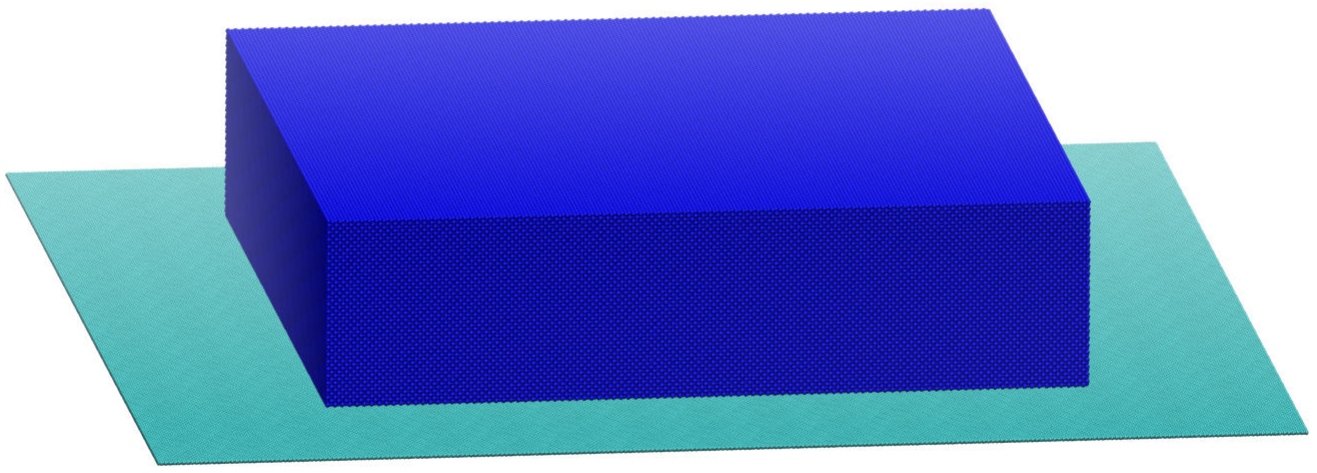}
  \caption{Initial atomic configuration of a simulation setup for the largest Al system size, the snapshot obtained with VMD~\cite{Humphrey1996}.}
  \label{fgr:init_slab}
\end{figure}

Initially, a cuboid slab of metal atoms arranged in the bulk FCC lattice is placed above the graphene sheet located in the $xy$-plane, with the $x$- and $y$-axes parallel to the zigzag and armchair edges of the substrate, cf. Fig.~\ref{fgr:init_slab} (the snapshots in this paper were done with VMD software~\cite{Humphrey1996}) and the electronic supplementary materials (the ESI$^\dag$ videos were produced using OVITO software~\cite{Stukowski2010}). The slab's height is approximately 1/3 of its width/length. The atomic plane (110) of the metal slab is initially parallel to the substrate. To keep the graphene sheet stationary in the vertical $z$-direction, the perimeter atoms of the substrate are held fixed throughout each simulation run. Ultrahigh vacuum conditions are maintained.

The initial size of the metal slab is controlled using the number of elementary cells in each direction. For example, the largest NP contained initially $104 \cdot 104 \cdot 26$ unit cells in the $x$, $y$ and $z$-directions, respectively. The lateral dimensions of the carbon substrate are 50\% larger than those of the slab and hence vary with the NP size. The total number of atoms $N_{0}$ ranges from 352 to 1280864, and the number of metal atoms $N$ in a nanoisland ranges from 64 to 1124864. The system's temperature $T$ is controlled using the Berendsen thermostat~\cite{Griebel2007}. The leapfrog method is used to integrate the equations of motion~\cite{Rapaport2004} with a time step of $\Delta t = 0.2$~fs.

The embedded-atom method (EAM) potential is used for the interatomic energy of metal atoms~\cite{Zhou2001}. Carbon atoms inside the graphene layer interact via a simple spring potential~\cite{Sasaki1996}. Our model uses a 6--12 Lennard-Jones (LJ) potential for the metal-carbon interactions. Different estimates exist for the binding energy and equilibrium distances of Al and Cu atoms with graphene, which can be used to approximate the LJ energy $\epsilon$ and distance $\sigma$ parameters, ranging between 35--38 meV and 3.01--3.72~\AA, respectively~\cite{Vanin2010,Ma2024}. As the values for both metals are relatively close, we used $\epsilon = 35$~meV and $\sigma = 3.0135$~\AA~for both Cu and Al~\cite{Ma2024}.

Our own MD code is based on the highly parallel graphics processing unit (GPU) computing platform NVIDIA CUDA, which allows treating systems of millions of atoms on a single GPU within a reasonable time. Computations were done using NVIDIA GeForce RTX 4070, NVIDIA GeForce RTX 4090, NVIDIA RTX 4000 Ada Generation, and NVIDIA RTX 6000 Ada Generation GPUs.

\subsection{Nanoparticle preparation}
\label{sec:prep}

\begin{figure}[h]
\centering
  \includegraphics[height=4.7cm]{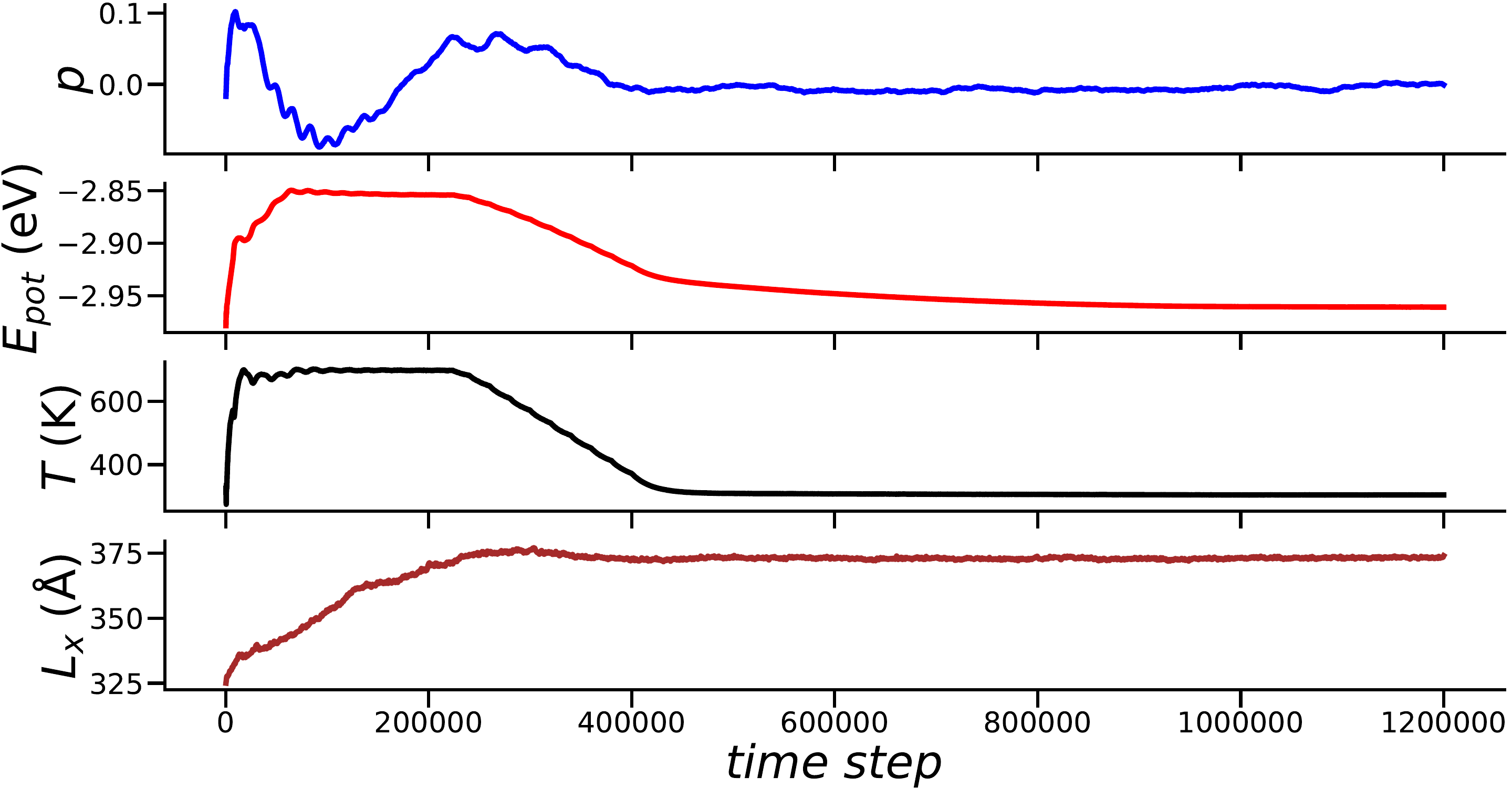}
  \caption{Time dependencies of the total impulse $p$, potential energy $E_{pot}$, temperature $T$, and lateral size $L_{x}$ of an Al NP containing 512000 atoms.}
  \label{fgr:time_dependencies}
\end{figure}

One important peculiarity of the current model is that the NPs are not prepared manually but rather obtained via a self-assembly procedure that mimics the final stage of thermal dewetting of thin metal films~\cite{Neek-Amal2009,Harsha2024,Kim2023,Badan2022nm}. Such an approach makes the model more realistic, eliminating the need to assume the surface roughness of the contact atomic layer, in contrast to macroscopic~\citep{Persson2005jpcm,Almqvist2011jmps,Pastewka2013,
Prodanov2014triblett} or some atomistic models~\cite{Gao2022fic,Oo2024}. Thermal dewetting involves depositing a thin metal film on a substrate, heating it until it melts, and then allowing it to form isolated nanoislands, which are finally cooled to form solid NPs~\cite{Badan2022nm,Harsha2024,Kim2023}. This process is reflected in the time evolution of the total impulse $p$, potential energy $E_{pot}$, system temperature $T$, and lateral size $L_{x}$ in the $x$-direction of an Al NP containing 512000 atoms, shown in Fig.~\ref{fgr:time_dependencies}.

\begin{figure*}[!h]
\centering
  \includegraphics[height=7.5cm]{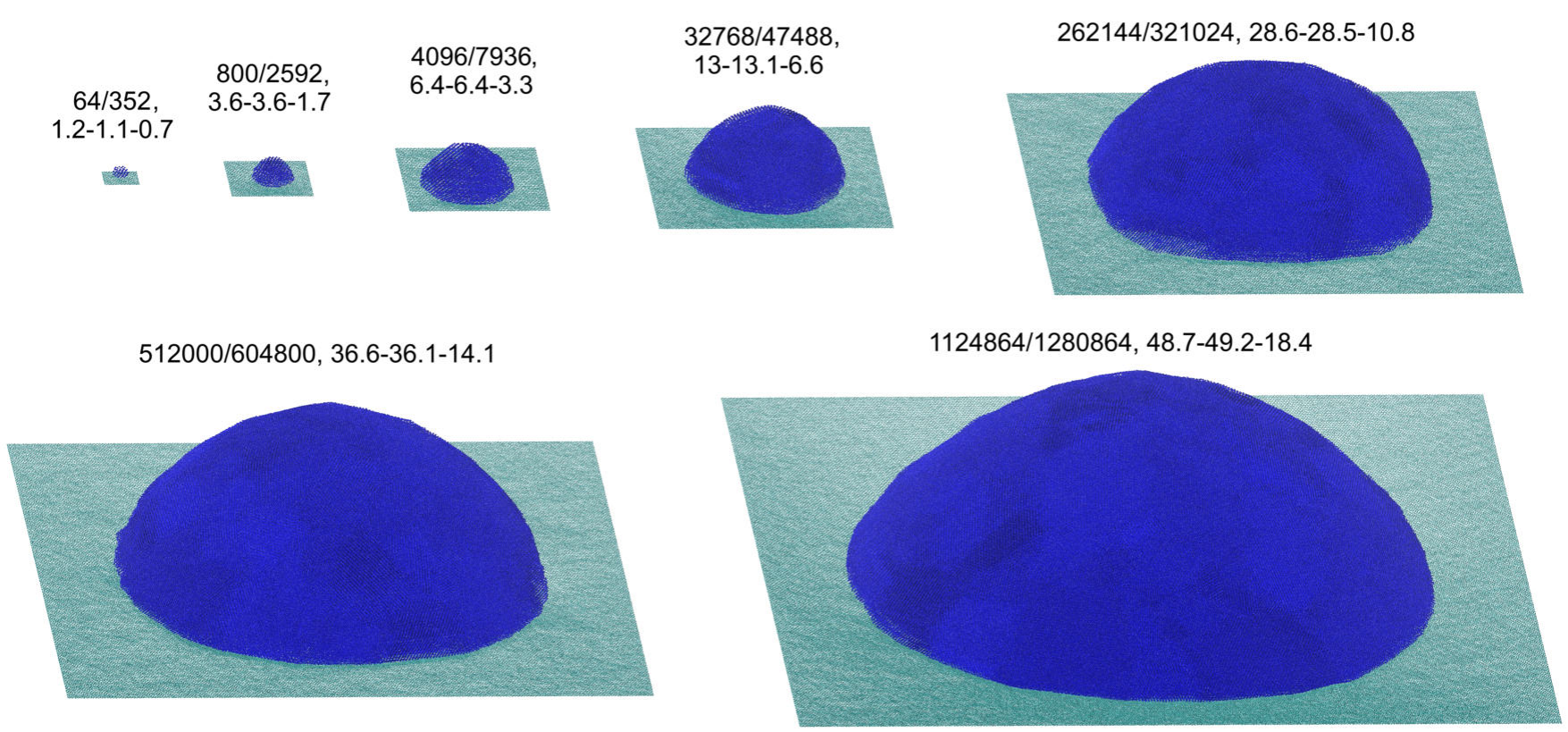}
  \caption{Al NPs of different sizes. The notations in the figure: number of metal atoms $N$/total number of atoms $N_{0}$, NP sizes $L_{x}$--$L_{y}$--$L_{z}$ in nm.}
  \label{fgr:al_scaling_side}
\end{figure*}

At first, the system is heated to a temperature $T$ in the range 700--1200~K (depending on the metal type and the NP size; Cu requires a higher $T$ because it has a higher bulk melting point). The constant high $T$ is maintained for some time, and then the system is cooled down back to about 300~K. Cooling begins based on the geometrical criterion that is when the molten droplet rises for more than 40--70~\% (depending on the NP size) of the original cuboid slab height. Heating and cooling are performed with gradual $T$ changes so that the temperature-time profile matches the experimental plots~\cite{Badan2022nm}. The shape and size of the NP can be controlled by the moment at which cooling starts: early cooling results in more square (close to the initial metal slab shape), while later cooling yields round nanoislands, suggesting the possibility of considering the shape-dependent properties of NPs~\cite{Min2008,Saleh2022,Pozzi2024,
Eker2024mol,Suhas2026}. Fig.~\ref{fgr:al_scaling_side} shows the size comparison of some round Al nanoislands used in our simulations.

After the $T$ has decreased and reached an equilibrium value of around 300~K, the system equilibrates until variations in the potential energy are less than $10^{-4}$~eV across several averaged data frames, where averaging is done over 250 time steps. The whole run typically varied between 600000 time steps for the smallest NPs, which equilibrate faster, and 1600000 for the larger ones, which equilibrate slower, for a total of 120~ps and 320~ps, respectively.

\subsection{Measurements}
\label{sec:measure}

\begin{figure}[h]
\centering
  \includegraphics[height=3.1cm]{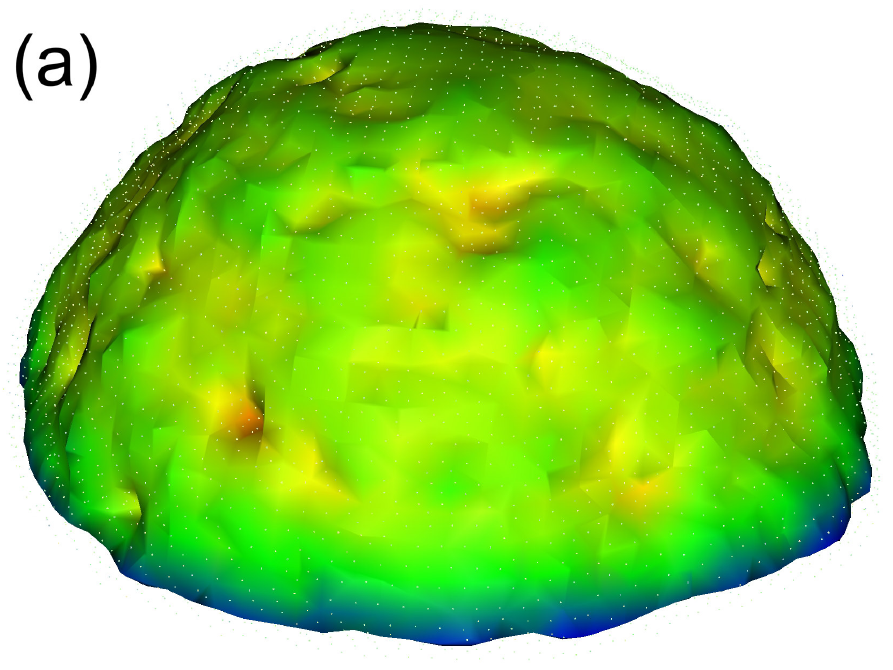}
  \includegraphics[height=2.8cm]{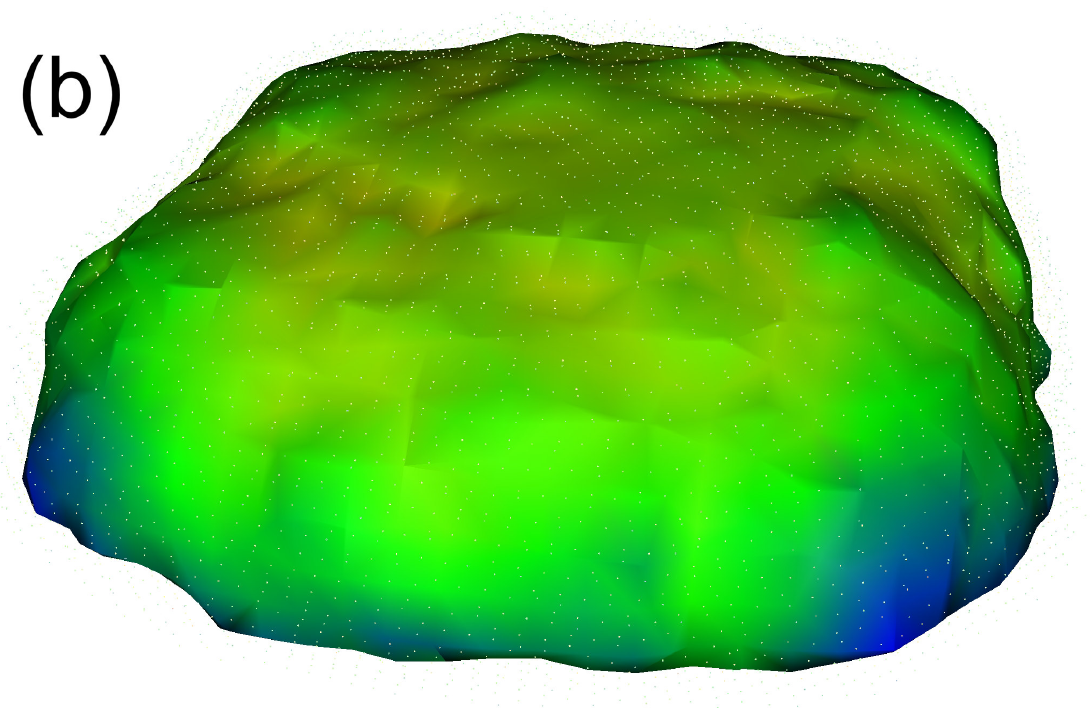}
  \caption{Example meshes used for evaluating $S$ and $V$ of Al (a), Cu (b) NPs containing 32768 atoms. Produced with CloudCompare~\cite{CloudCompare}.}
  \label{fgr:cu_al_mesh}
\end{figure}

In addition to the quantities mentioned in Sec.~\ref{sec:prep}, standard parameters, such as the total energy, the position and velocity of the NP's center of mass, and the corresponding root-mean-square (RMS) deviations, were calculated during each simulation run~\cite{Frenkel2002,Rapaport2004,Griebel2007}. Additionally, we computed morphological and CM quantities primarily using CloudCompare~\cite{CloudCompare}. The total surface area $S$ and volume $V$ of each NP were calculated using the corresponding surface mesh approximations obtained using the Poisson surface reconstruction algorithm~\cite{Kazhdan2006}, where each atom normal was estimated using the point cloud library wrapper with 10 nearest neighbors~\cite{CloudCompare}. Mesh values of $S$ and $V$ obtained this way are underestimated; typically, it was not feasible for the mesh to include surface atoms, as this resulted in manifolded edges, which do not allow a more precise evaluation of $V$. Nevertheless, such a methodology should properly reflect the trends in $S$, $V$, and the surface area-to-volume ratio $SA/V = S/V$. Example surface reconstructions of NPs containing 32768 atoms are displayed in Fig.~\ref{fgr:cu_al_mesh}. Note that the CM parameters are actually dynamic, because NPs with up to 373248 atoms (linear size of $\sim 30$~nm) could diffuse during the equilibration period (cf. ESI$^\dag$). We computed them only for a single atomic configuration ``snapshot'' of the last averaged data frame of each simulation run.

Distances between the surfaces of the NP and the substrate, i.e., the \textit{interfacial separations} (or \textit{gaps}) $u$, were evaluated using the local modeling technique, the least-squares plane point cloud-to-cloud distance calculation algorithm~\cite{Diaz2024}. Namely, $u$ for the given NP atom is the distance to the least-square best fitting plane that goes through the nearest carbon atom and its neighbors in the substrate. This is the true minimal distance from a metal atom to the corresponding approximated substrate plane, and not just the distances from a NP atom to the nearest carbon atom. Note that the equilibrated NPs had a crystalline structure. This made filtering out the bottom (contact) atomic layer of NPs easy: atoms with interfacial gaps smaller than the lattice constants of 4~\AA~for Al and 3.6~\AA~for Cu were considered to be in the contact layer. The height $h(x, y)$ of the bottom layer of a NP is just the values of the $z$-coordinates of the corresponding atoms.

\begin{figure}[h]
\centering
  \includegraphics[height=3.9cm]{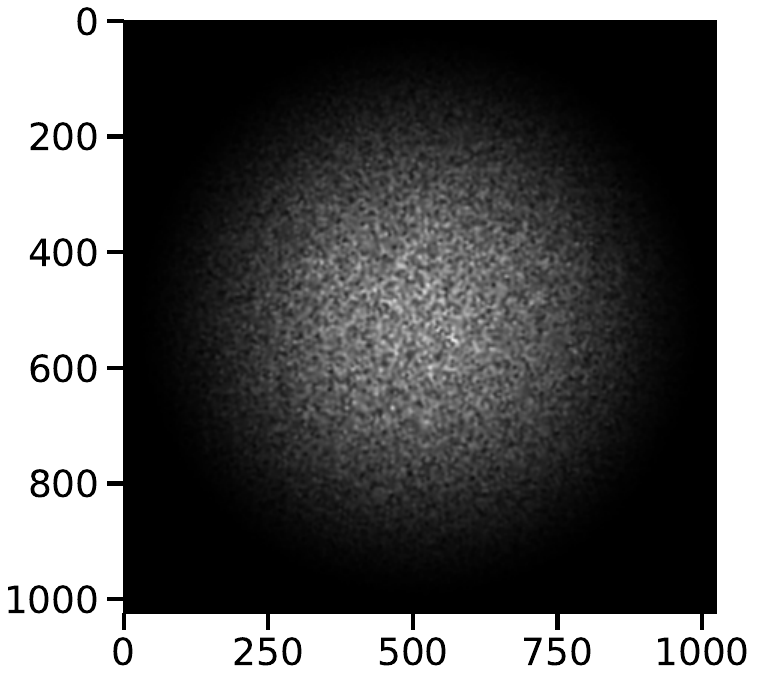}
  \includegraphics[height=3.9cm]{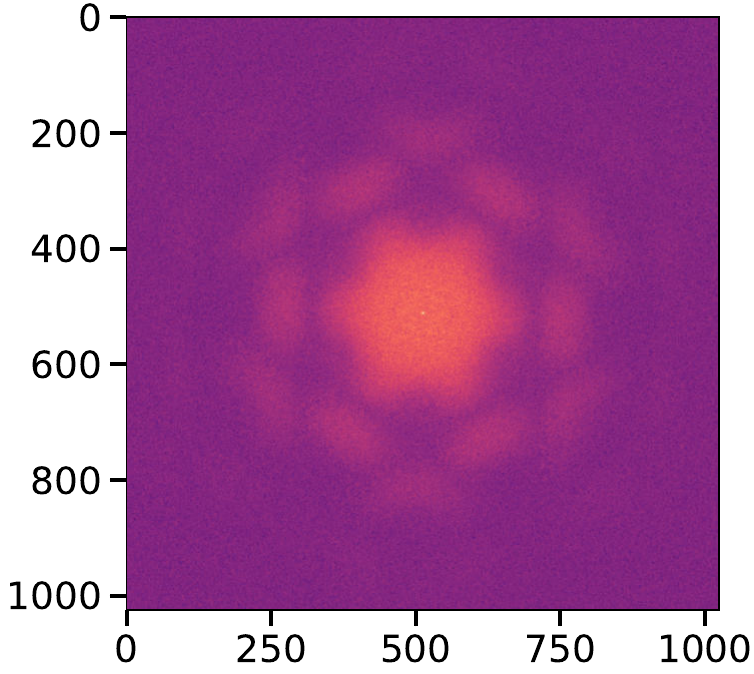}
  \caption{Windowed grey-scale image depicting the rasterized interfacial separation $u$ values and the corresponding 2D PSD $C_{u}(q_{x}, q_{y})$ for the largest Al NP. The axes depict the pixel (grid) number.}
  \label{fgr:window_fft}
\end{figure}

\begin{figure*}[!h]
\centering
  \includegraphics[height=6.5cm]{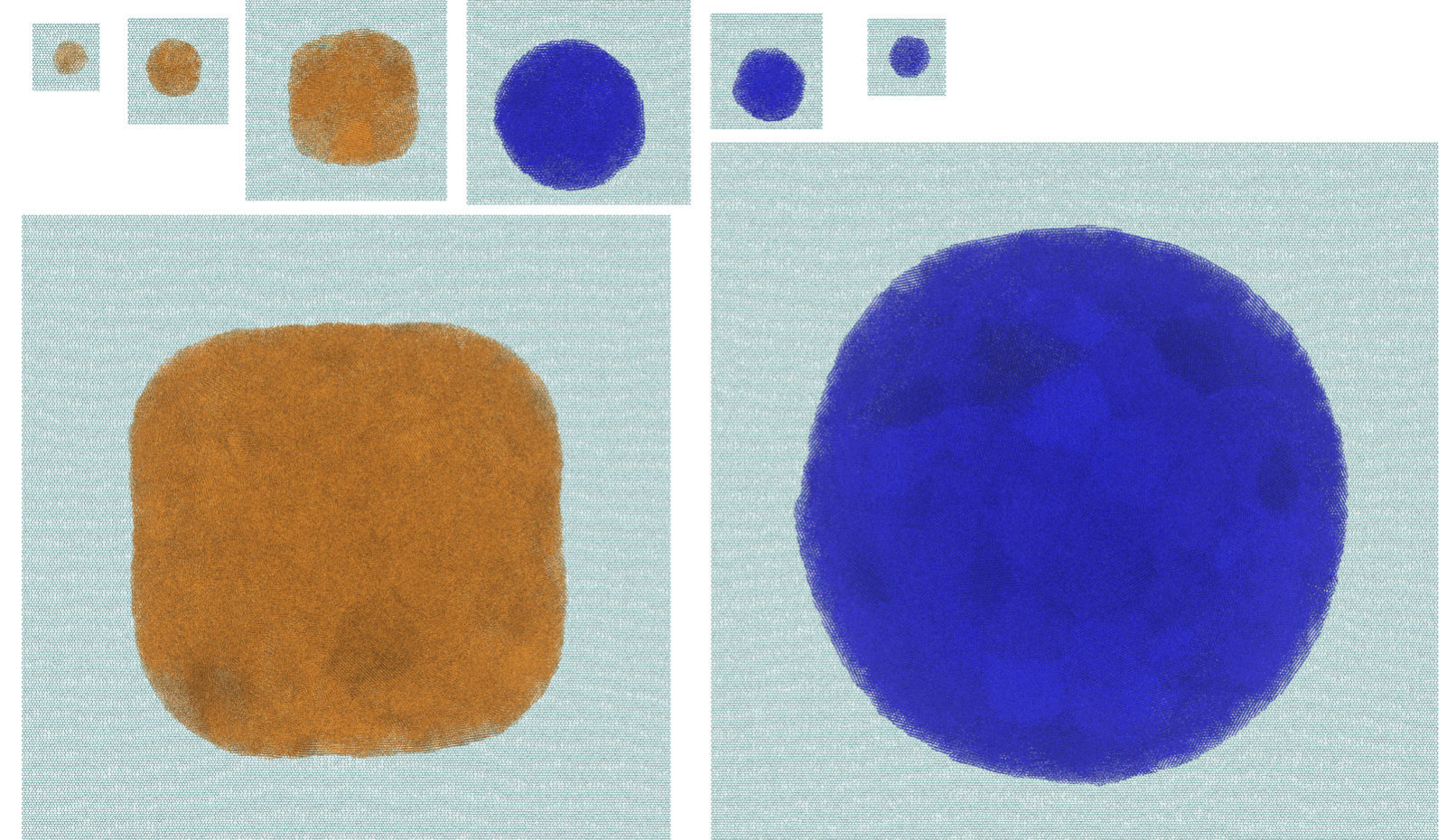}
  \caption{A top-down view of nanoislands by relative size. Top row: Cu and Al NPs containing 800, 4096, 32768 metal atoms. Bottom row: the largest NPs consisting of 1124864 metal atoms.}
  \label{fgr:cu_al_top_view}
\end{figure*}

\begin{figure*}[h]
\centering
  \includegraphics[height=4cm]{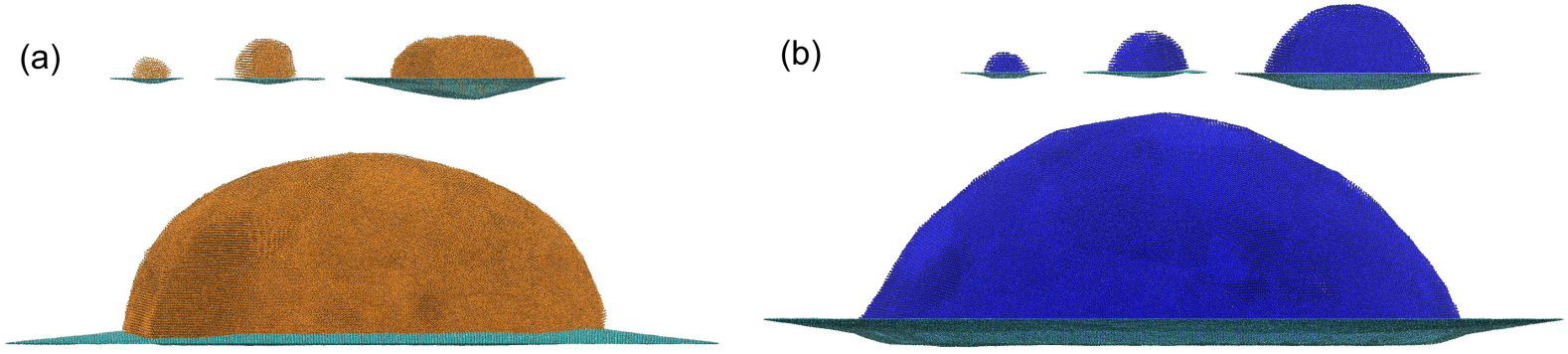}
  \caption{The side view of Cu (a), Al (b) NPs from Fig.~\ref{fgr:cu_al_top_view} (the smallest Al NPs are shown in the reverse horizontal order compared to Fig.~\ref{fgr:cu_al_top_view}).}
  \label{fgr:cu_al_side_view}
\end{figure*}

CloudCompare was also used to calculate the apparent contact area $A_{0}$ of the NPs: all NP atoms were projected onto the $xy$-plane and then fit to a 2D polygon facet. $A_{0}$ is the area, while the apparent NP perimeter is the perimeter of this polygon. The contact area $A_{bot}$ of the bottom atomic layer was also estimated as the area of the corresponding 2D polygon facet fit.

Spectral analysis was done using the 2D PSDs $C_{h}(q_{x}, q_{y})$, and $C_{u}(q_{x}, q_{y})$ of the atomic heights $h(x, y)$ and gaps $u(x,y)$ of the bottom atomic NP layer, respectively~\cite{Persson2005jpcm,Jacobs2017stmp,Jacobs2022mrsbul}. An example of the $C_{u}(q_{x}, q_{y})$ heatmap is shown in Fig.~\ref{fgr:window_fft} (right). A PSD of a 2D function, e.g., $h(x, y)$, is defined as follows~\cite{Jacobs2017stmp}:
\begin{equation}
  C_{h}(q_{x}, q_{y}) = (L_{x} L_{y})^{-1} \bigl\lvert \tilde{h}(x, y) \bigr\rvert^{2},
  \label{eq:psd_h_x_y}
\end{equation}
where $q_{x}$, $q_{y}$ are the spatial frequencies and $L_{x}$, $L_{y}$ are the linear sizes of the region of interest in the $x$- and $y$-direction, respectively, and $\tilde{h}(x, y)$ is the Fourier transform of the height profile:
\begin{equation}
  \tilde{h}(x, y) = \frac{L_{x} L_{y}}{N_{x} N_{y}} \sum_{x,y} h(x, y) e^{-i(q_{x} x + q_{y} y)},
  \label{eq:ft_h_x_y}
\end{equation}
where $N_{x}$, $N_{y}$ are the numbers of grid points in the $x$- and $y$-direction, respectively. The same expressions hold for the gap where $u$ substitutes $h$. In this work, $q_{x} = 1/\lambda_{x}$ and $q_{y} = 1/\lambda_{y}$ denote the frequencies, not the wave-vectors; the factor $2 \pi$ is omitted for convenience in transforming frequencies into the corresponding wavelengths $\lambda_{x}$ and $\lambda_{y}$. The pixel frequencies were rescaled, and the frequency unit $1.76 \cdot 10^{-3}$~\AA$^{-1}$~was chosen so that a frequency of 400 corresponds to the graphene covalent bond length of 1.42~\AA, regardless of NP size.

In practice, the standard discrete Fourier transform (or the Fast Fourier transform (FFT) that we used) requires a regular grid of values, while the surface atoms of the NP form irregular configurations \cite{Khom_PhysRevE_19}. To overcome this difficulty, we rasterized the scalar fields $h(x,y)$ and $u(x,y)$ into a grey-scale image, which was transformed using the FFT. Each pixel of the image is a point of the input grid, and its color is the interpolated value of the original $h(x, y)$ and $u(x, y)$ obtained in the MD simulations. Before the interpolation, the average value of the scalar field was subtracted from each point to remove the constant signal contribution to the spectrum. Each such picture had an odd width and height of $N_{x} = N_{y} =$1025 pixels. Besides the regular grid, FFT requires periodic input that doesn't have sharp edges. Otherwise, artifacts, e.g., ringing, can be observed in the transformed output. Windowing is a standard technique used to make the input image appear periodic to the FFT. We utilized Hann windowing~\cite{Prabhu2013}. Note that $L_{x}$, $L_{y}$ in Eqs.~(\ref{eq:psd_h_x_y}) -- (\ref{eq:ft_h_x_y}) are the lateral sizes of the NP, which are almost identical to the sizes of the rasterized contact layer. Fig.~\ref{fgr:window_fft}~(left) exemplifies a windowed rasterized image of $u(x, y)$ values of the largest Al NP, while Fig.~\ref{fgr:window_fft}~(right) is its 2D PSD heatmap.

In this study, we also use $C_{h}(q)$ and $C_{u}(q)$, which are analogs of the isotropic PSD~\cite{Persson2005jpcm,Jacobs2017stmp}. These are calculated as histograms of $C_{h}(q_{x}, q_{y})$ and $C_{u}(q_{x}, q_{y})$ values that correspond to the radial spatial frequency $q = \sqrt{q_{x}^{2} + q_{y}^{2}}$. Even though 2D PSDs are not isotropic in general, $C(q)$ still may give valuable insight into the PSD spectrum behavior, as in most cases 2D PSDs of NPs turn out to be radially symmetric.

Finally, we note that the classical MD approach does not compute exact trajectories in phase space~\cite{Frenkel2002}. Additionally, the statistical spread of surface roughness can exist, as in macroscopic surfaces without a roll-off vector in their surface-height PSD, so ensemble averaging is needed in such cases~\cite{Persson2005jpcm,Rodriguez2025triblett}. To provide some statistics, we ran 3--4 simulations for each NP size. The spread of the values is reflected in the corresponding error bars in the scaling plots. Missing error bars means that the spread is smaller than the symbol size for that value.

\section{Results and discussions}

\subsection{Morphological properties}
\label{sec:morphology}

Even though all the NPs were produced using the same procedure described in Sec.~\ref{sec:prep}, Al and Cu, in general, have different shapes, as is seen in Fig.~\ref{fgr:cu_al_top_view} and Fig.~\ref{fgr:cu_al_side_view}, which depict Cu and Al NPs containing 800, 4096, 32768, and 1124864 metal atoms. Al NPs are round, while the Cu nanoislands are close to the ``square'' shape in the $xy$-plane. As expected, Al NPs are larger than the Cu ones with the same number of atoms because of the larger Al lattice constant $a_{Al} = 4.08$~\AA~compared to $a_{Cu} = 3.62$~\AA. The ``square'' shape of Cu NPs may be attributed to the higher surface density of Cu atoms and hence higher interaction energy density and a stronger surface interaction of Cu NPs. This can make the rearrangement of molten Cu interface atoms slower than that of Al, even with the same interaction energy with graphene. As a result, the Cu NPs retained a shape closer to the original rectangular metal slab configuration.

\begin{figure}[h]
\centering
  \includegraphics[height=3.8cm]{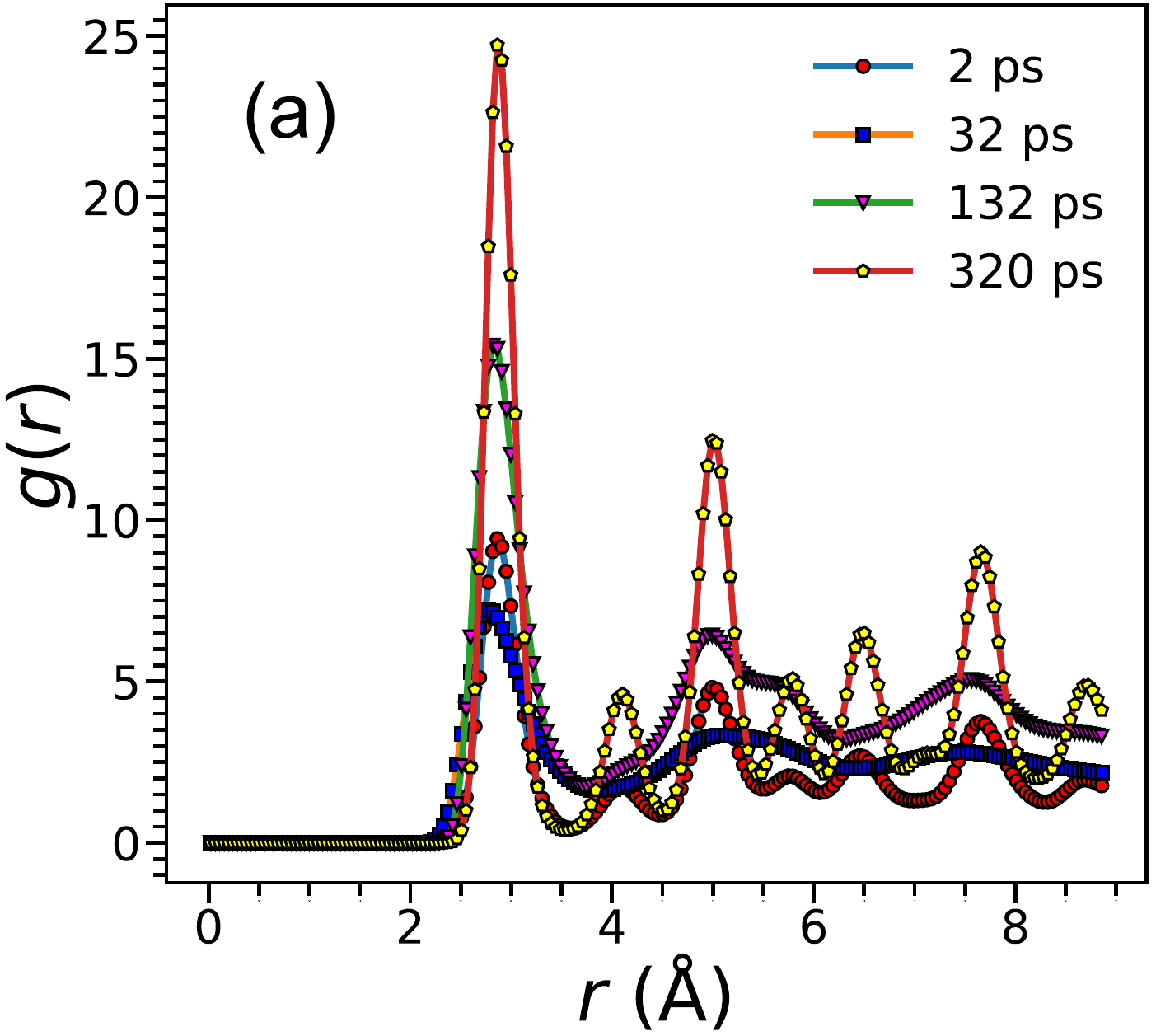}
  \includegraphics[height=3.8cm]{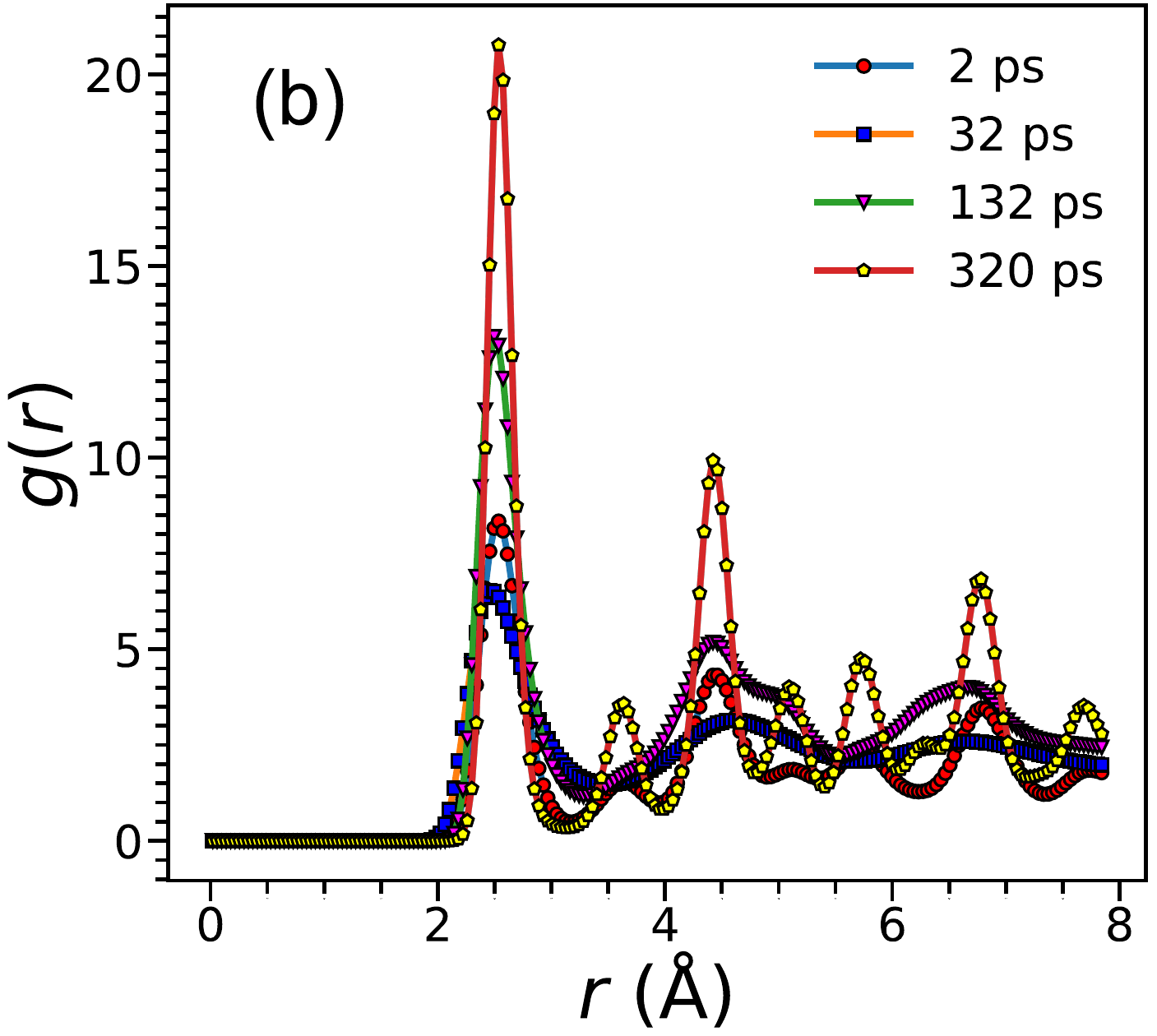}
  \caption{Time evolution of RDFs of the largest Al (a), Cu (b) NPs.}
  \label{fgr:rdfs}
\end{figure}

\begin{figure}[!h]
\centering
  \includegraphics[height=3.75cm]{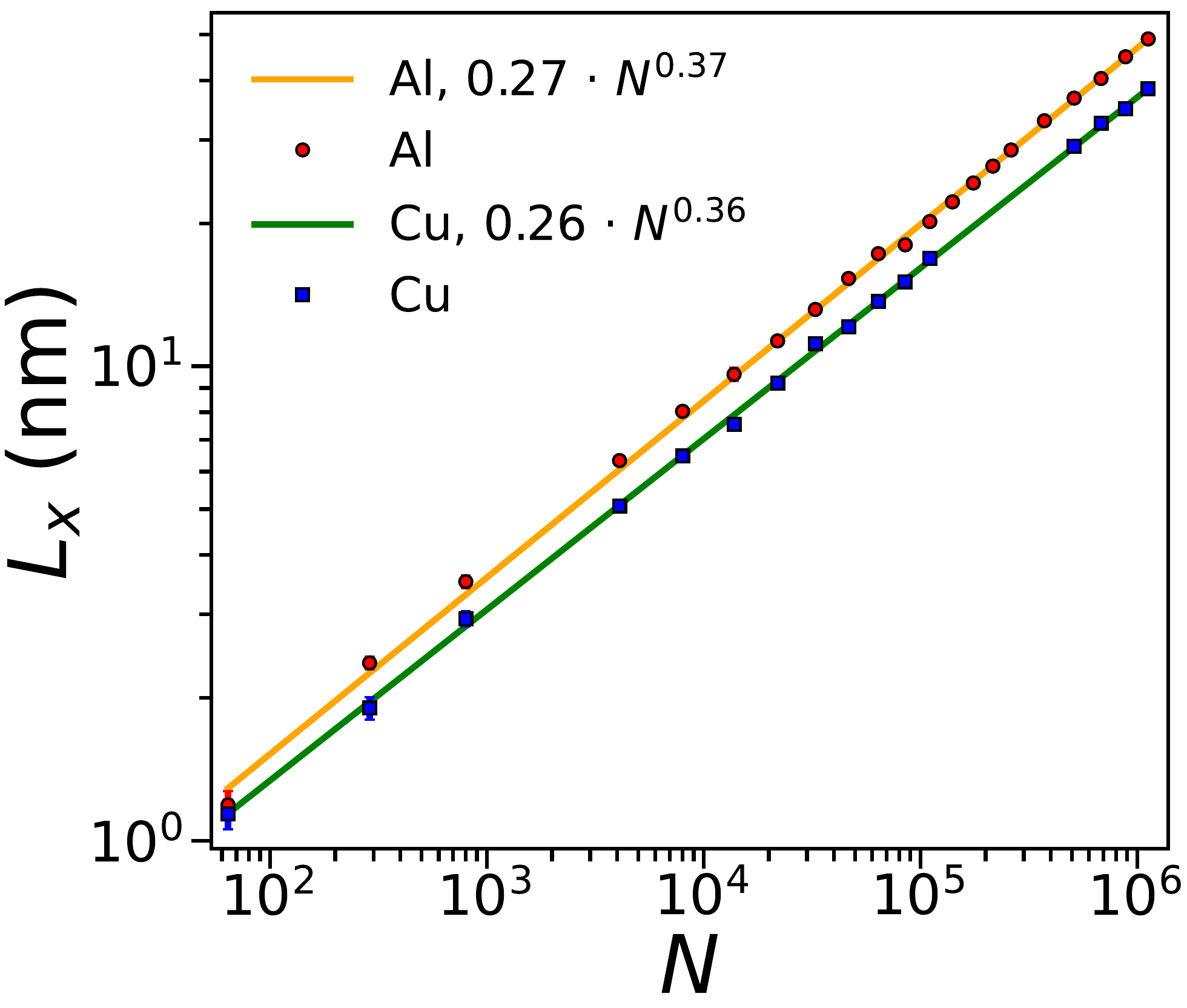}
  \includegraphics[height=3.75cm]{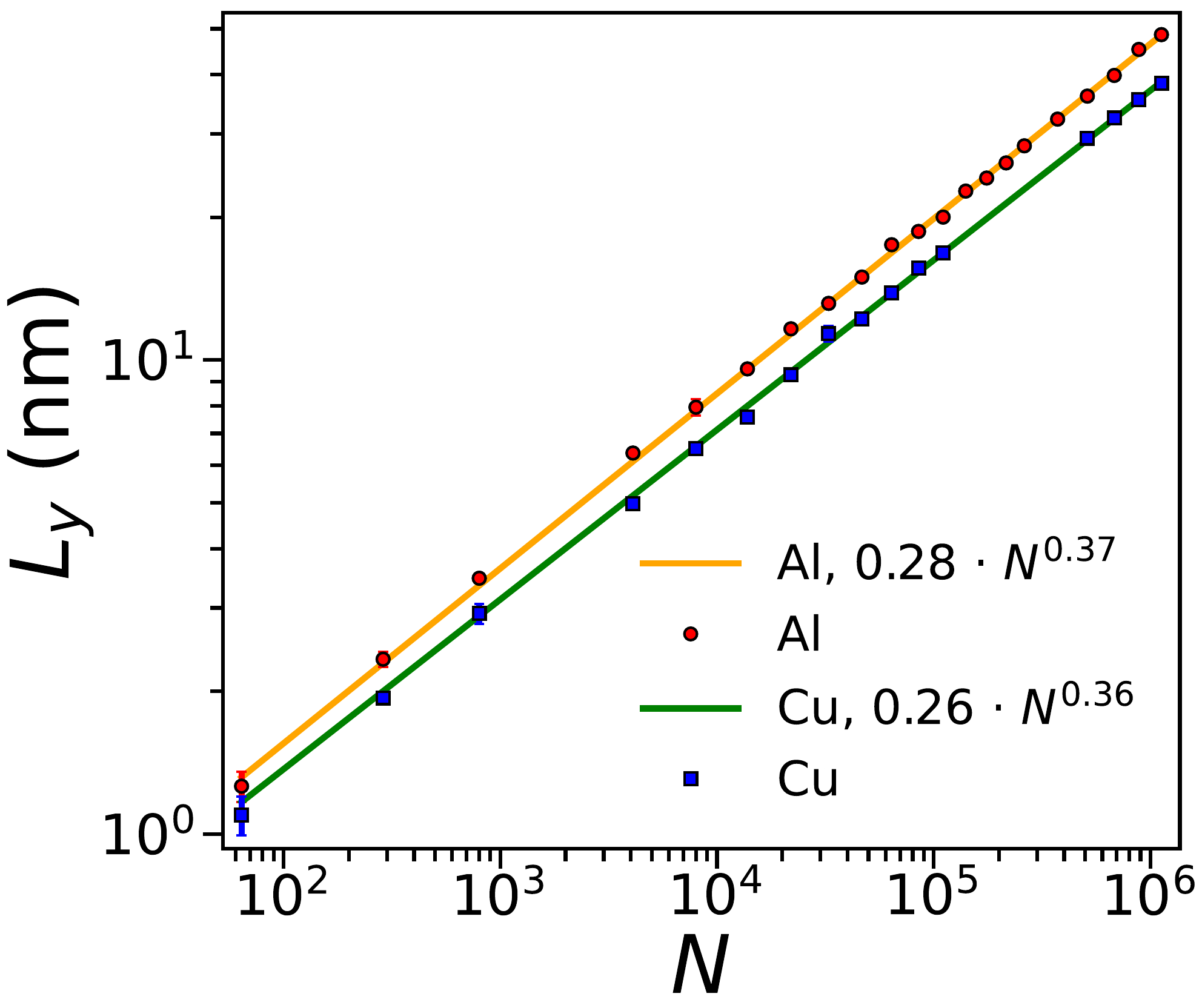}
  \caption{Lateral NP sizes along the $x$- and $y$-directions vs the number $N$ of metal atoms of NPs, with corresponding power-law fits.}
  \label{fgr:size_xy_vs_atoms_count}
\end{figure}

\begin{figure}[!h]
\centering
  \includegraphics[height=3.75cm]{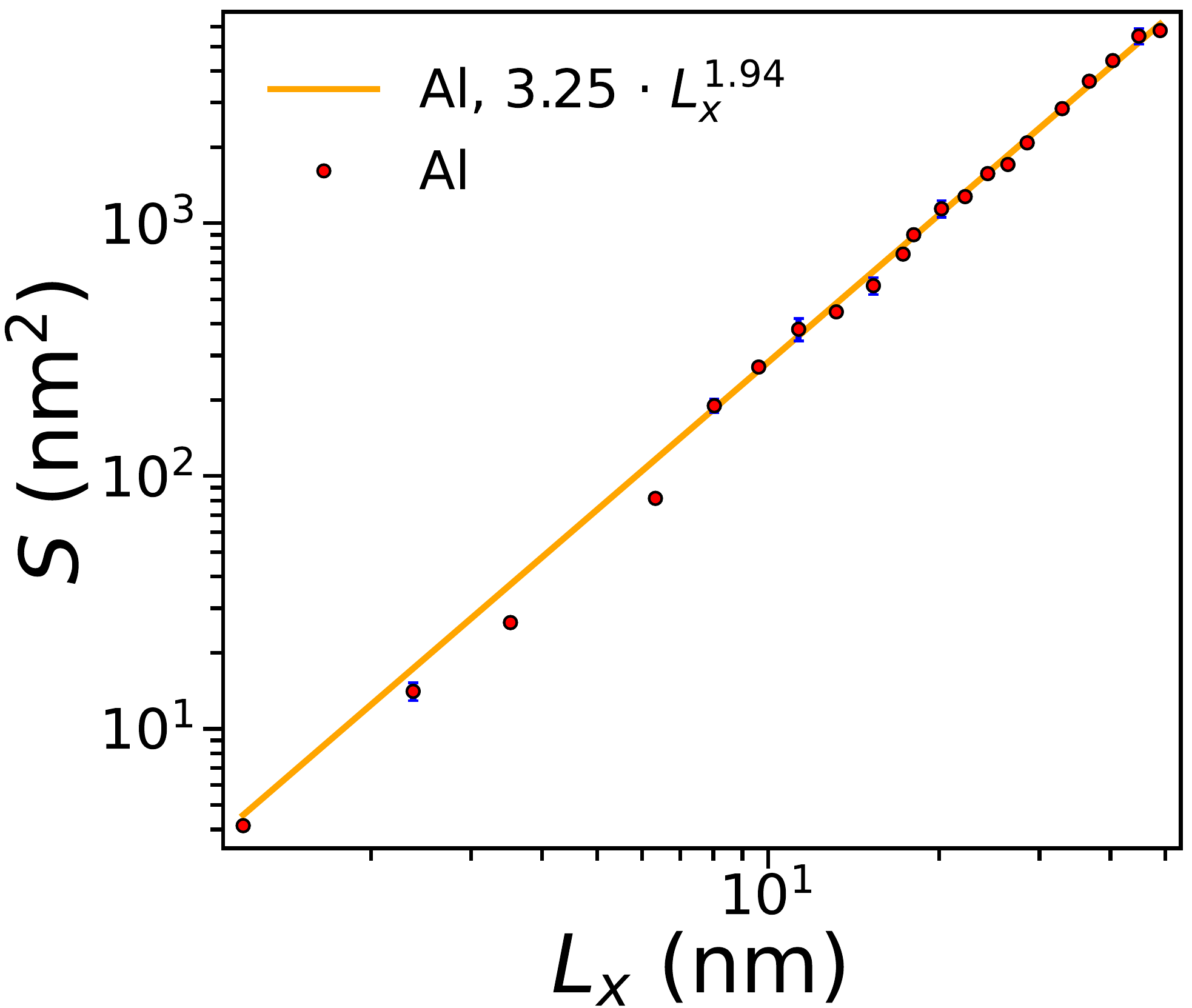}
  \includegraphics[height=3.75cm]{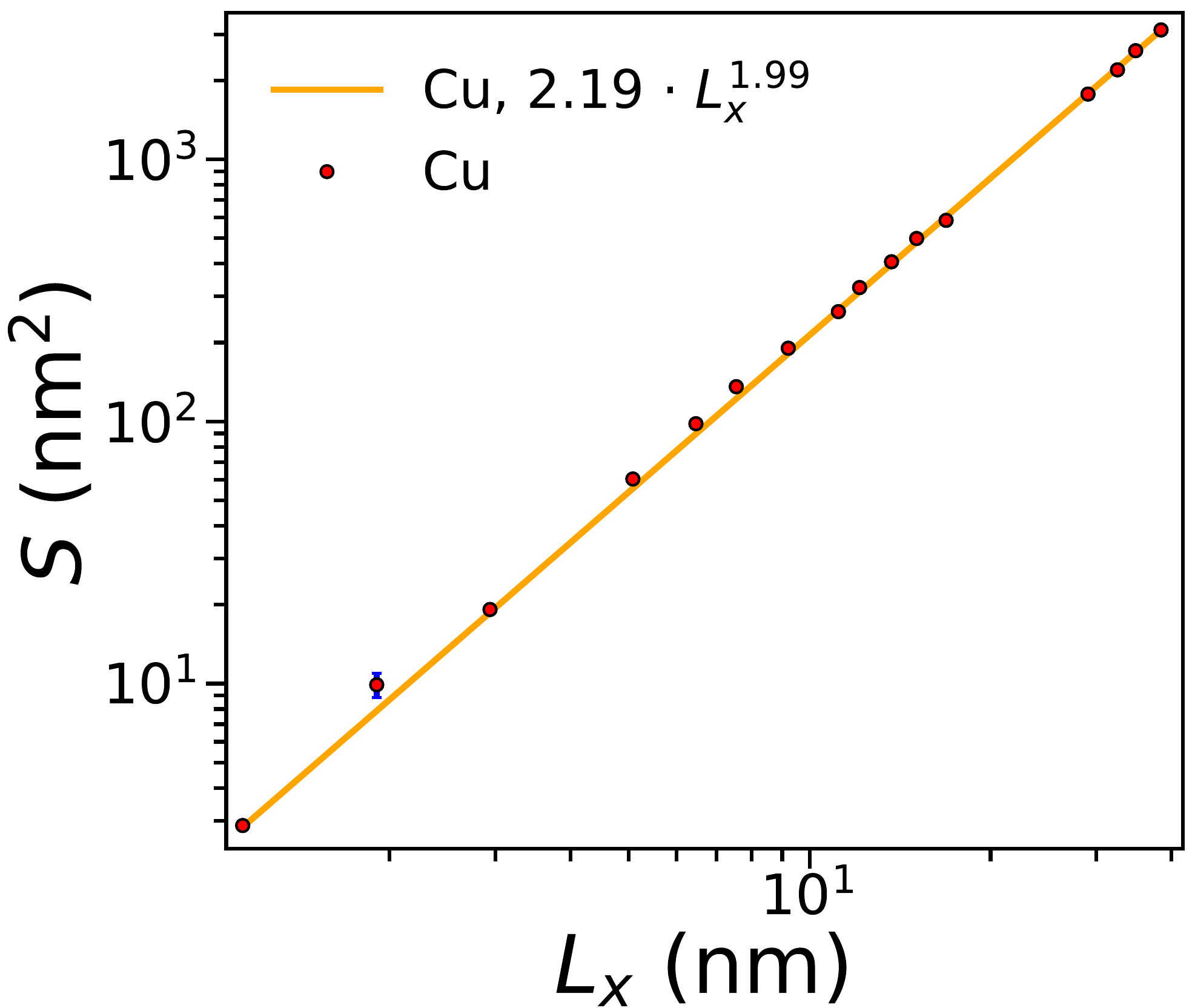}
  \caption{Total surface $S$ of NPs vs $L_{x}$, with corresponding power-law fits.}
  \label{fgr:surface_area_vs_size_x}
\end{figure}

\begin{figure}[!h]
\centering
  \includegraphics[height=3.75cm]{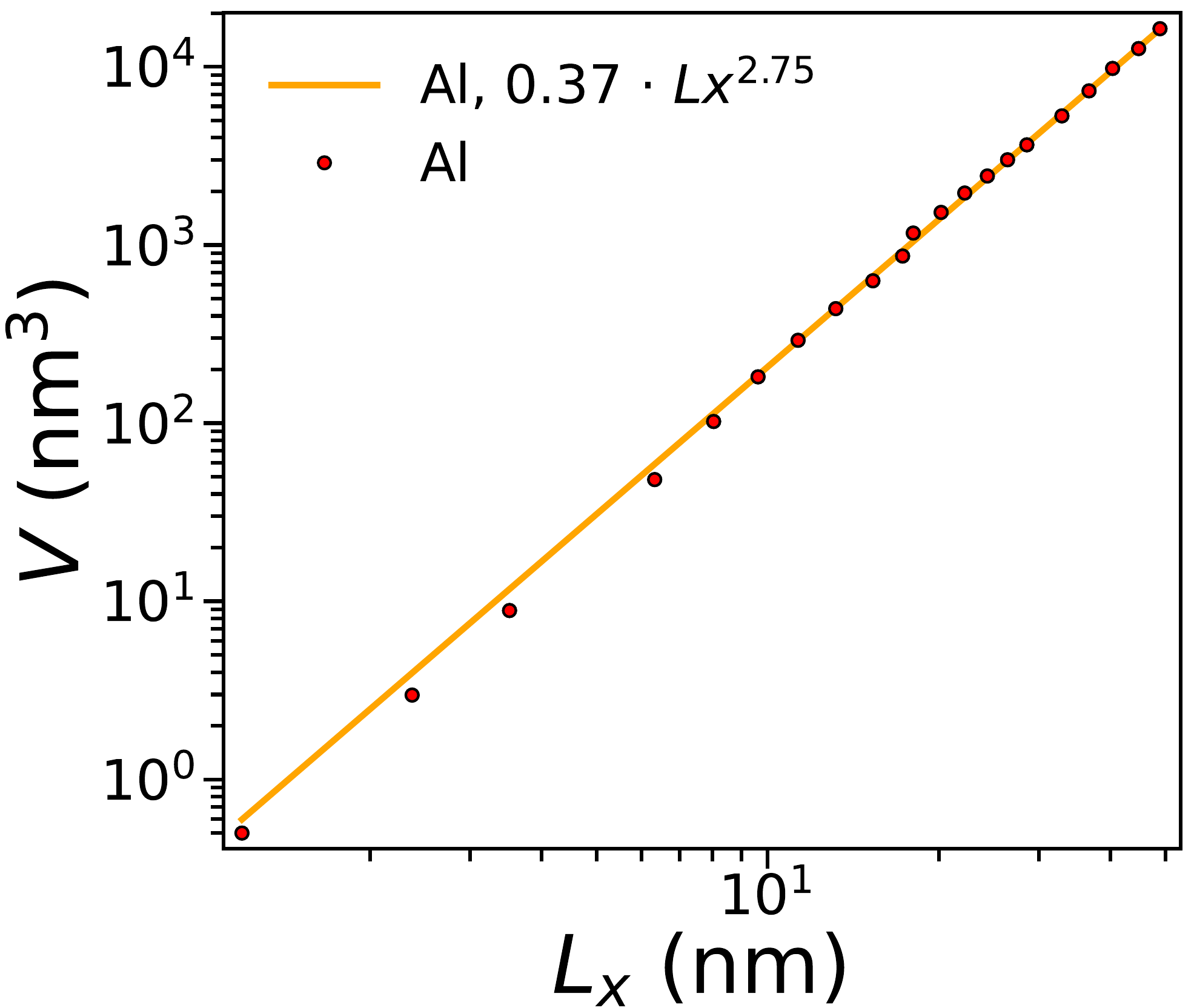}
  \includegraphics[height=3.75cm]{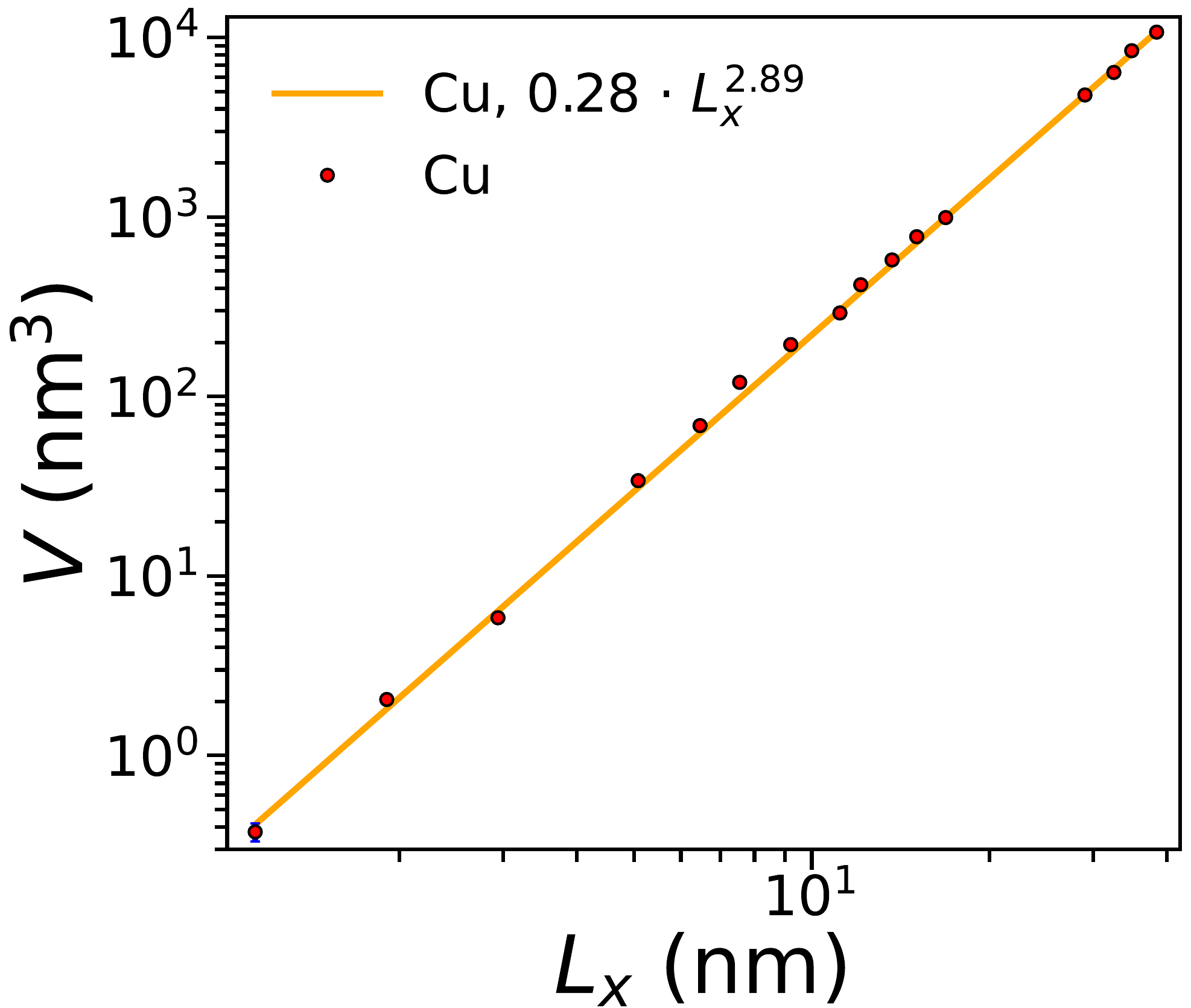}
  \caption{Total volume $V$ of NPs vs $L_{x}$, with corresponding power-law fits.}
  \label{fgr:v_vs_size_x}
\end{figure}

\begin{figure}[!h]
\centering
  \includegraphics[height=3.7cm]{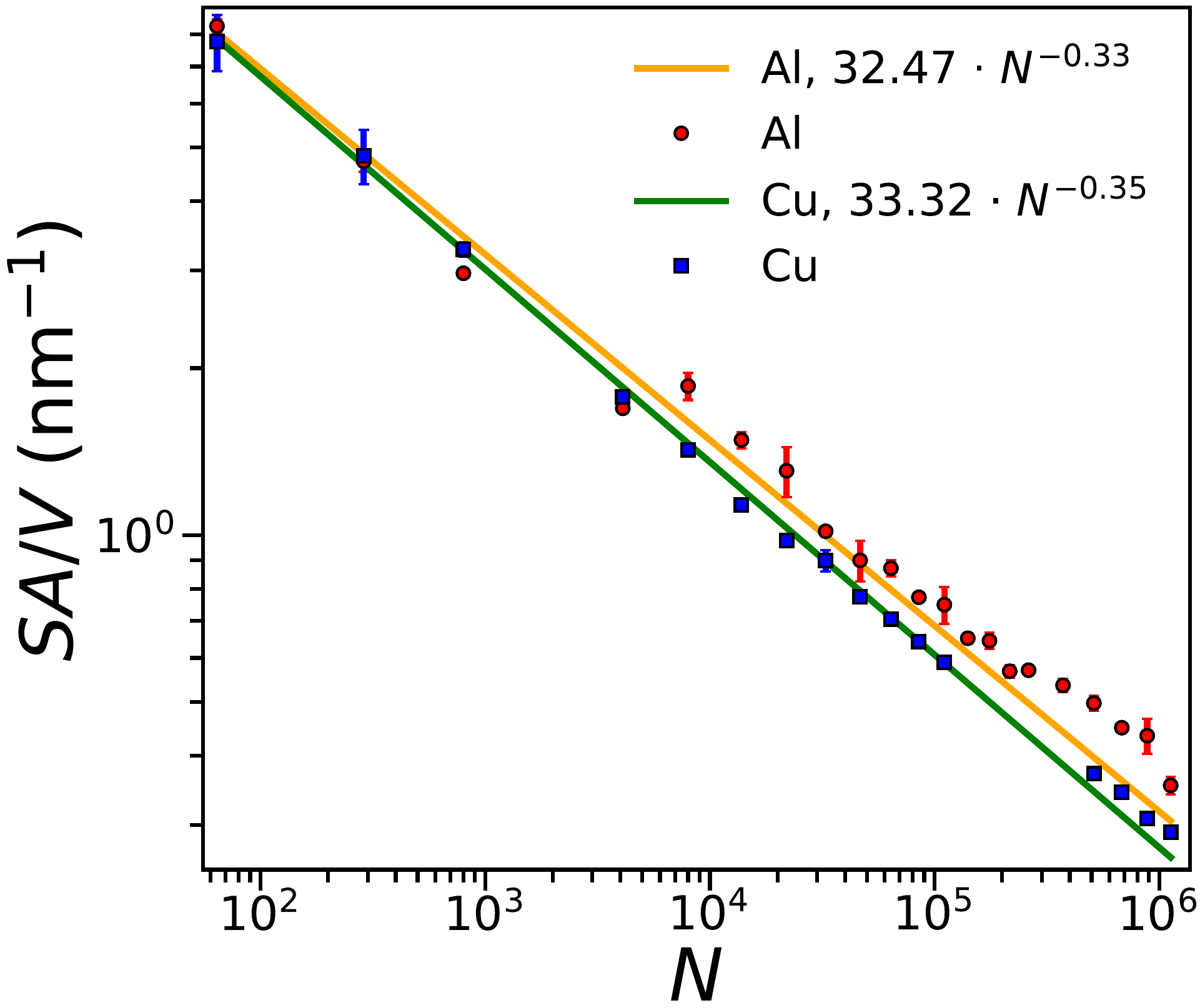}
  \includegraphics[height=3.7cm]{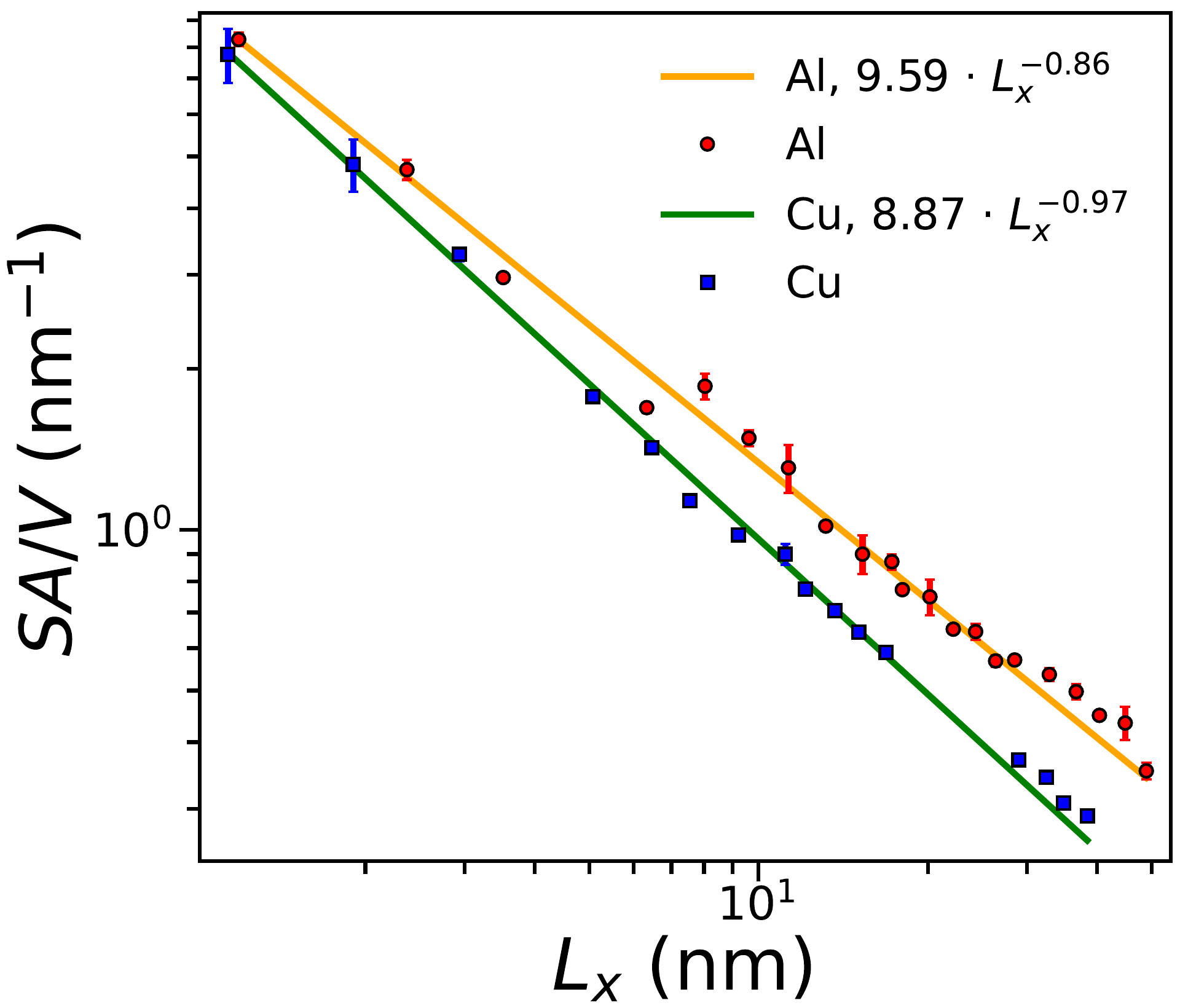}
  \caption{$SA/V$ vs $N$, $L_{x}$, with corresponding power-law fits.}
  \label{fgr:sav_vsn_size_x}
\end{figure}

Equilibrated NPs have a polycrystalline atomic structure, which can be inferred from their radial distribution functions (RDFs). Examples of time evolution of RDFs of the largest Al and Cu NPs are shown in Fig.~\ref{fgr:rdfs}. At 32~ps, the NPs are already molten since their RDFs are smeared, corresponding to the liquid state~\cite{Dong2003jpcm}. After cooling, the smooth RDF maxima transform into spikes that become clear at 320~ps (the time instance of the last averaged data frame of the MD run). The first RDF spike corresponds to the atomic nearest neighbor distance of 2.886~\AA~and 2.556~\AA~for Al and Cu, respectively~\cite{Zhou2001}.


\begin{figure*}[!h]
\centering
  \includegraphics[height=7.6cm]{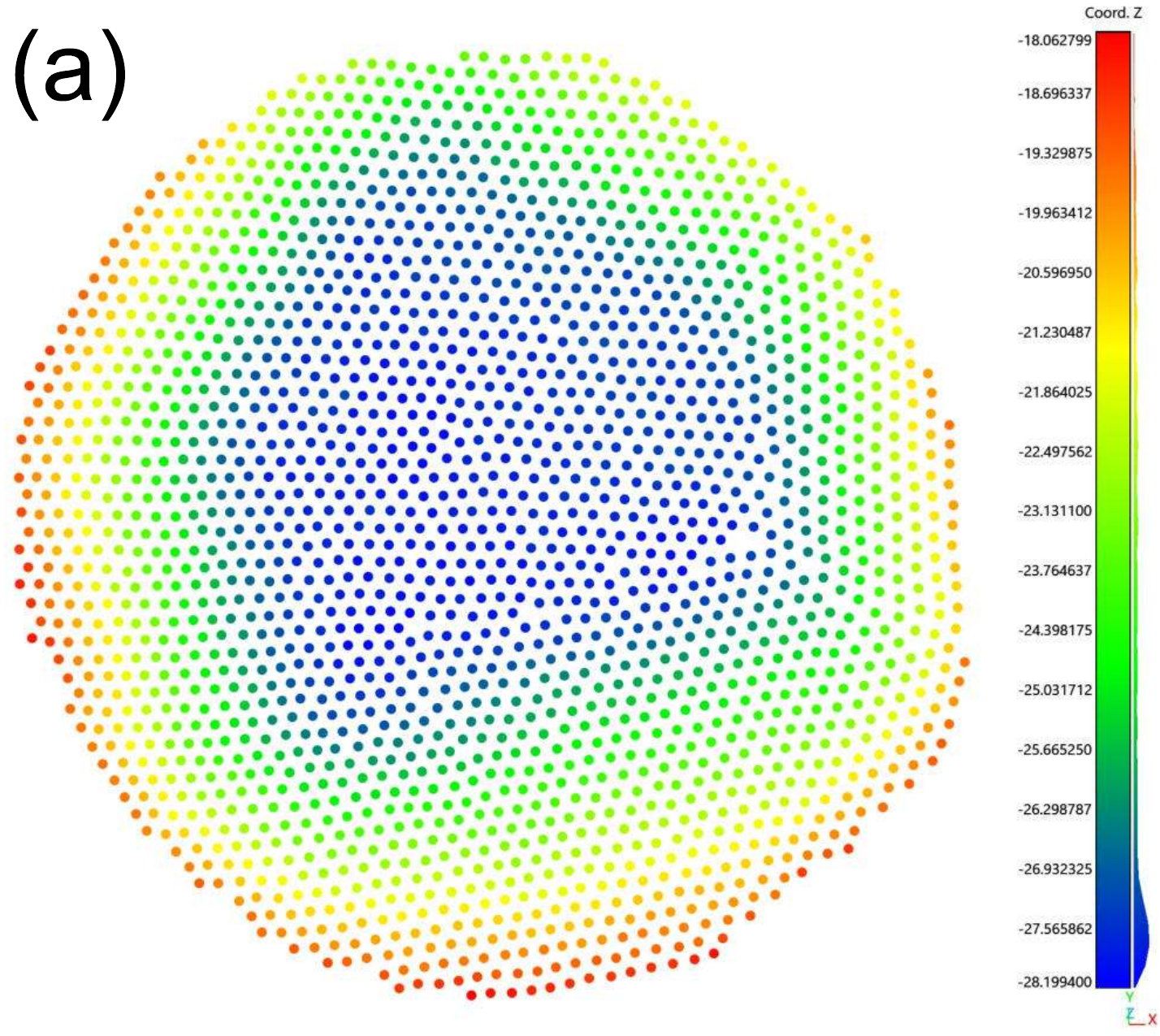}
  \includegraphics[height=7.6cm]{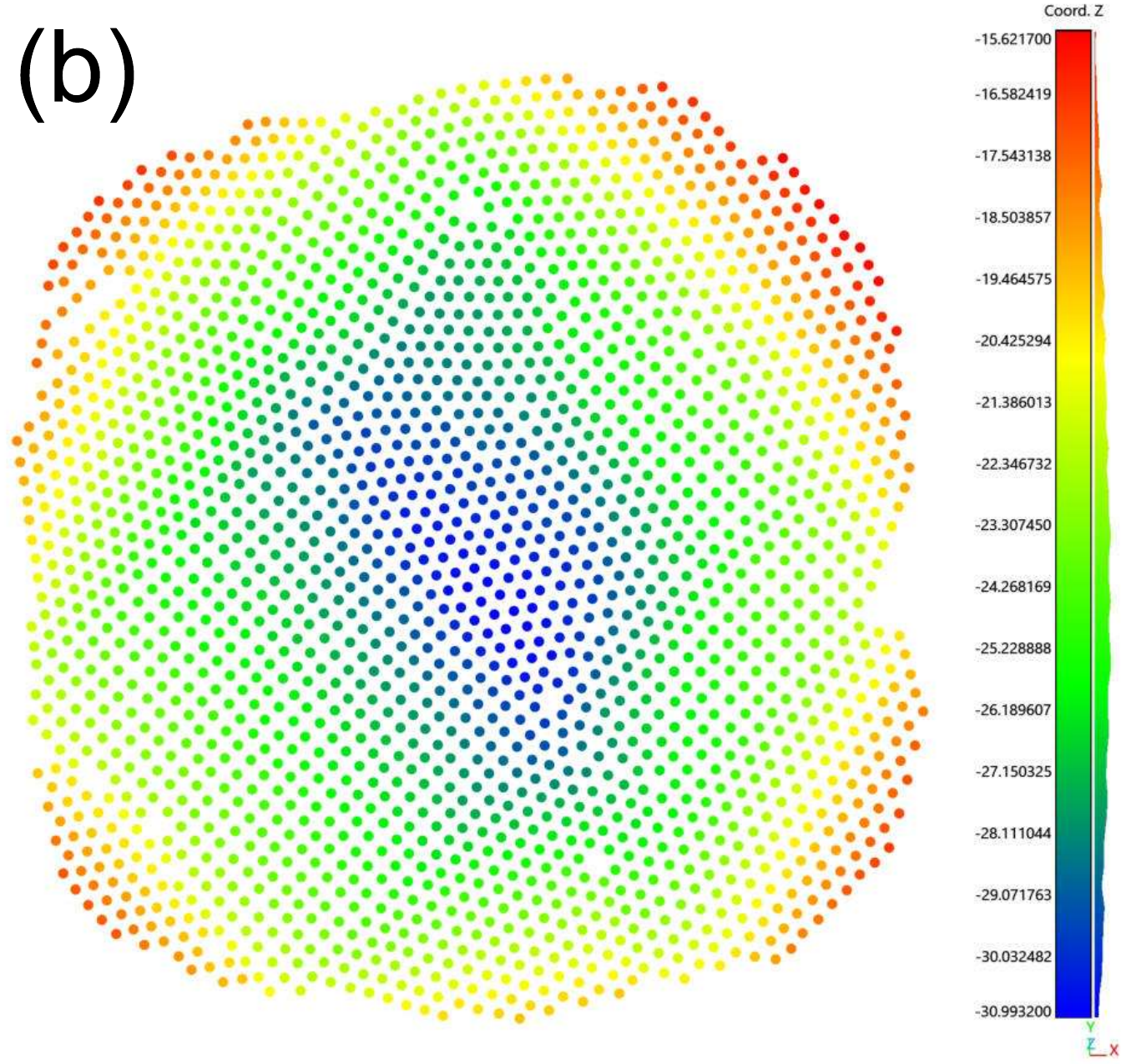}
  \includegraphics[height=1cm]{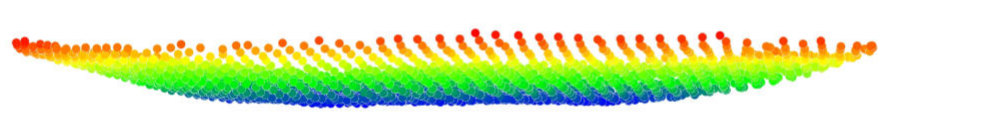}
  \includegraphics[height=1cm]{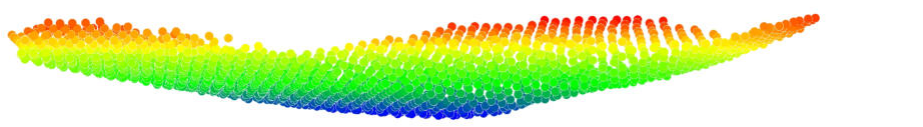}
    \caption{Top and side views of the $h(x, y)$ heat maps for Al (a), Cu (b) NPs with 32768 atoms (the largest NPs in the top rows in Fig.~\ref{fgr:cu_al_top_view} and Fig.~\ref{fgr:cu_al_side_view}). The blue atoms have the smallest height, the red ones have the largest $h(x, y)$.}
  \label{fgr:side_top_view_bot_height_map}
\end{figure*}

\begin{figure*}[!h]
\centering
  \includegraphics[height=7.6cm]{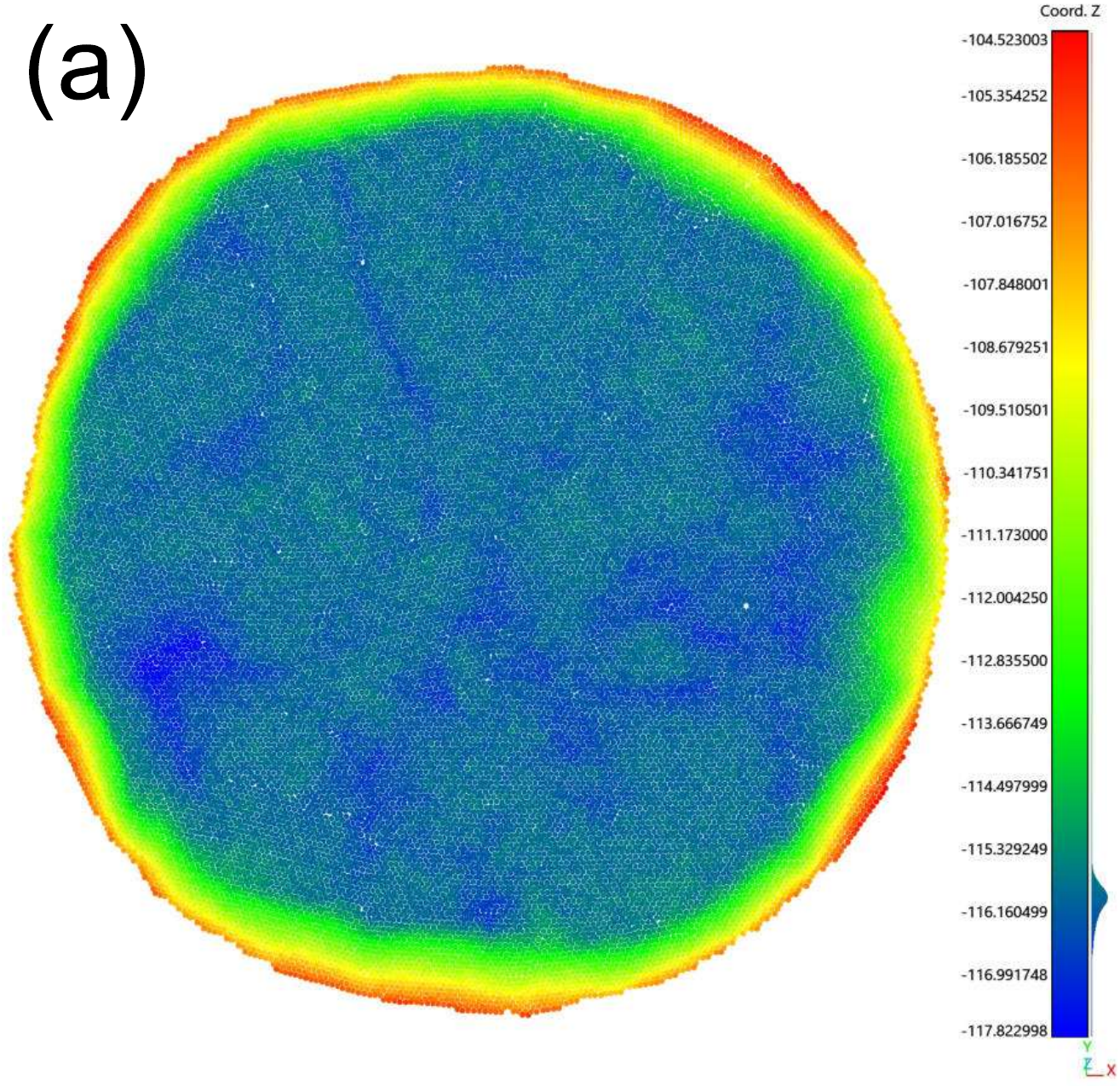}
  \includegraphics[height=7.6cm]{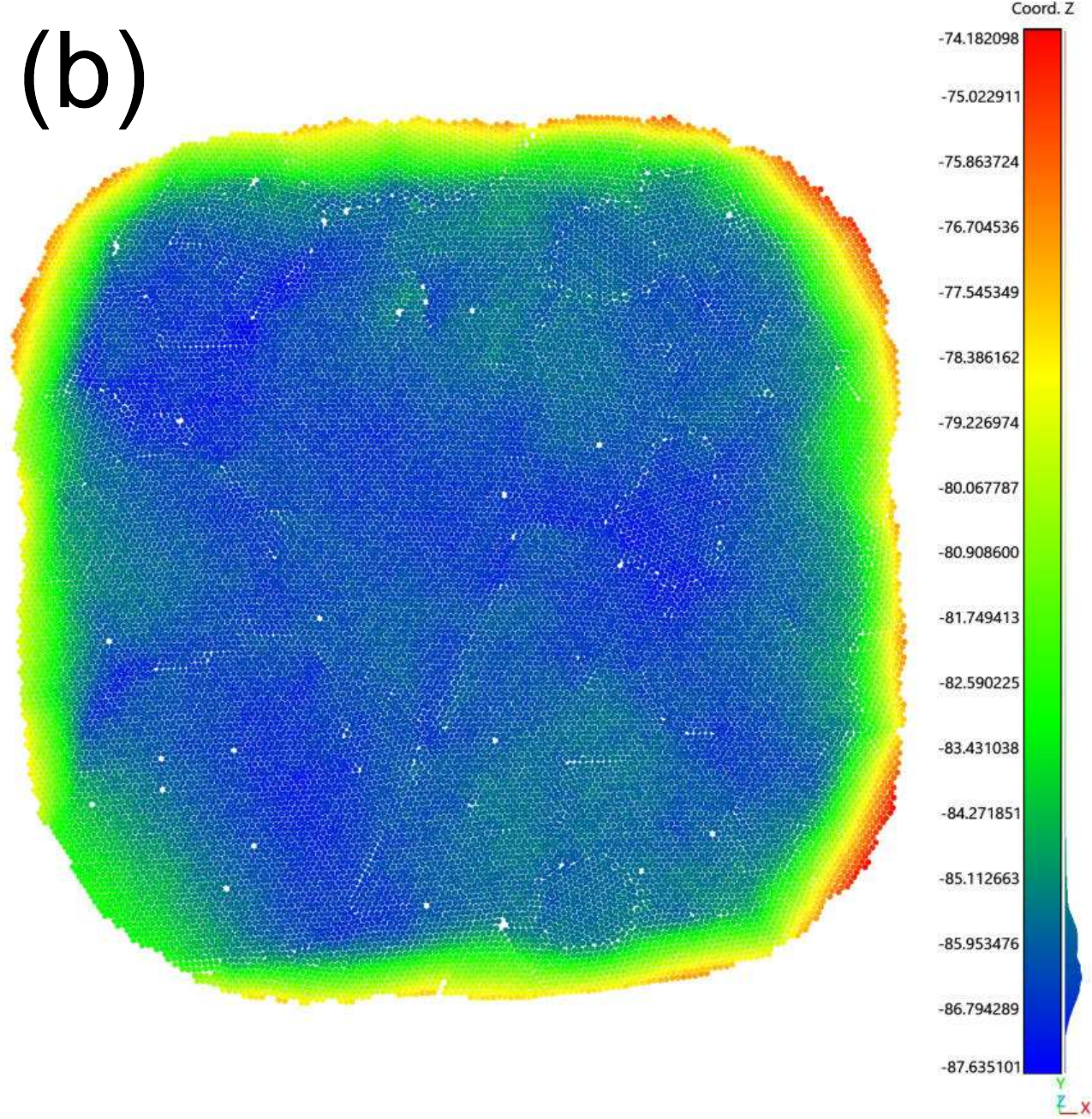}
  \includegraphics[height=0.45cm]{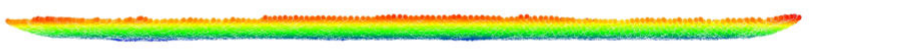}
  \includegraphics[height=0.4cm]{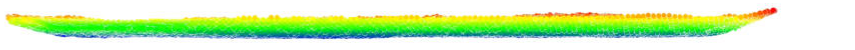}
    \caption{Height heat maps of the bottom atomic layer of the largest Al and Cu NPs. Top and side views. The blue color depicts the smallest value, the red color shows  the largest one.}
  \label{fgr:side_top_view_bot_height_map_largest}
\end{figure*}

The lateral NP sizes $L_{x}$, $L_{y}$ scale with $N$ approximately as a power of 1/3, see Fig.~\ref{fgr:size_xy_vs_atoms_count}. Namely, the power-law fits give a power of around 0.36--0.37. As $L_{x}$ and $L_{y}$ are approximately equal and scale almost identically, we will further use $L_{x}$ as a characteristic linear size in other scaling relationships. In particular, the NP total surface area $S$ and volume $V$ of most NP sizes are close to the quadratic and cubic dependencies on $L_{x}$, respectively. More precisely, the power-law fitting yields powers of 1.94 and 2.75 for Al, 1.99 and 2.99 for Cu $S$ and $V$, respectively. The exceptions are the smaller NPs whose size is $\lesssim 10^4$ atoms. Namely, for Al, $S$ and $V$ scale slower than the mentioned powers for $L_{x} \lesssim 8$~nm. For $S$ of the Cu nanoisland, the scaling discrepancy is observed for $L_{x} \lesssim 3$~nm, while the $V$ scaling difference is negligible. As expected, $SA/V$ scales approximately inversely with $L_{x}$ with fitting powers of $-0.86$ and $-0.97$ for Al and Cu, respectively, cf. Fig.~\ref{fgr:sav_vsn_size_x}. The apparent perimeter of all the NPs scales linearly with $L_{x}$ (not shown here).

\begin{figure*}[!h]
\centering
  \includegraphics[height=7.6cm]{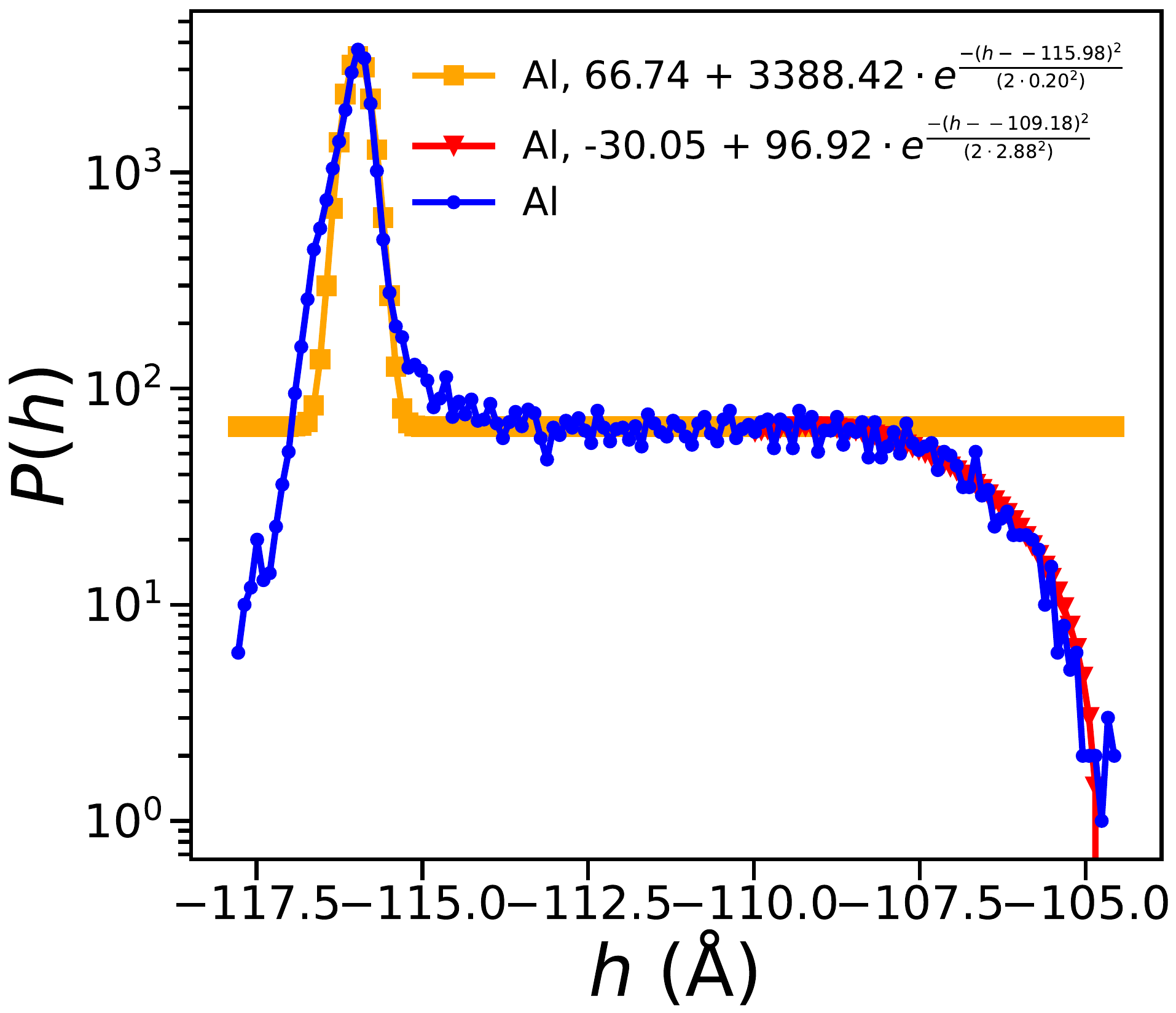}
  \includegraphics[height=7.6cm]{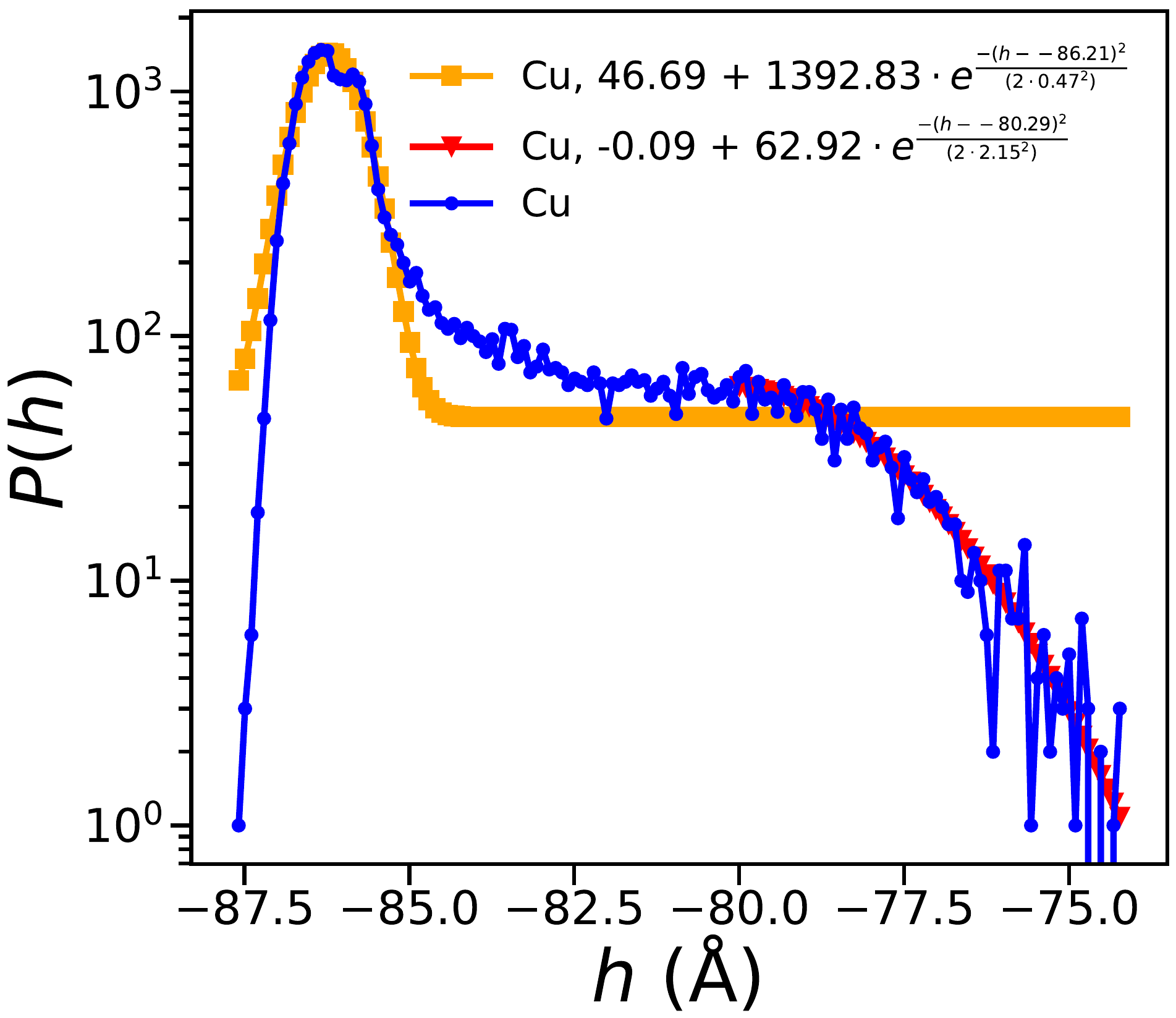}
  \caption{Height distributions and corresponding Gaussian fits for the largest nanoislands.}
  \label{fgr:h_distributions_and_fits}
\end{figure*}

\begin{figure}[h]
\centering
  \includegraphics[height=3.8cm]{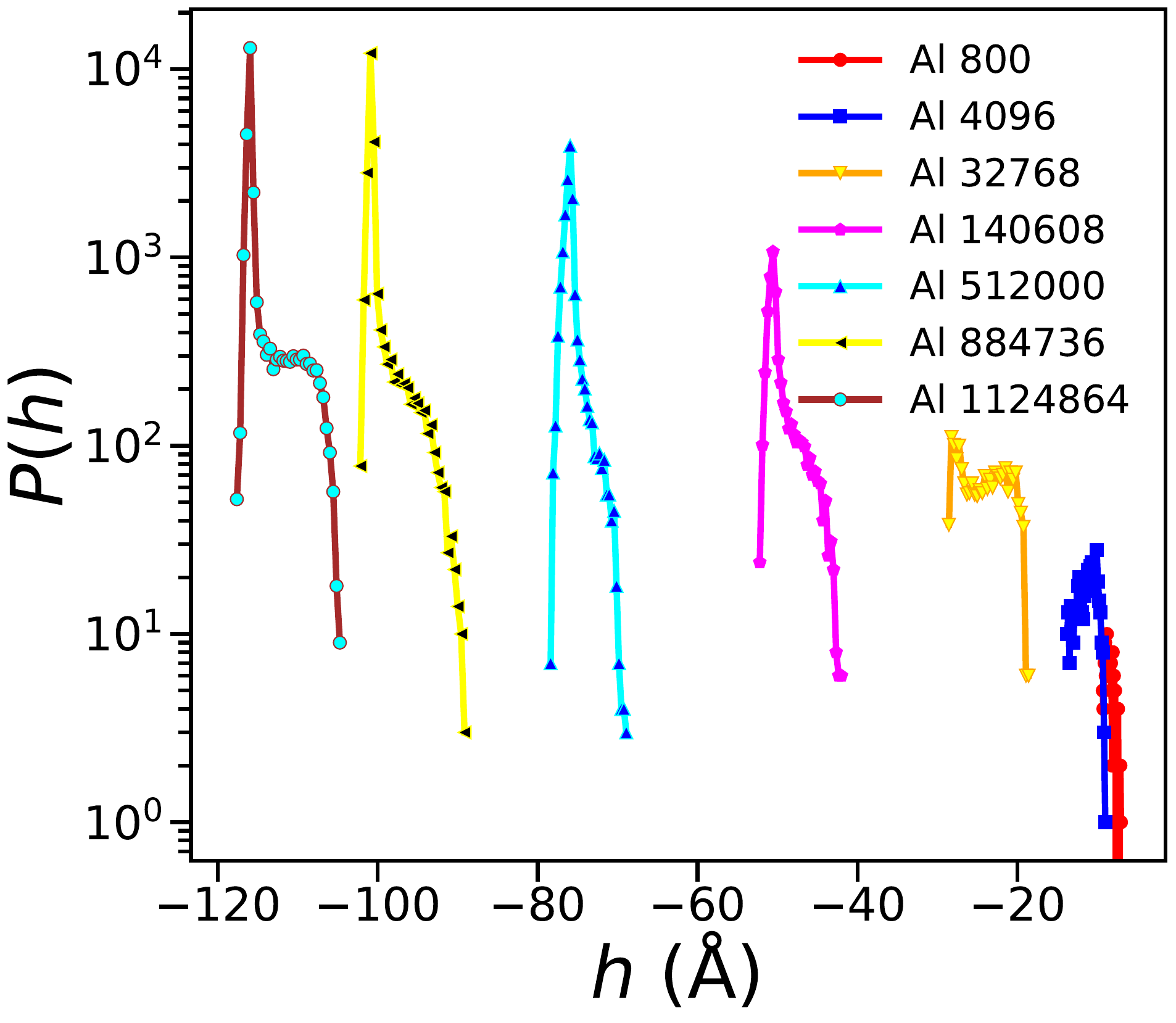}
  \includegraphics[height=3.8cm]{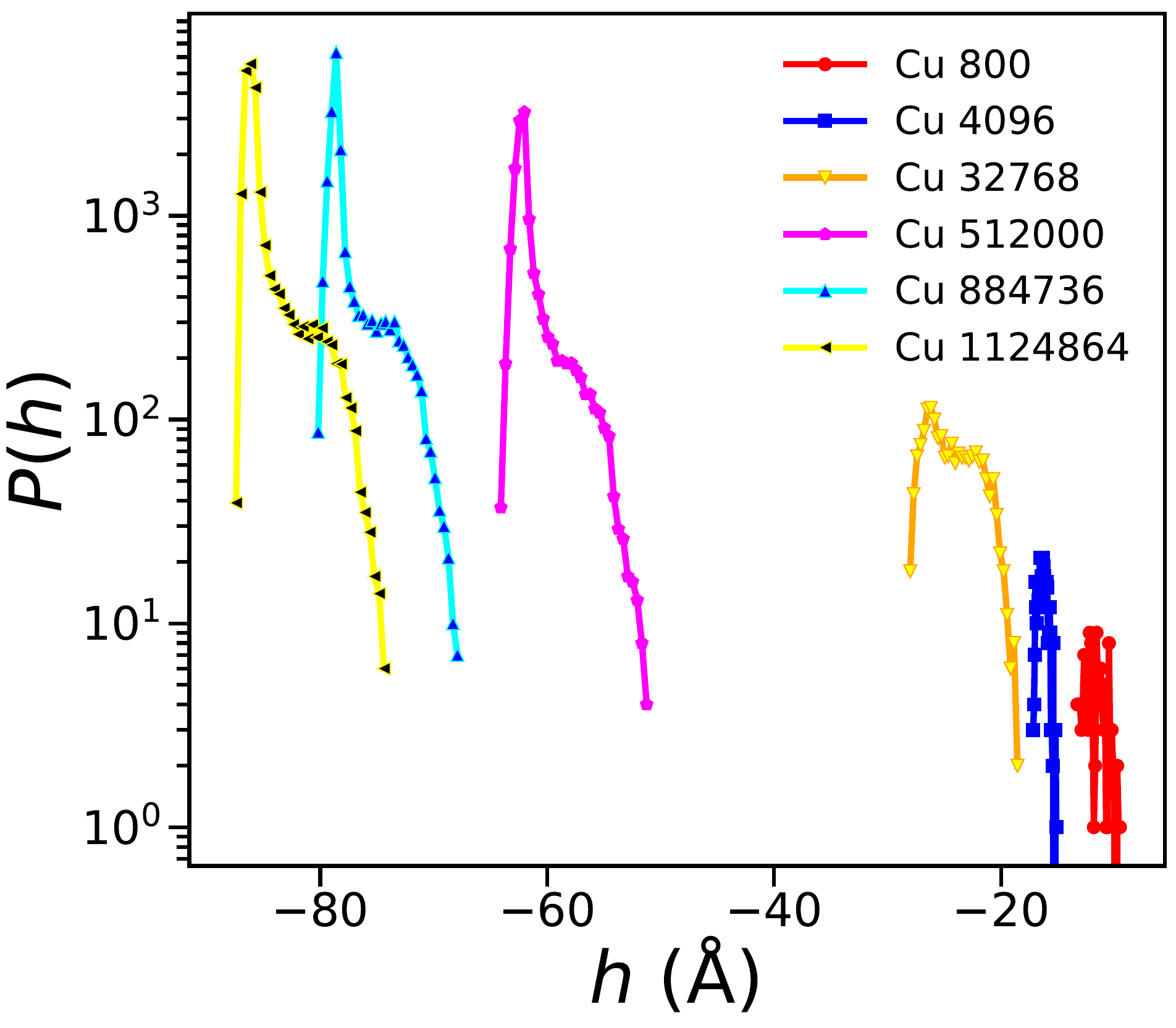}
  \caption{Height distributions of the bottom layer for Al and Cu NPs of different sizes (the legends depict the corresponding number of atoms).}
  \label{fgr:bot_h_distribution}
\end{figure}

\begin{figure}[!h]
\centering
  \includegraphics[height=3.9cm]{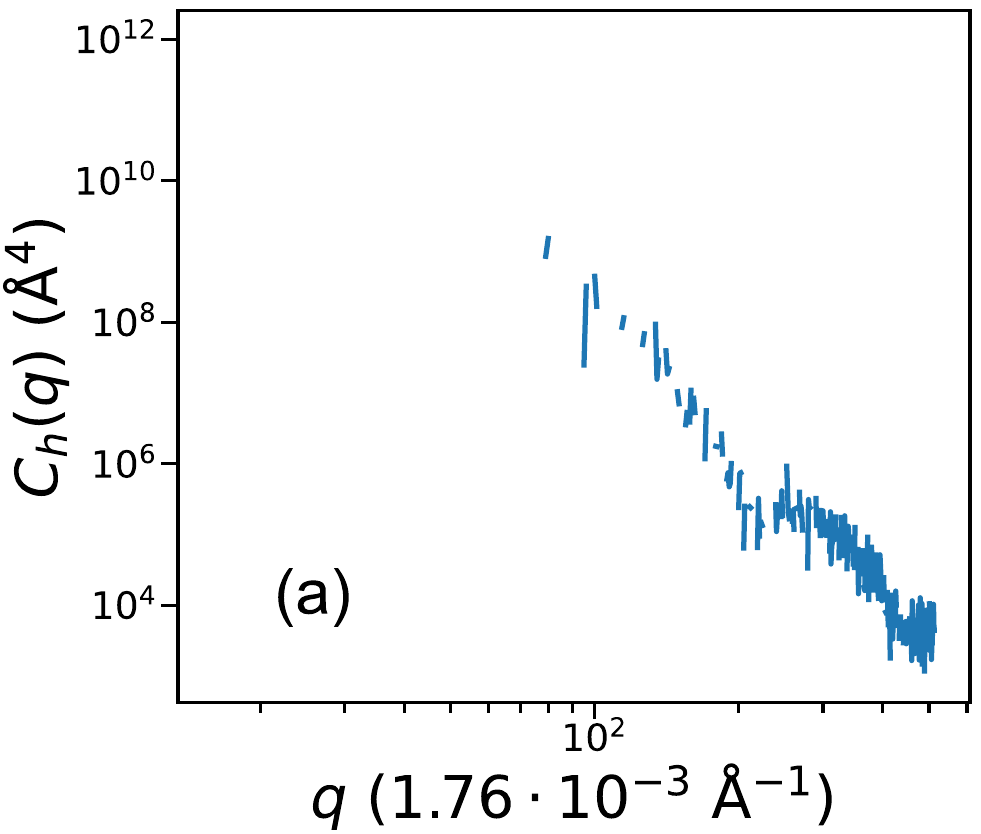}
  \includegraphics[height=3.9cm]{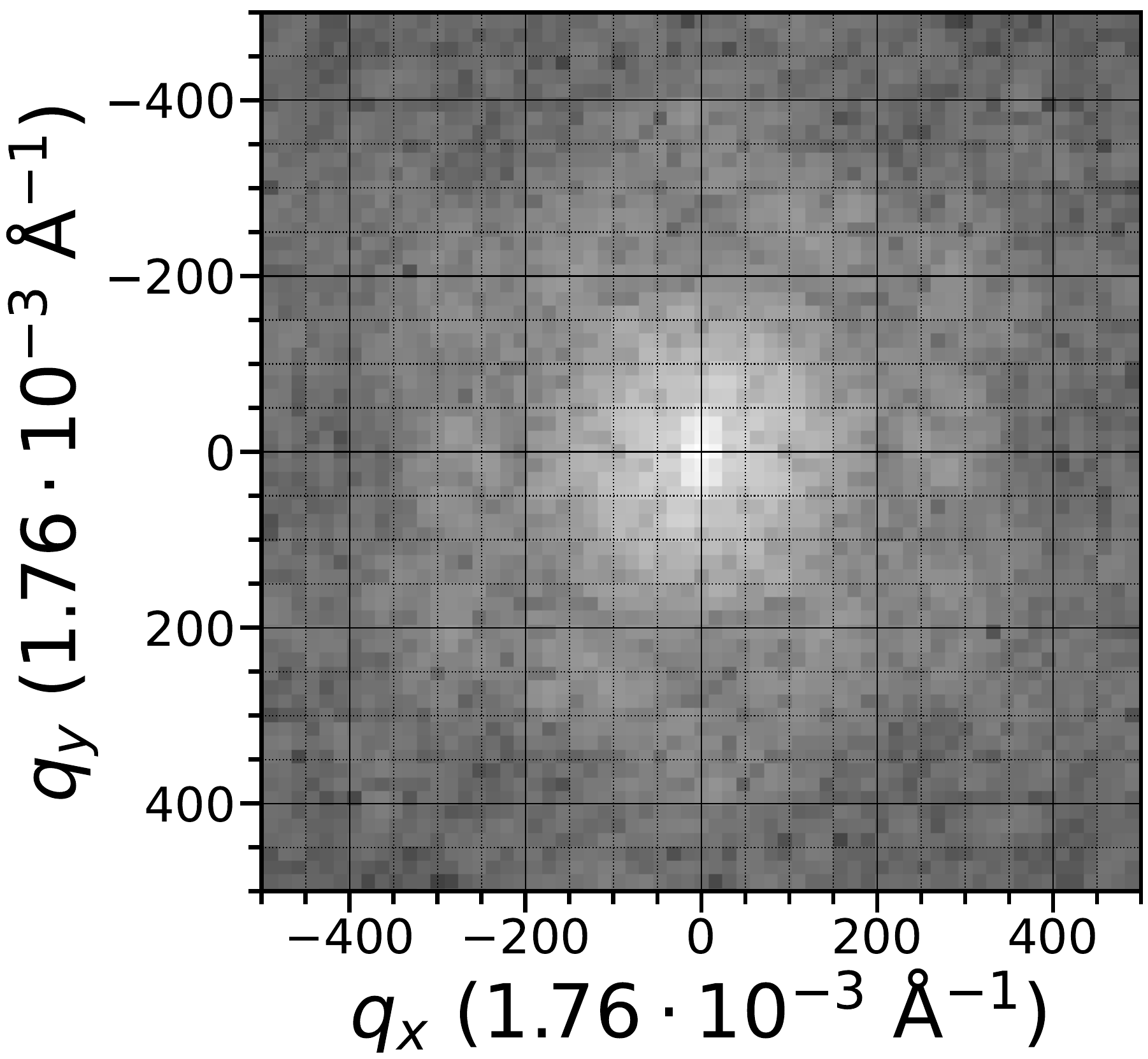}
  \includegraphics[height=3.9cm]{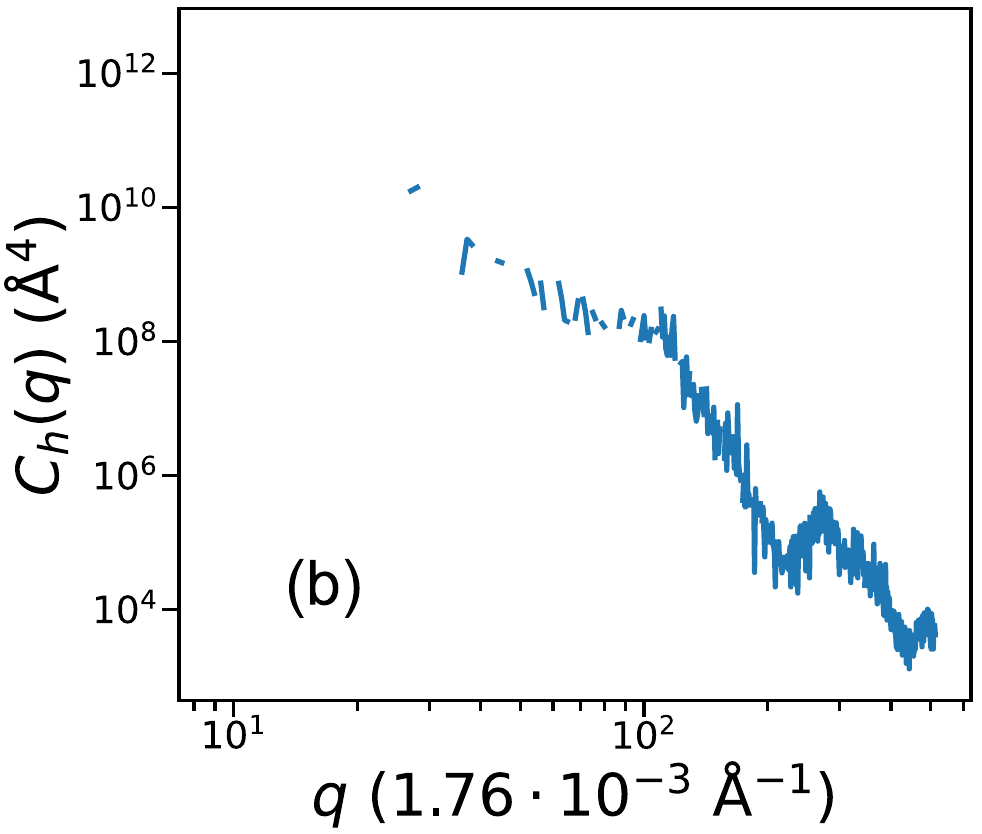}
  \includegraphics[height=3.9cm]{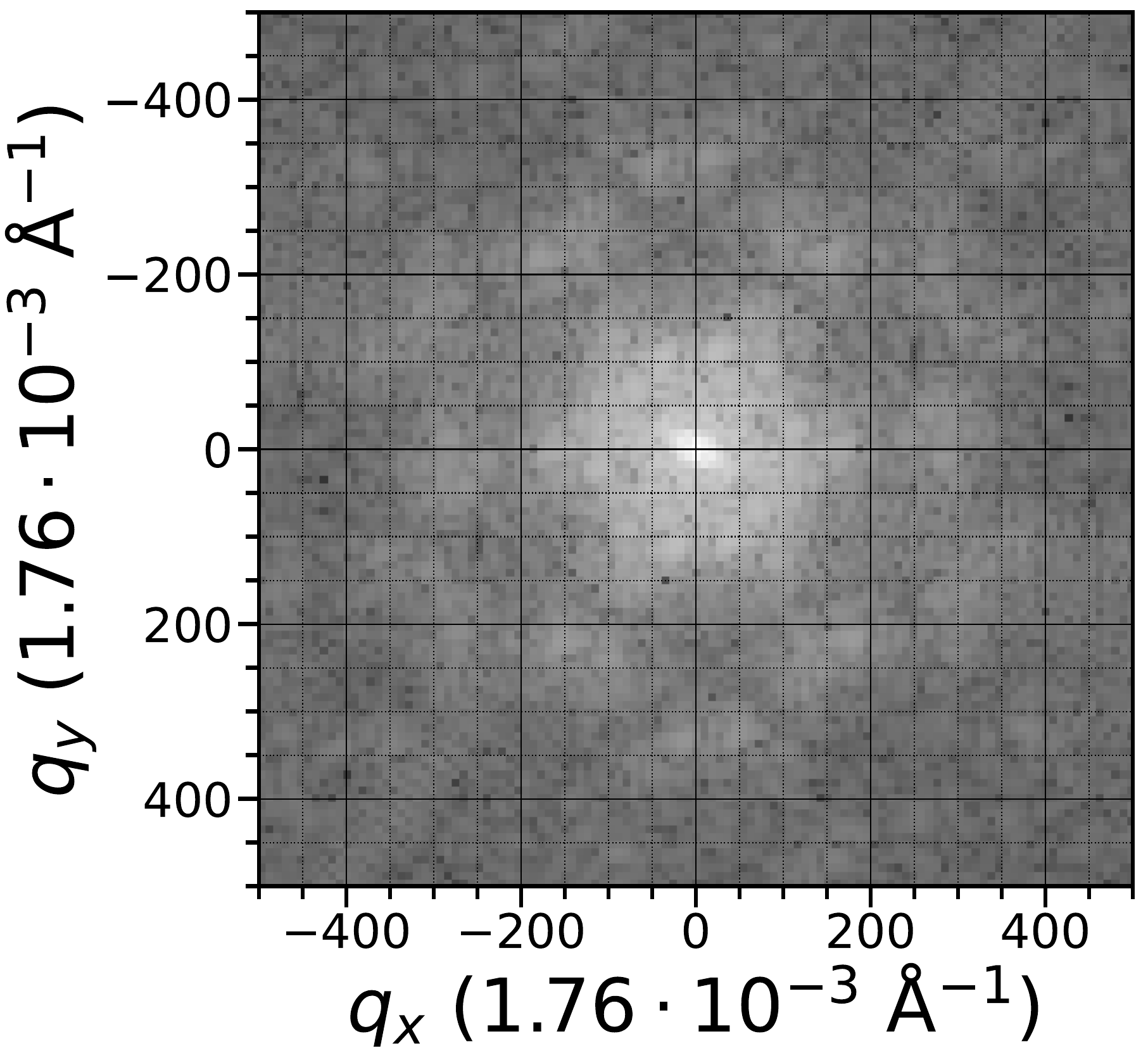}
  \includegraphics[height=3.9cm]{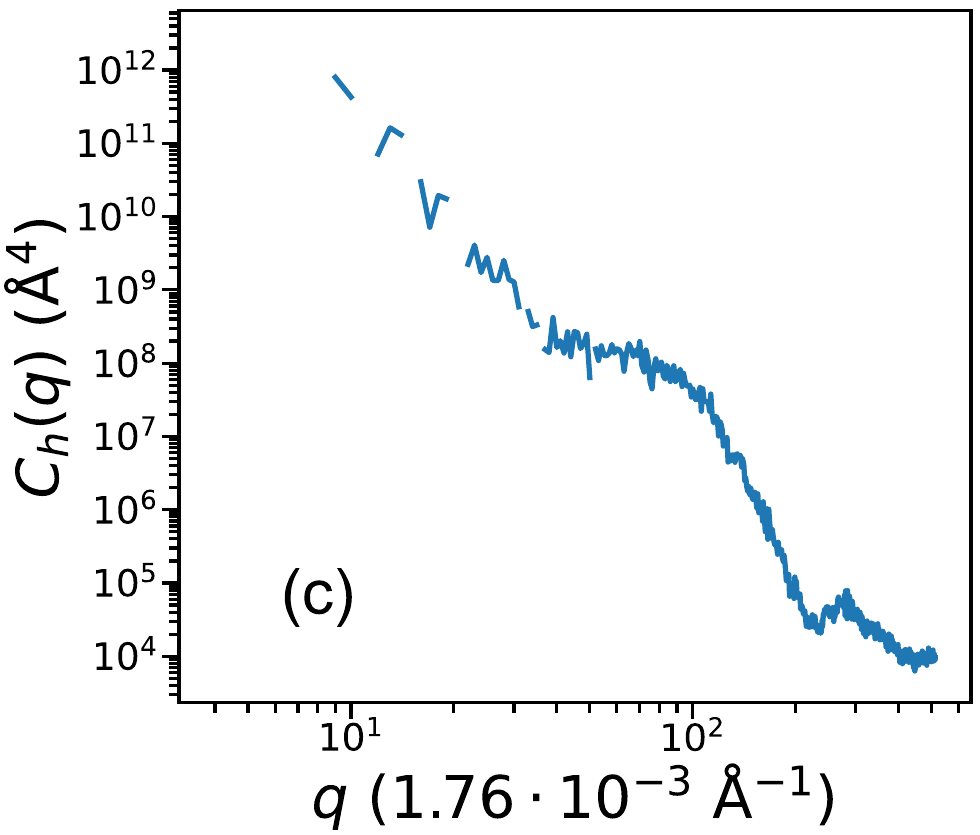}
  \includegraphics[height=3.9cm]{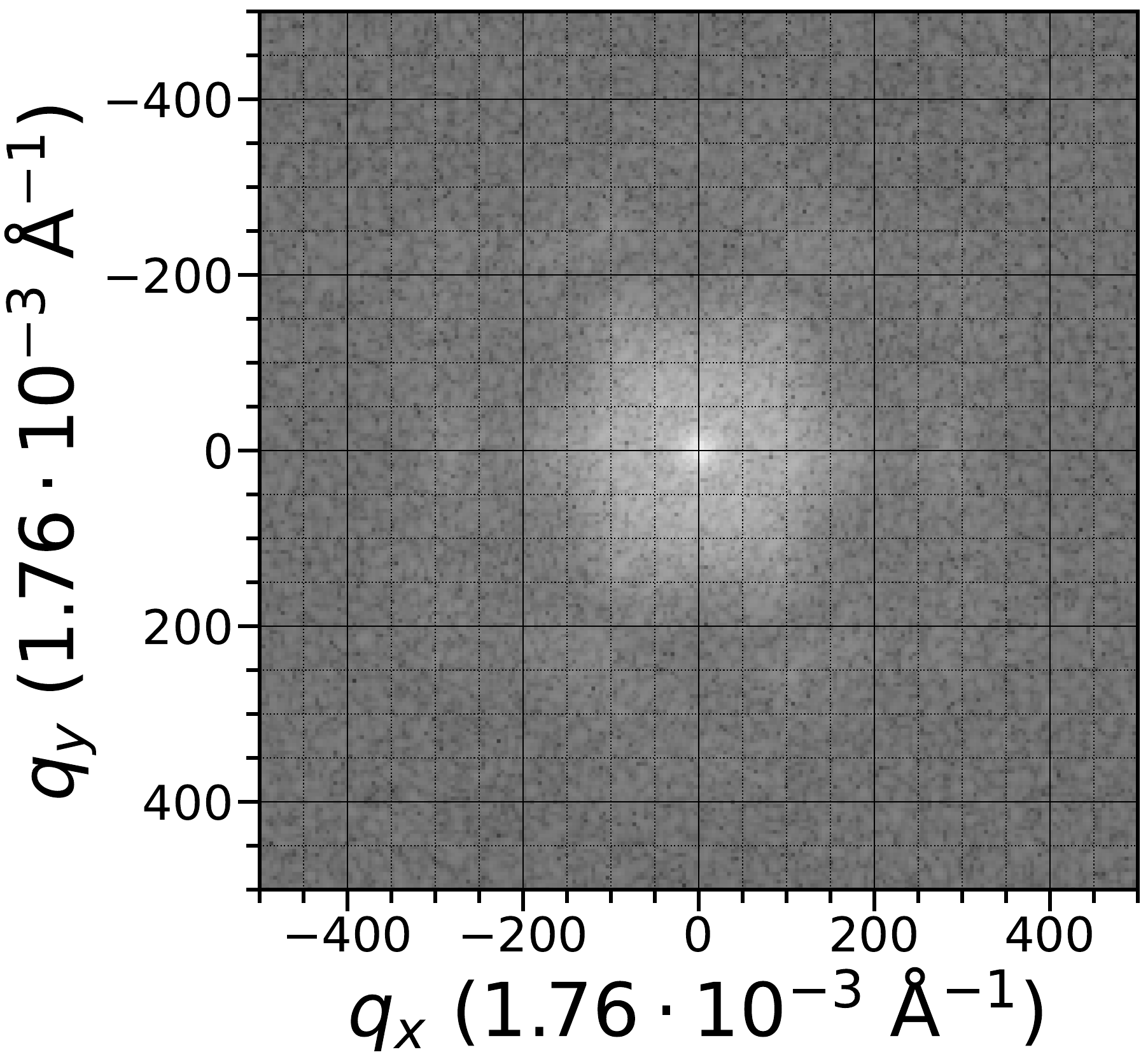}
  \includegraphics[height=3.9cm]{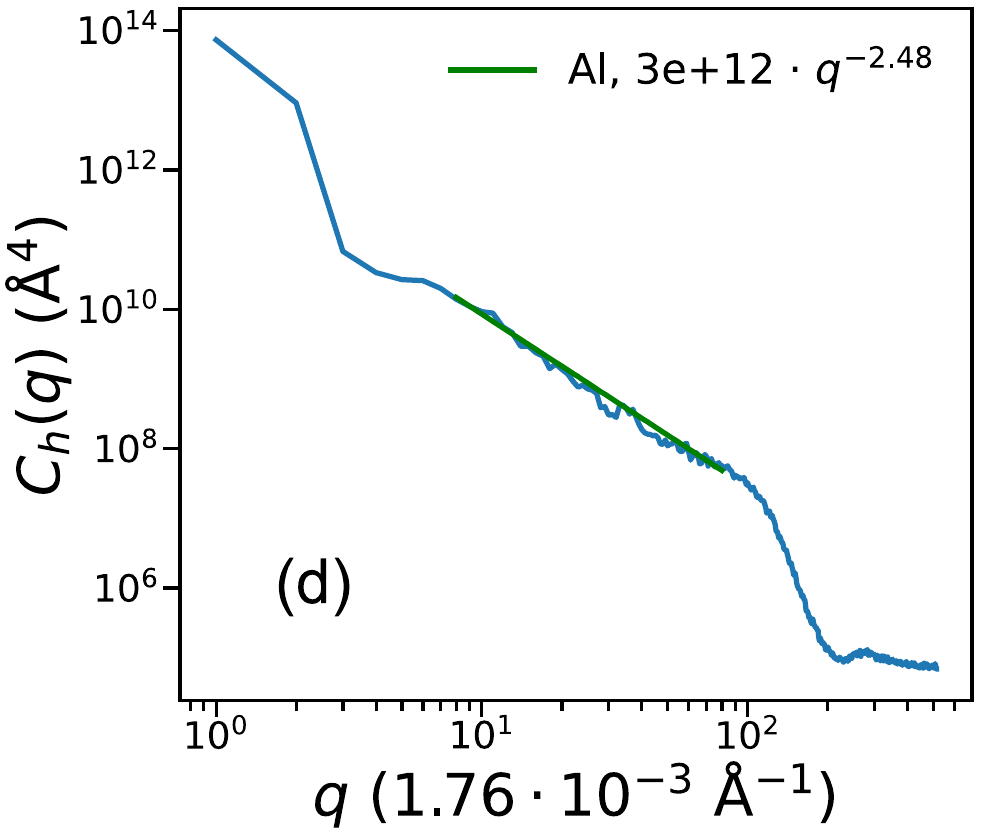}
  \includegraphics[height=3.9cm]{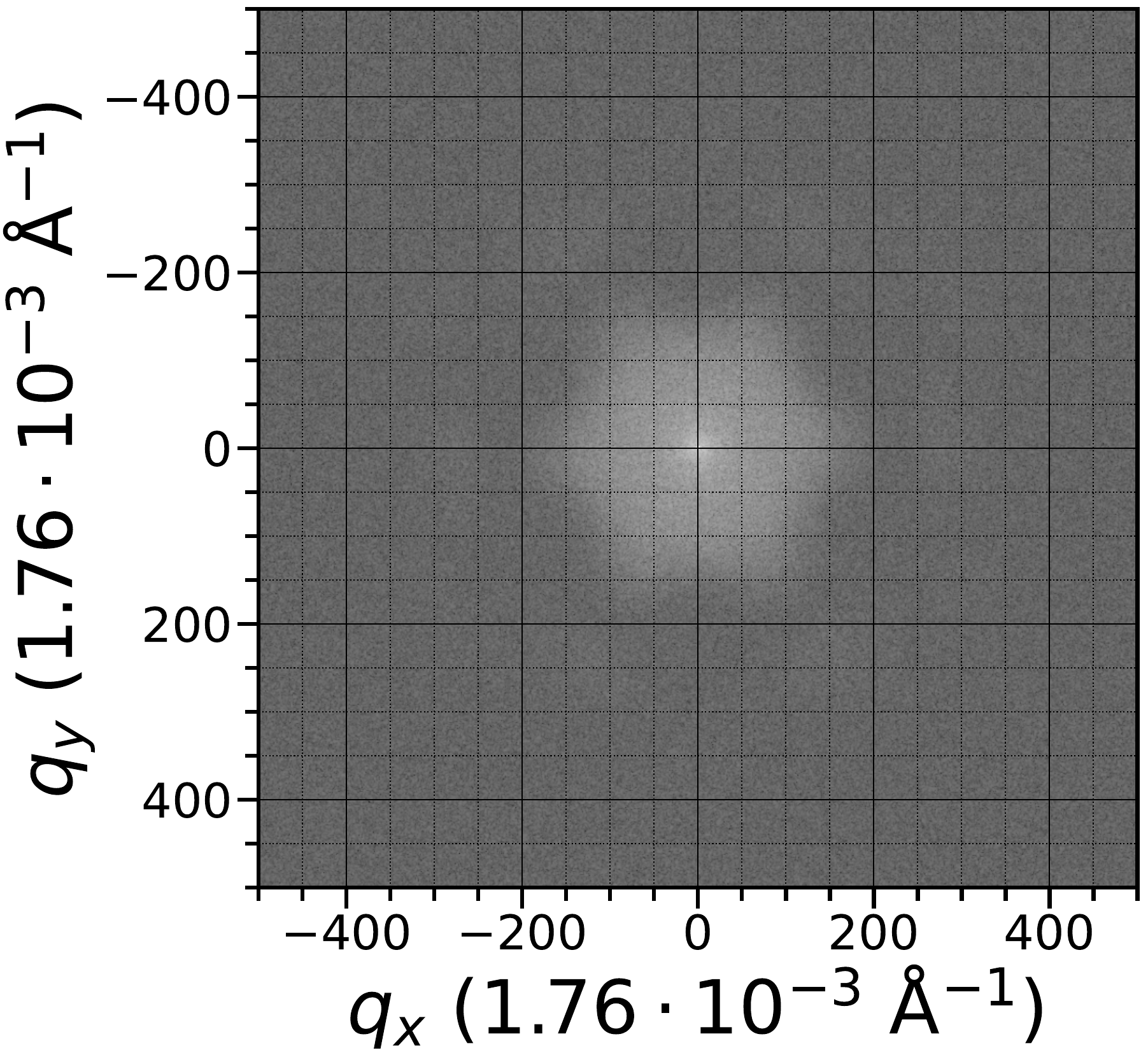}
  \caption{$C_{h}(q)$ log-log plots and full 2D PSD $C_{h}(q_{x}, q_{y})$ heatmaps for Al NPs with 800 (a), 4096 (b), 32768 (c), 1124864 (d) atoms. The legend in (d) shows the power law fit $\propto q^{-2(1 + H)}$ with $H$ = 0.24.}
  \label{fgr:h_bottom_psd_al}
\end{figure}

\begin{figure}[!h]
\centering
  \includegraphics[height=3.9cm]{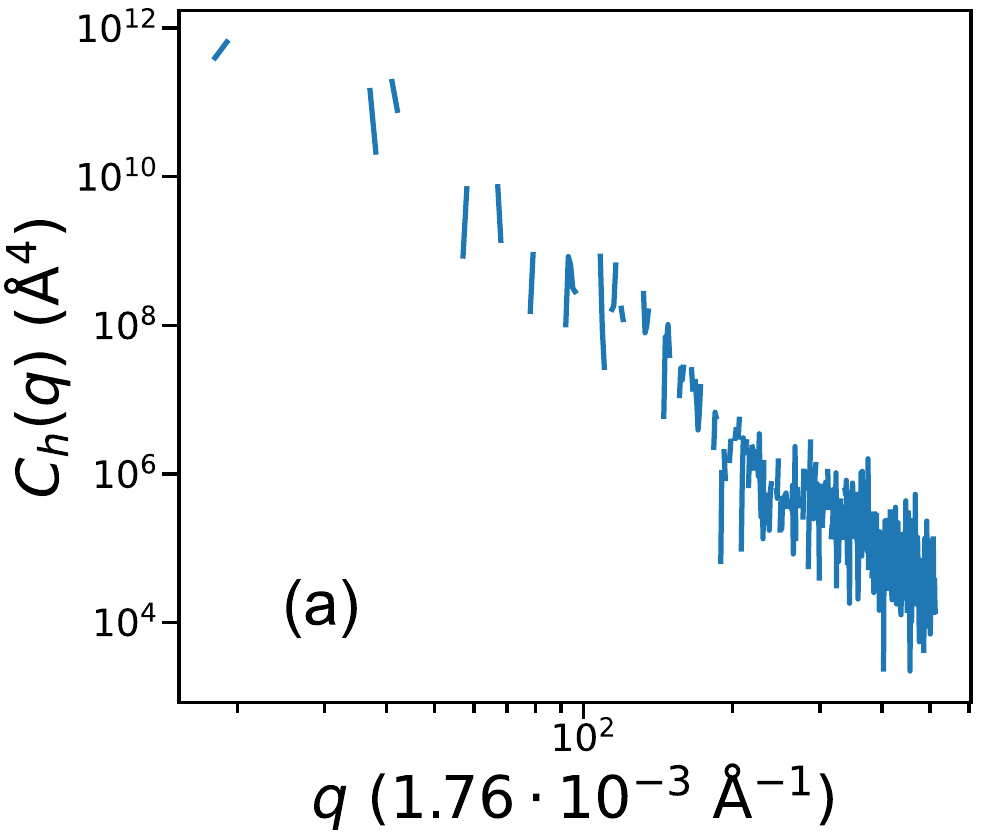}
  \includegraphics[height=3.9cm]{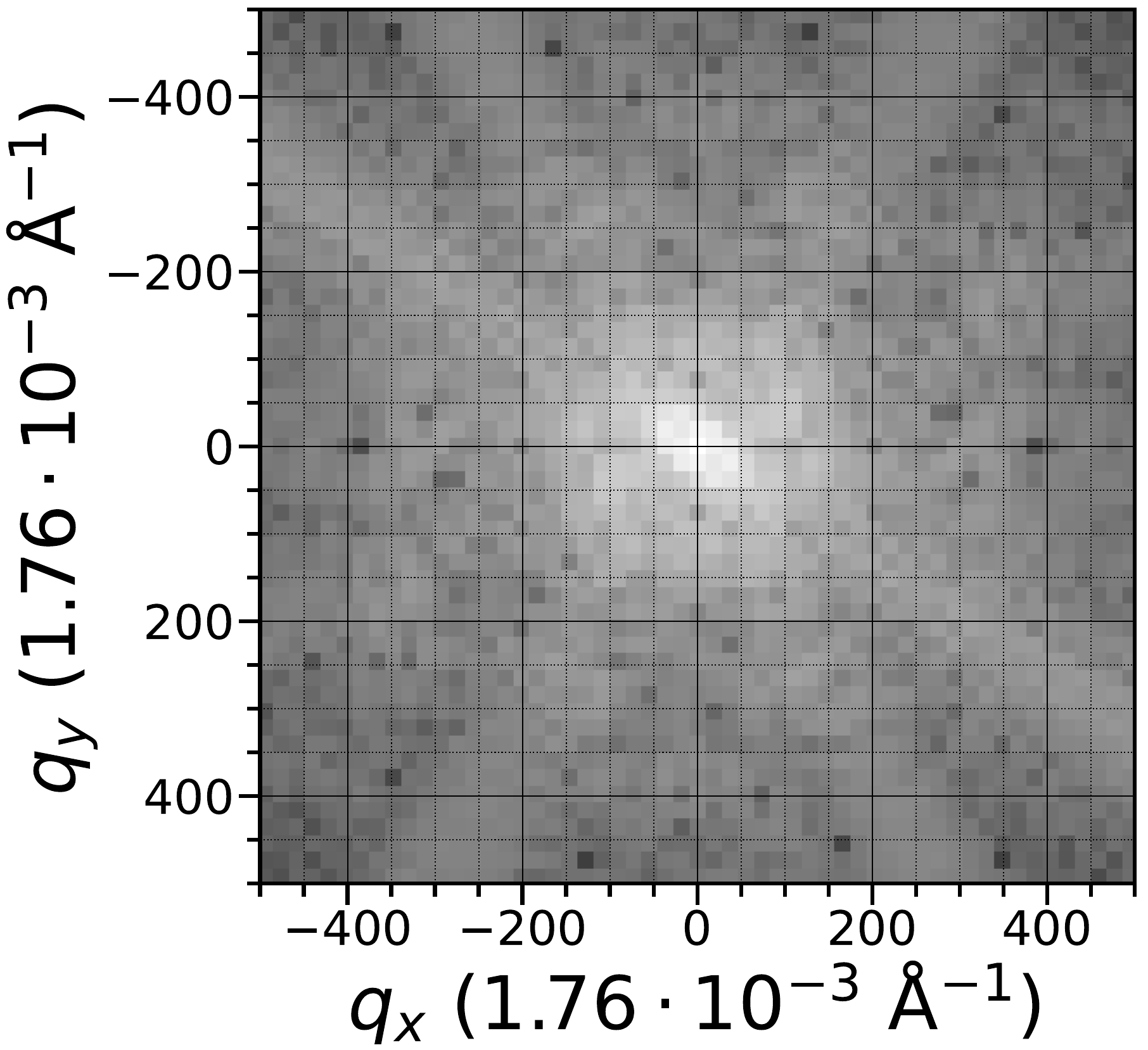}
  \includegraphics[height=3.9cm]{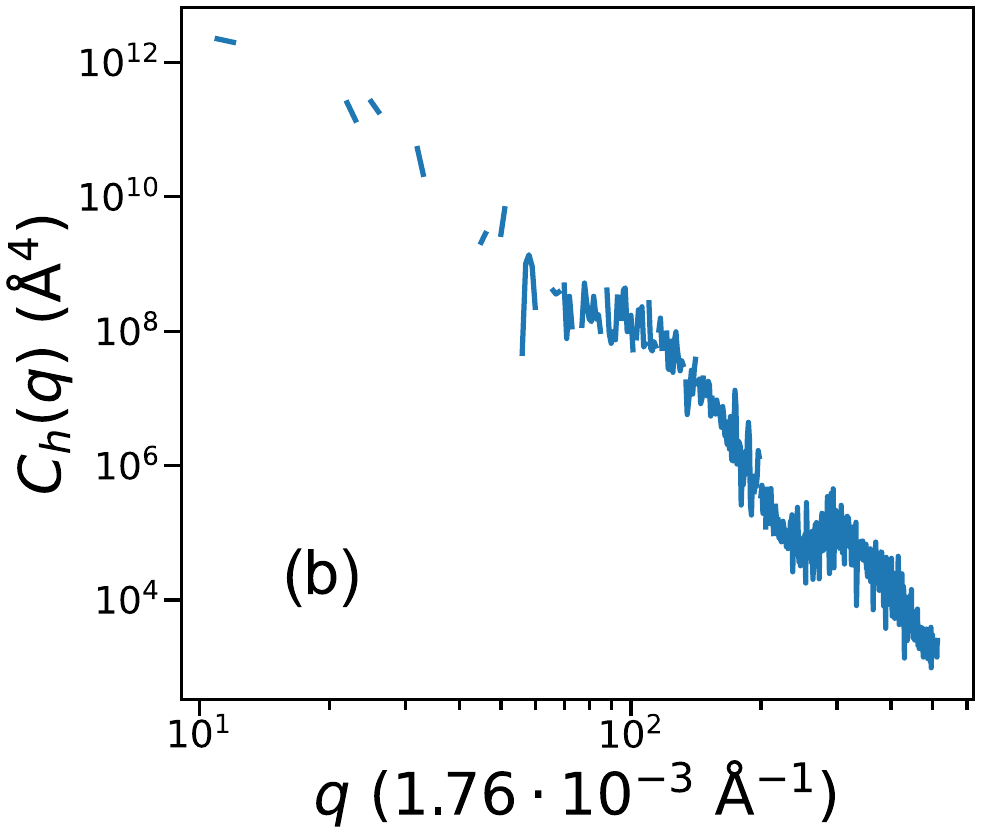}
  \includegraphics[height=3.9cm]{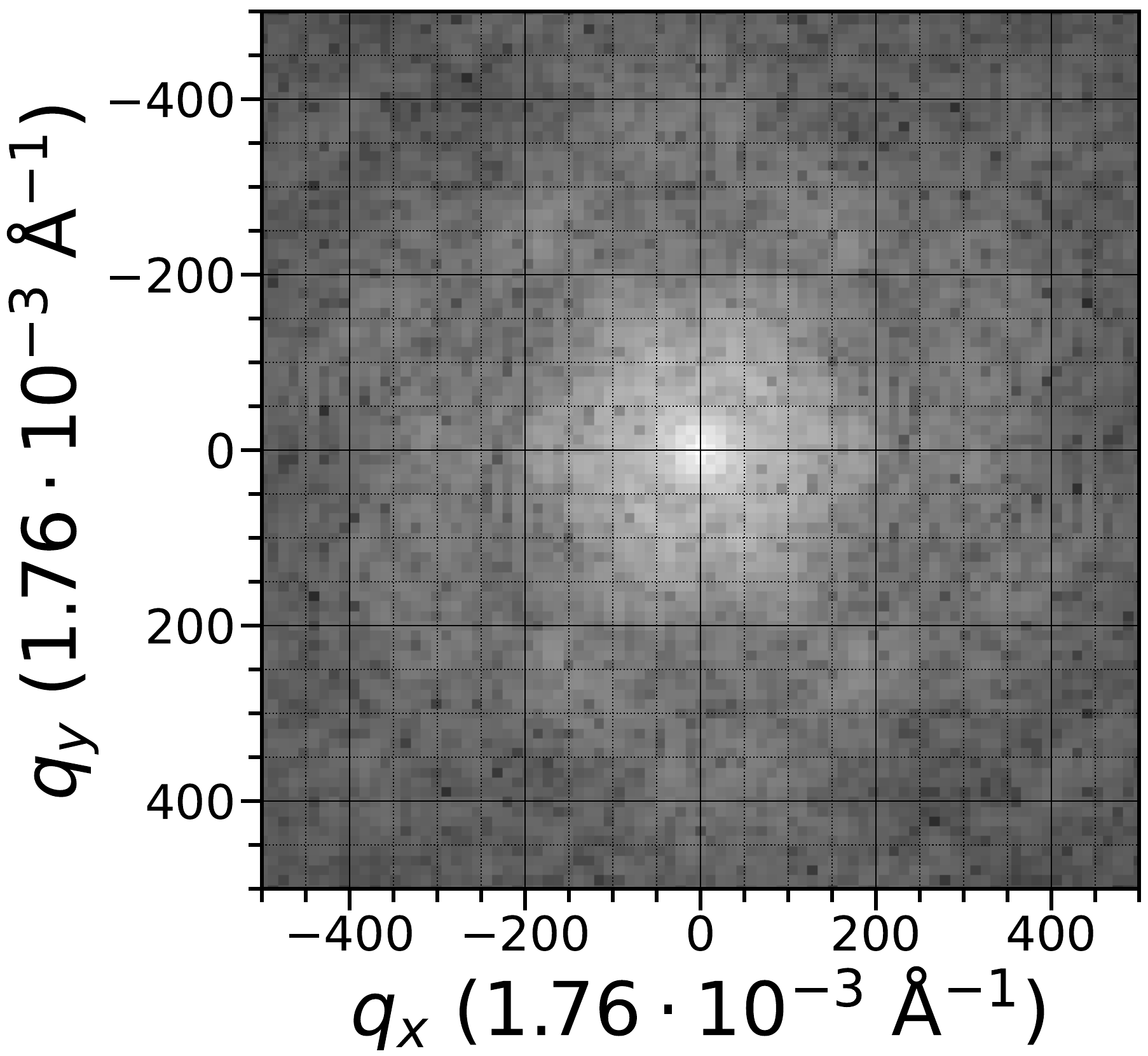}
  \includegraphics[height=3.9cm]{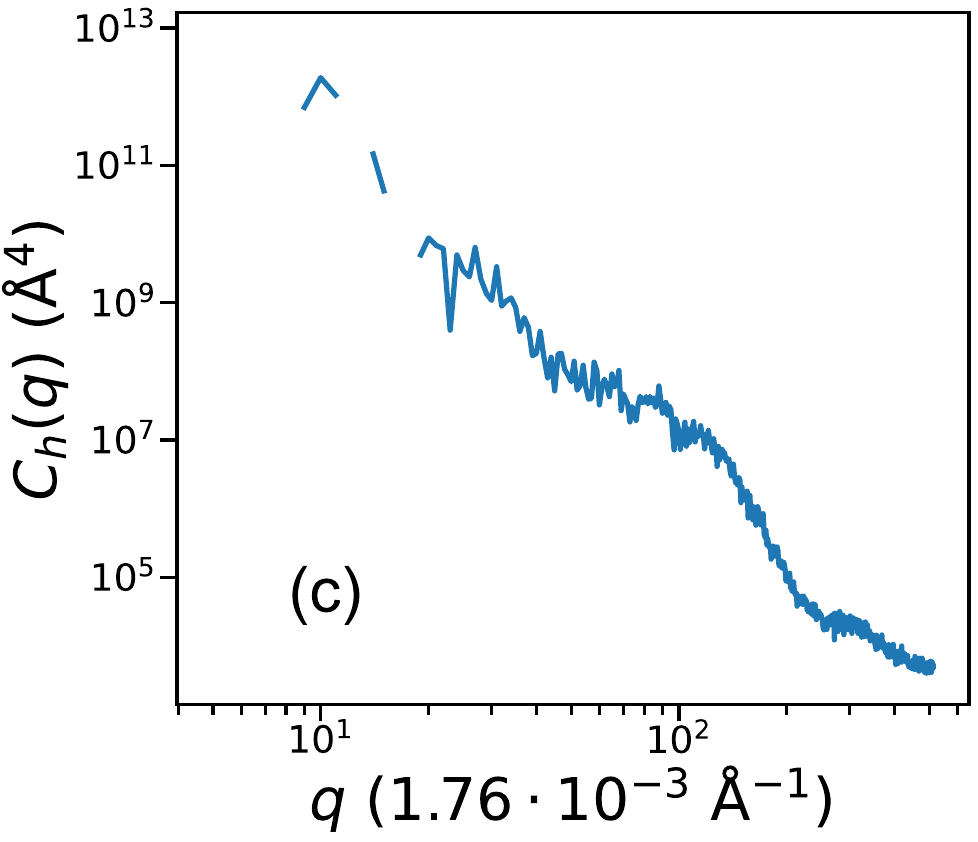}
  \includegraphics[height=3.9cm]{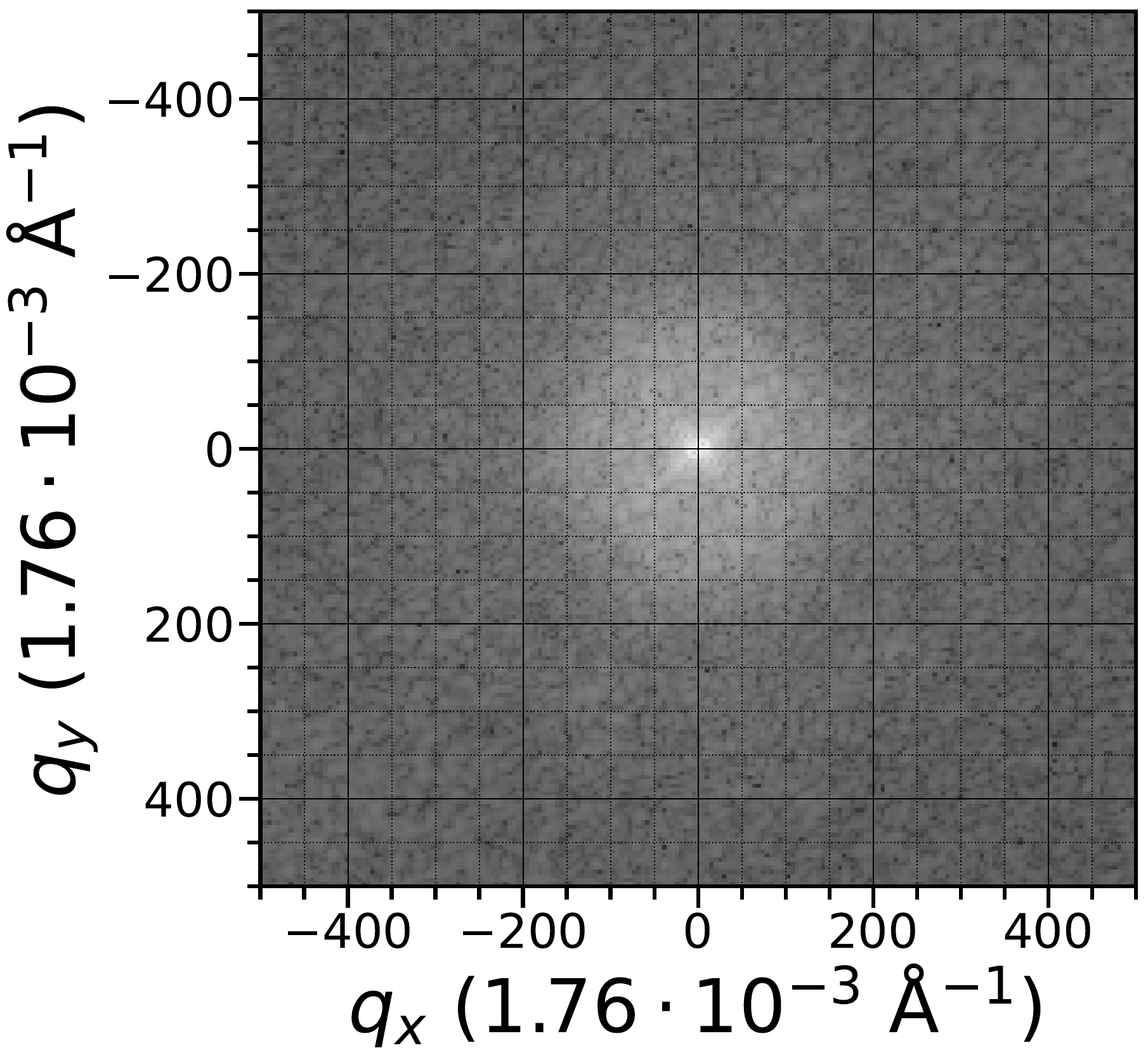}
  \includegraphics[height=3.9cm]{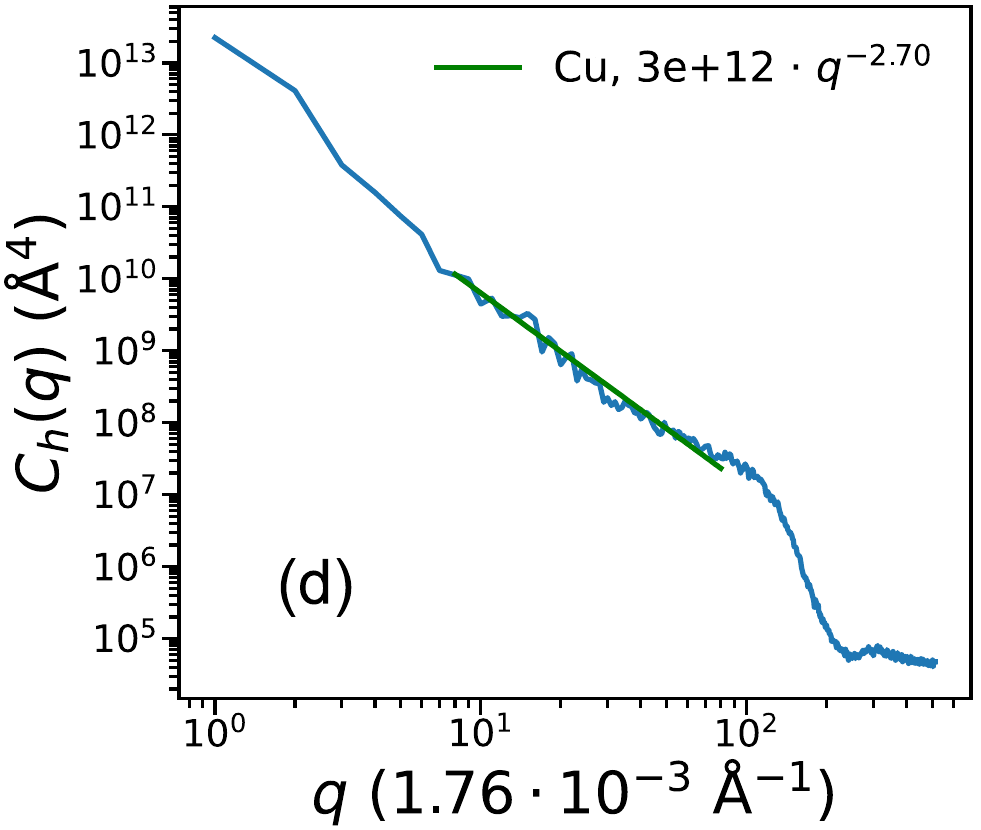}
  \includegraphics[height=3.9cm]{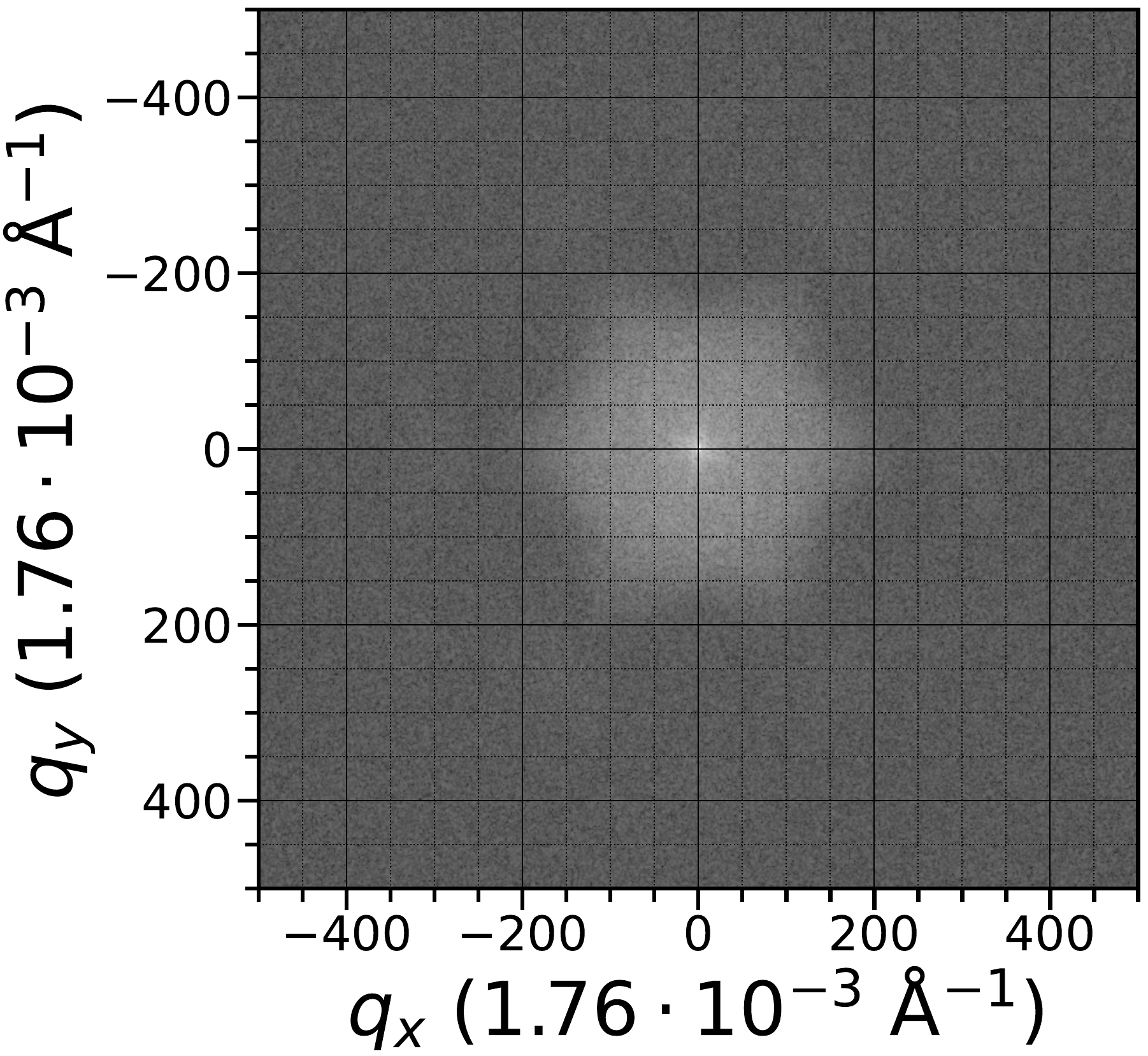}
  \caption{$C_{h}(q)$ log-log plots and $C_{h}(q_{x}, q_{y})$ heatmaps for Cu NPs with 800 (a), 4096 (b), 32768 (c), 1124864 (d) atoms. The legend in (d) shows the power law fit $\propto q^{-2(1 + H)}$ with $H$ = 0.35.}
  \label{fgr:h_bottom_psd_cu}
\end{figure}

The interface atomic layer, i.e., the metal atoms of the bottom layer of a NP, in most Al nanoislands have a hexagonal arrangement, see Fig.~\ref{fgr:side_top_view_bot_height_map} showing the $h(x, y)$ heat map of NPs containing 32768 atoms. The contact atomic layer corresponds to (111) atomic plane, which is consistent with the experimentally observed surface structure of NPs~\cite{Badan2022nm,Chung2022}. For example, oxide-free copper NPs had a (111) surface plane with an interatomic distance~\cite{Chung2022} of around 2.5~\AA, and the same hexagonal symmetry was reported for Ag NPs~\cite{Badan2022nm}. Such a structure is generally the most energetically favorable for FCC metals. It has the lowest surface energy because it is the most closely packed plane, so that fewer surface atomic bonds are broken compared to other atom plane orientations. Note that this atomic arrangement is different from the original (110) cuboid slab orientation, meaning that the molten NP atoms typically solidified into a more energetically favorable state.

\begin{figure*}[!h]
\centering
  \includegraphics[height=7.2cm]{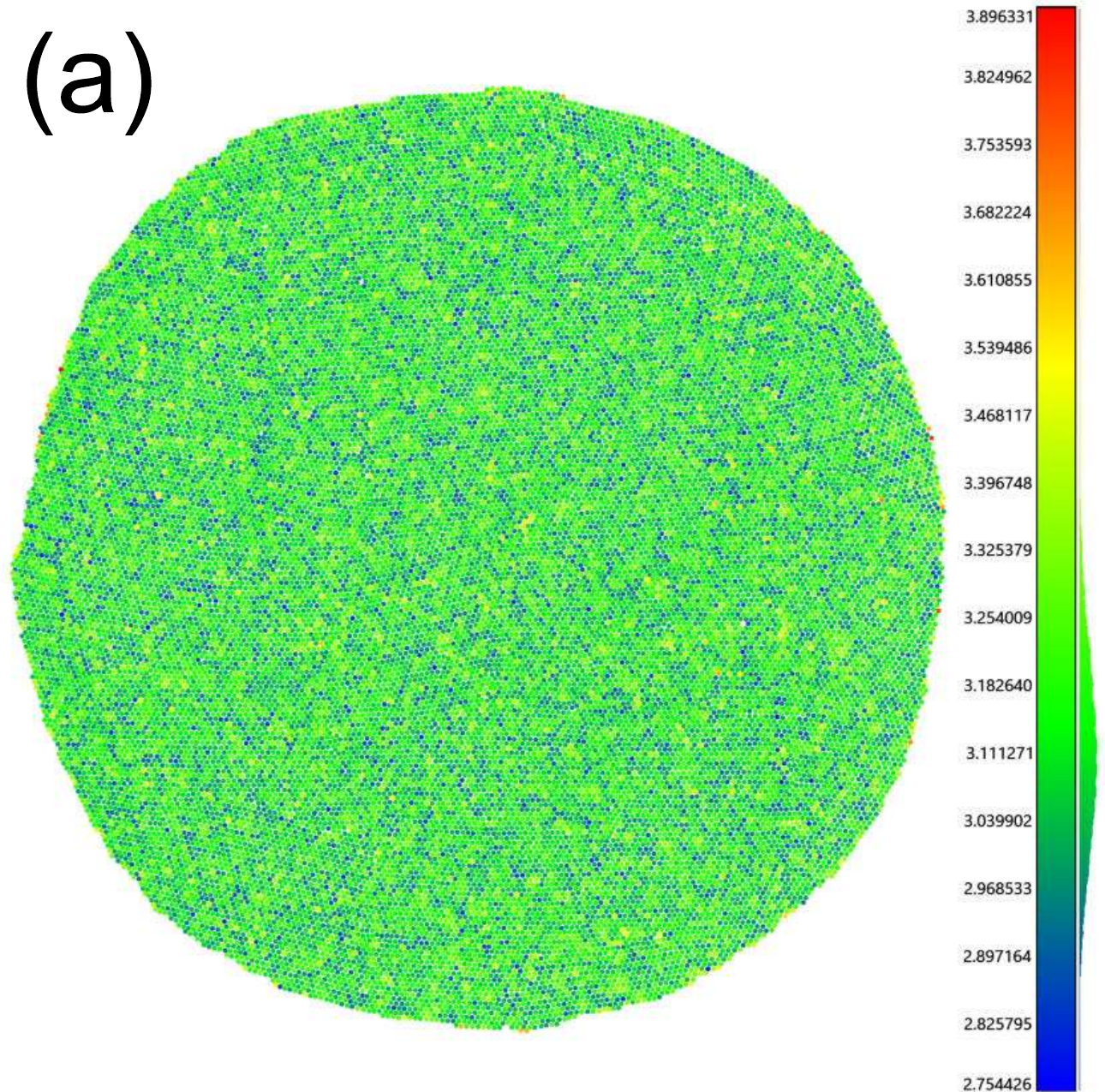}
  \includegraphics[height=7.2cm]{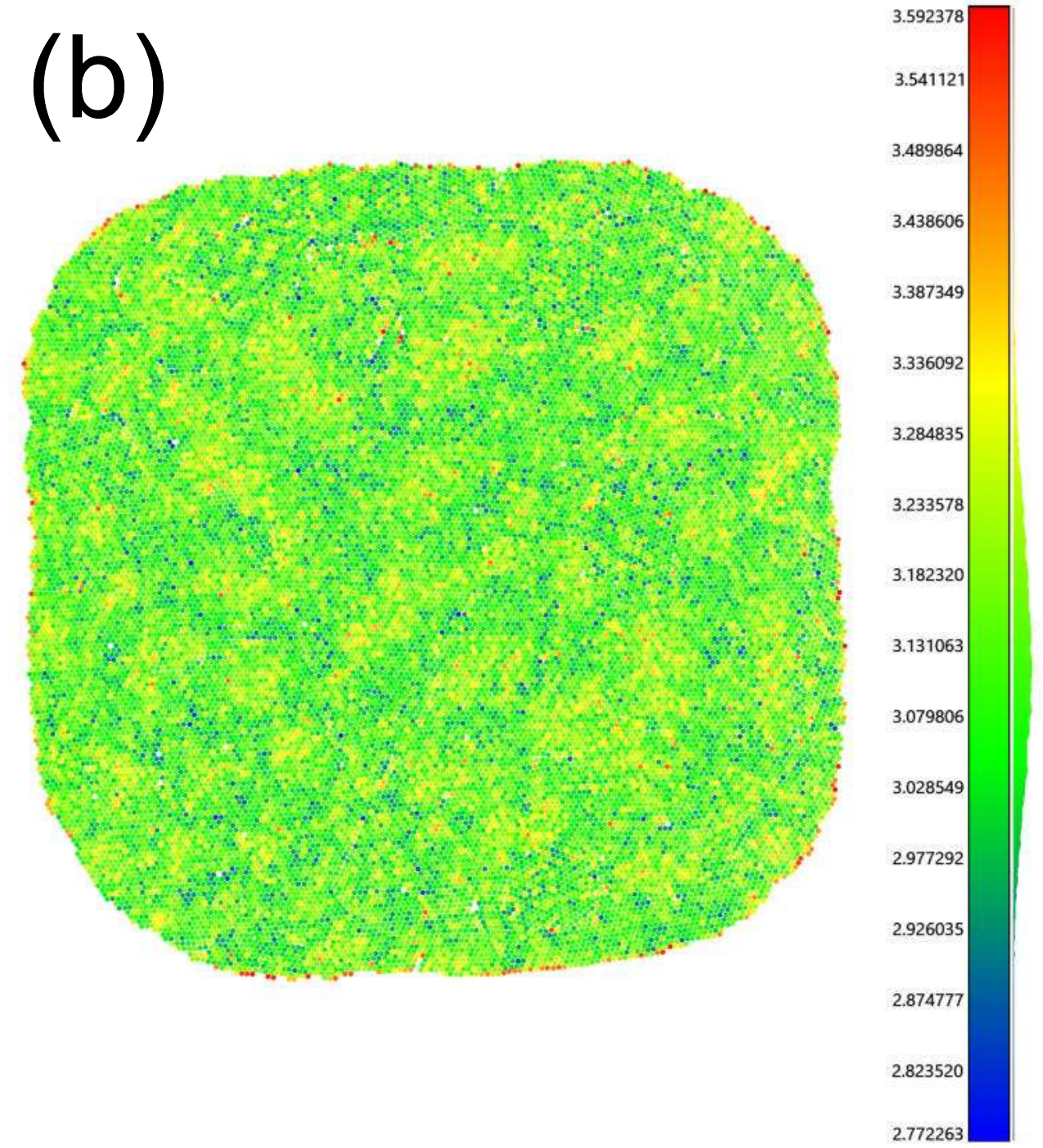}
    \caption{A top-down view of the interfacial separation $u$ heatmaps for the largest Al (a) and Cu (b) NPs.}
  \label{fgr:side_top_view_bot_gap_map}
\end{figure*}

However, not all NPs exhibit hexagonal symmetry of the interface atoms, as shown in the height heat map of the Cu NP in Fig.~\ref{fgr:side_top_view_bot_height_map}. In particular, some interface Cu atoms are arranged into separate domains with atoms having a hexagonal or a square arrangement. Again, such behavior of Cu NPs can be attributed to the slower rearrangement of interface atoms in the molten NP, due to stronger interactions with the substrate.

From Figs.~\ref{fgr:side_top_view_bot_height_map},~\ref{fgr:side_top_view_bot_height_map_largest} (depicting the height heat map of selected NPs), it is obvious that both surfaces are not atomically flat but rather have some atomic-scale corrugation, which is also confirmed by the corresponding height distributions $P(h)$ in Figs.~\ref{fgr:h_distributions_and_fits},~\ref{fgr:bot_h_distribution}. Each $P(h)$ can be roughly divided into three parts: a narrow spike, a region with approximately uniform heights, and a smooth, decaying tail. The spike suggests that many atoms have values of $h$ close to each other. Starting from $N \gtrsim 140608$ atoms in a NP, the spike can be fit to a Gaussian:
\begin{equation}
  P(x) = P_{0} + P_{1} e^{-\frac{(x - x_{0})^{2}}{2 \sigma^{2}}}.
  \label{eq:gaussian_fit}
\end{equation}
The Gaussians' parameters of the spikes in Fig.~\ref{fgr:h_distributions_and_fits} (orange lines with square symbols) are $P_{0}=66.74$, $P_{1}=3388.42$, $x_{0}=-115.98$~\AA, $\sigma=0.2$~\AA~and $P_{0}=46.69$, $P_{1}=1392.83$, $x_{0}=-86.21$~\AA, $\sigma=0.47$~\AA, for the Al and Cu NPs, respectively. Noticeably, the tail of $P(h)$ can also be fit to a Gaussian, Eq.~(\ref{eq:gaussian_fit}). This is shown in Fig.~\ref{fgr:h_distributions_and_fits} (red lines with triangle symbols), where the corresponding fitting parameters are listed.

Macroscopic (or multiscale) randomly-rough surfaces whose PSD has a roll-off frequency, exhibit a Gaussian $P(h)$, which is defined as~\cite{Persson2005jpcm,Rodriguez2025triblett}:
\begin{equation}
  P(h) = \frac{1}{(2\pi)^{1/2}h_{\mathrm{rms}}}e^{-(h/h_{\mathrm{rms}})^{2}/2},
  \label{eq:gaussian_p_h}
\end{equation}
where $h_{\mathrm{rms}}$ is the RMS roughness. Assuming that Eq.~(\ref{eq:gaussian_p_h}) holds for the atomic-scale roughness and comparing Eq.~(\ref{eq:gaussian_fit}) with Eq.~(\ref{eq:gaussian_p_h}) gives $\sigma=h_{\mathrm{rms}}$. Therefore, for the largest NPs, $h_{\mathrm{rms}}$ is 0.2~\AA~and 0.47~\AA~in the spikes, and 2.88~\AA~and 2.15~\AA~in the distribution tails for Al and Cu, respectively. Our simulations show that the topography of larger NPs can feature regions of random roughness with RMS value of~$\sim$\AA~and less.

The PSDs $C_{h}(q)$ and $C_{h}(q_{x}, q_{y})$ in Fig.~\ref{fgr:h_bottom_psd_al} for Al and in Fig.~\ref{fgr:h_bottom_psd_cu} for Cu exhibit size-dependent behavior. For the smallest NPs containing 800 atoms (and smaller, not shown here), $C_{h}(q_{x}, q_{y})$ is smeared without a clear structure for both metals, cf. Fig.~\ref{fgr:h_bottom_psd_al}~(a) and Fig.~\ref{fgr:h_bottom_psd_cu}~(a). Starting from about 4096 atoms or $L_{x}$ of $\sim$ 6~nm for Al and $\sim$ 5~nm for Cu (as can be inferred from Fig.~\ref{fgr:size_xy_vs_atoms_count}), the sixfold anisotropic symmetry of $C_{h}(q_{x}, q_{y})$ and clearly visible discrete power areas start to emerge. Smaller NPs' PSDs contain high-frequency regions in Figs.~(b). These regions decrease in power for the larger NPs, Figs.~(c)--(d). Also note that Fig.~\ref{fgr:h_bottom_psd_cu} (c) corresponds to the Cu NP with 32768 atoms in Fig.~\ref{fgr:side_top_view_bot_height_map} (b). It does not have a clear hexagonal structure, reflecting the presence of separate domains of atoms with different, in particular, the square atomic arrangements. One can notice that the high-frequency contribution to $C_{h}(q)$ is negligible for the biggest NPs. The largest contribution to the height power comes from the frequencies $q_{Al} \lesssim 199$ and $q_{Cu} \lesssim 215$, which correspond to wavelengths roughly equal to the nearest-neighbor distances $a_{Al}$ and $a_{Cu}$ ($\lambda_{Al} = 1/q_{Al} >~ 2.9$~\AA~and $\lambda_{Cu} = 1/q_{Cu} >~ 2.6$~\AA), located in the central hexagonal ``snowflake'' region.

\begin{figure*}[!h]
\centering
  \includegraphics[height=7.2cm]{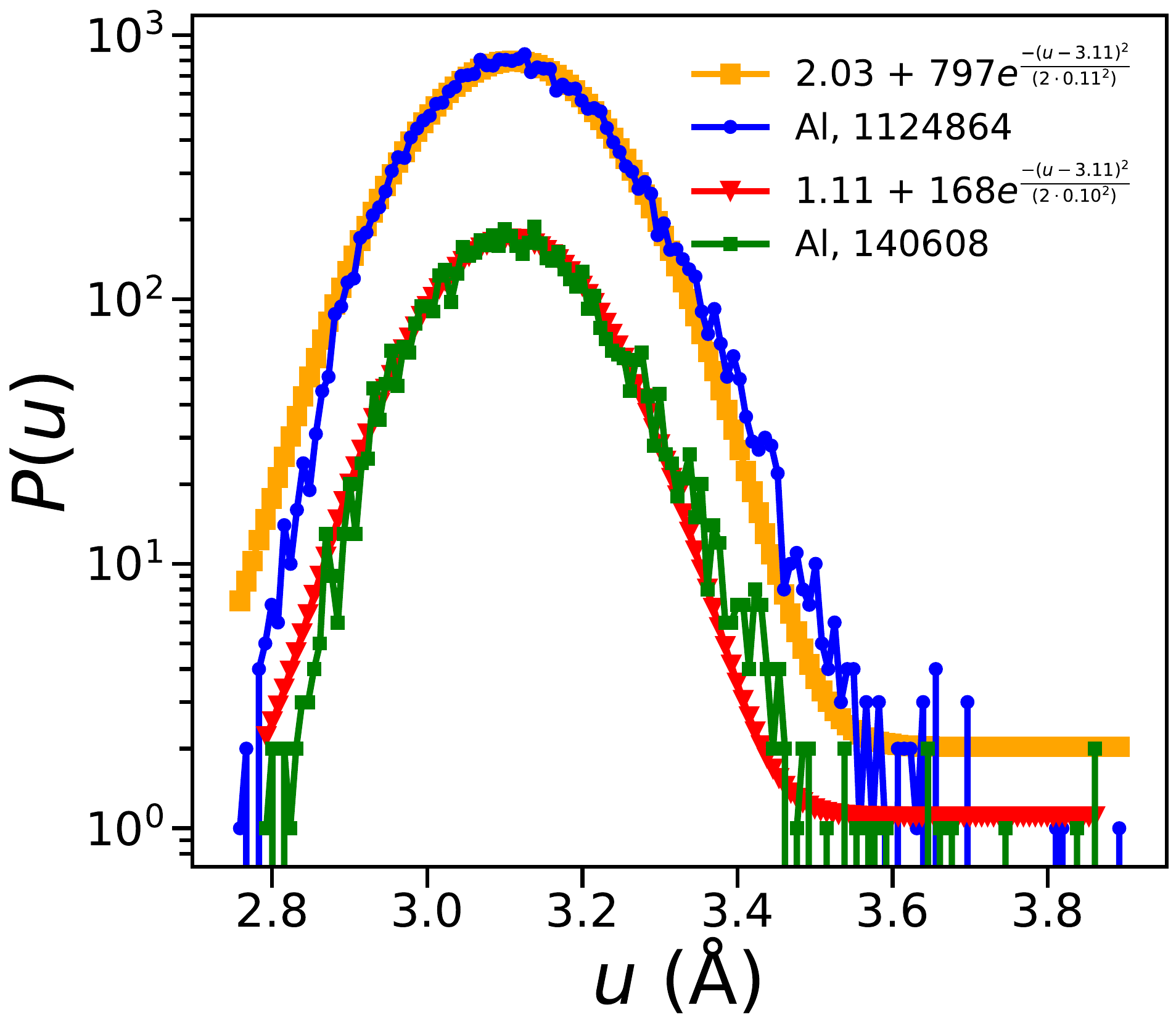}
  \includegraphics[height=7.2cm]{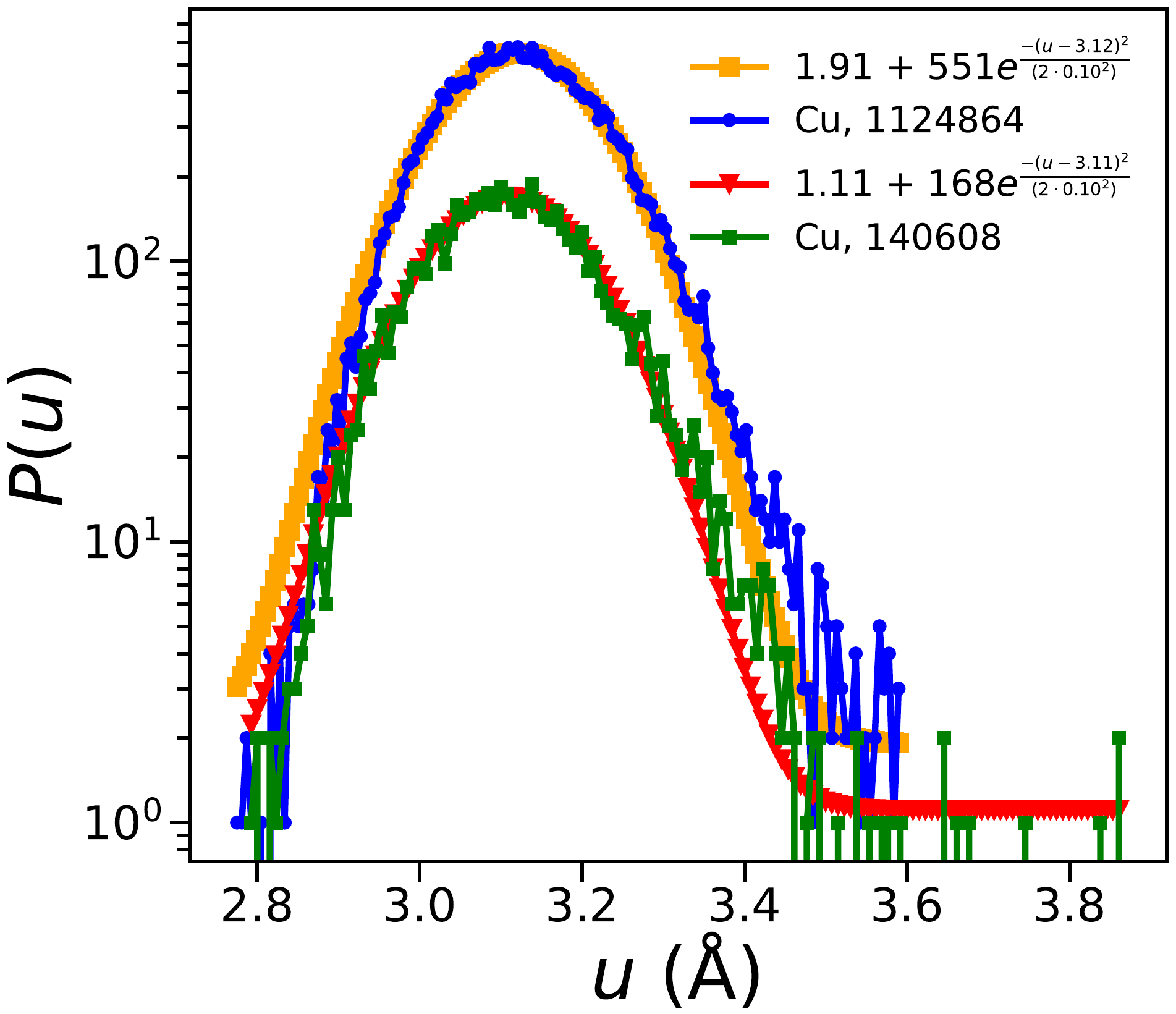}
  \caption{Interfacial separations distributions and corresponding Gaussian fits for selected NPs.}
  \label{fgr:gap_distributions_and_fits}
\end{figure*}

\begin{figure}[!h]
\centering
  \includegraphics[height=3.8cm]{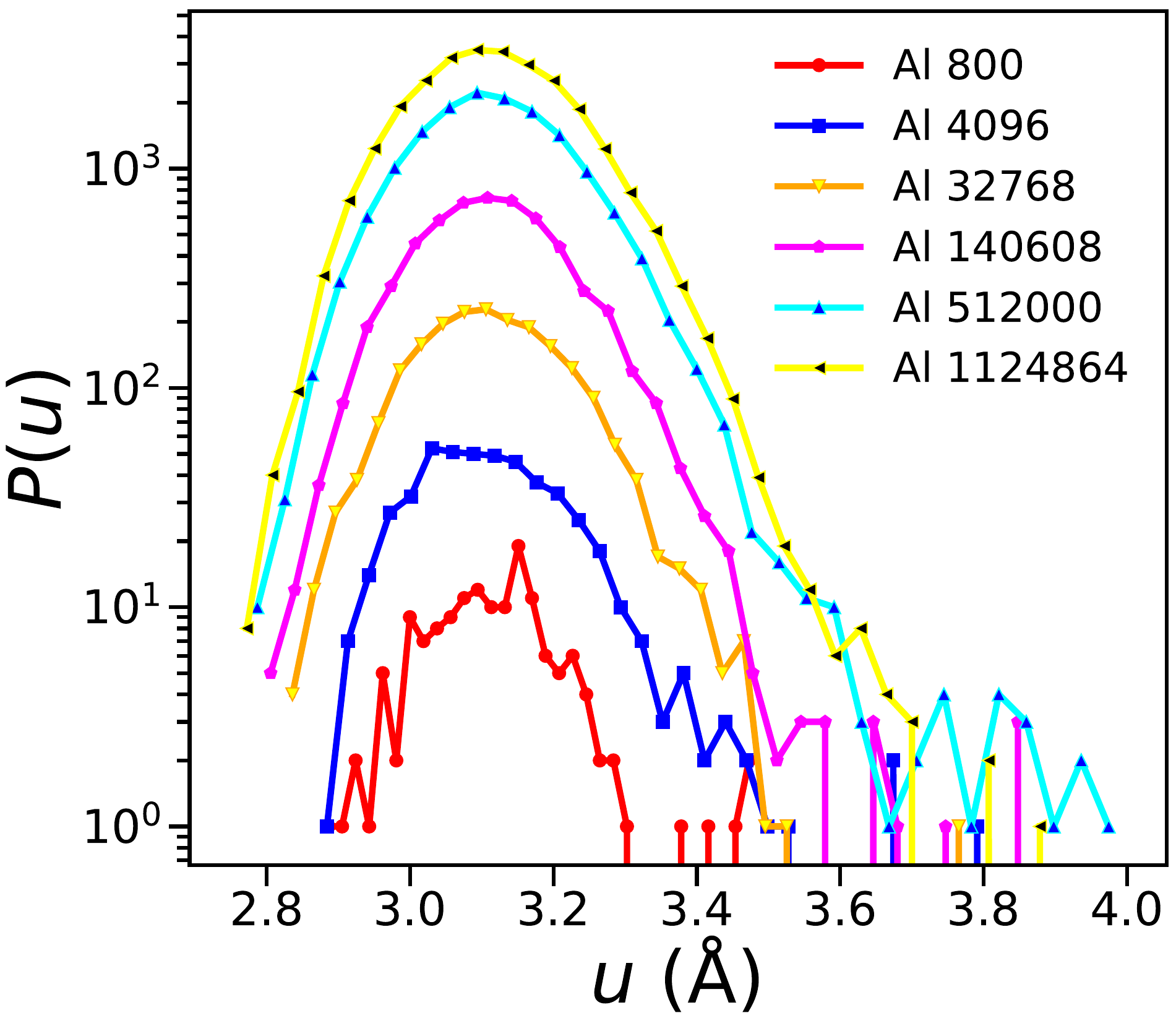}
  \includegraphics[height=3.8cm]{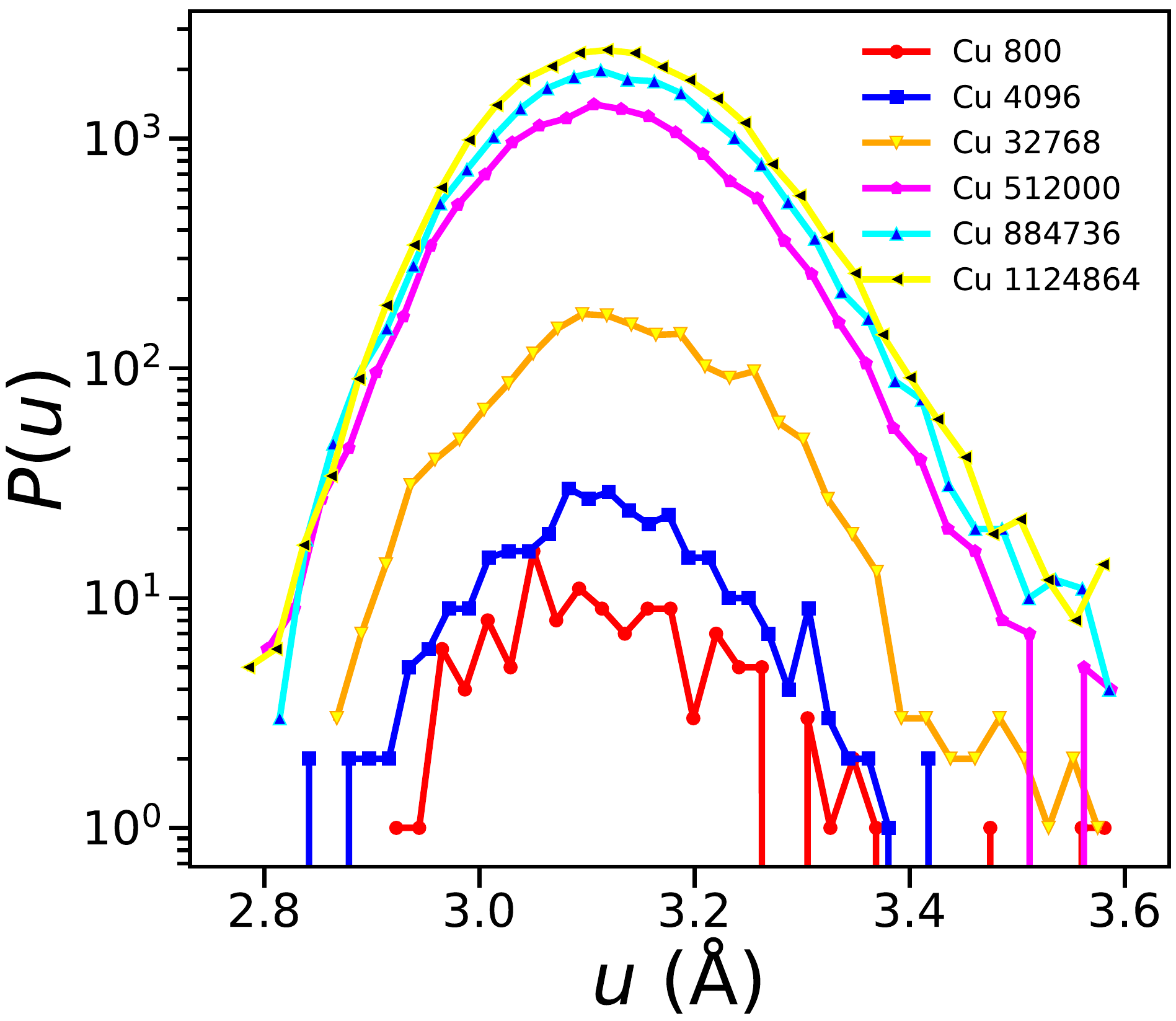}
  \caption{Bottom atomic-layer interfacial separation $u$ distributions for several NPs.}
  \label{fgr:bot_gap_distribution}
\end{figure}

\begin{figure}[!h]
\centering
  \includegraphics[height=3.6cm]{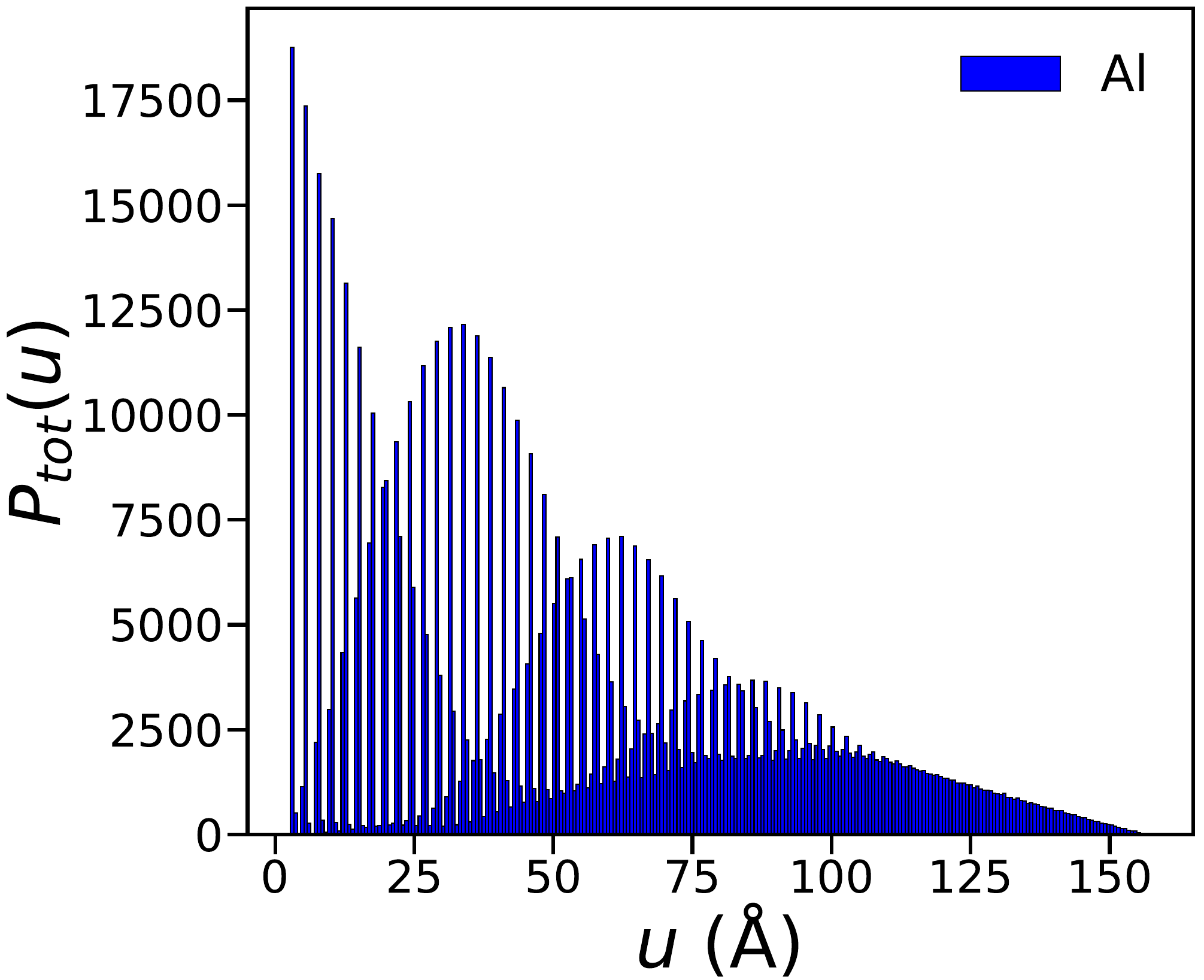}
  \includegraphics[height=3.6cm]{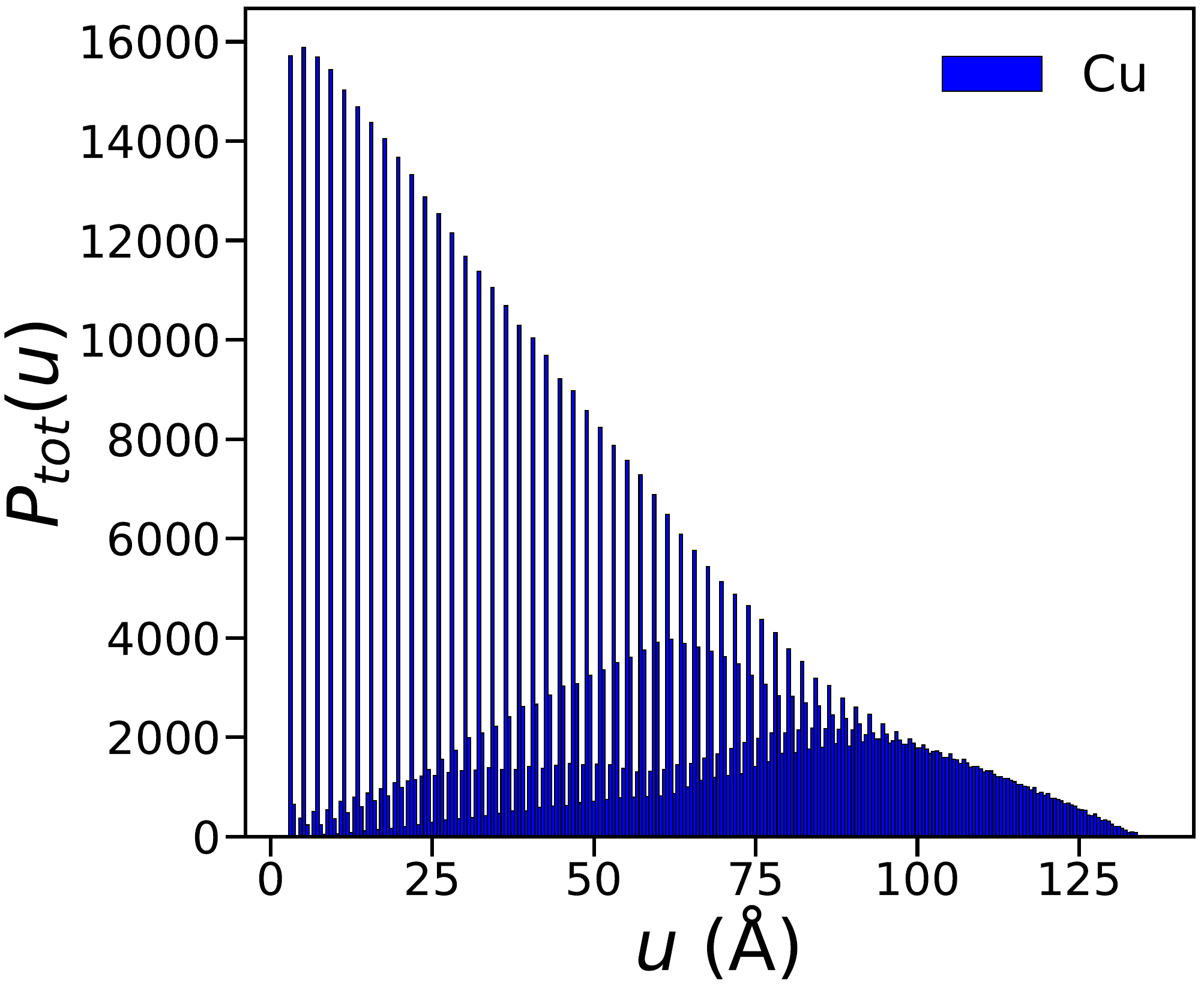}
  \caption{Histograms of separations $u$ for all the metal atoms in Al and Cu NPs containing 681472 atoms.}
  \label{fgr:interfacial_distances_all}
\end{figure}

The isotropic PSD $C_{h}(q)$ of NPs with $\gtrsim$ 175616 atoms or $L_{x} \gtrsim$ 20--25~nm starts to exhibit regions that change with frequency $q$ as a power law. Some of these regions correspond to the self-affine scaling~\cite{Persson2005jpcm} $C_{h}(q) \propto q^{-\beta}$, with $\beta = 2(1 + H)$. Power-law fitting can yield different results for $\beta$ depending on the frequency interval used for fitting. We obtained Hurst exponents $H$ in the range $\sim$~0.1--0.56 for different NP sizes. Self-affine regions in our results span $\lesssim 1$ decade of wavelengths, even for the largest NPs (that span 1.5~decades), in contrast to macroscopically rough surfaces, where self-affine scaling can extend over more than 9 decades of length scale~\cite{Jacobs2022mrsbul}. Self-affine $C_{h}(q)$ scaling of NPs' roughness may be related to their Gaussian $P(h)$ regions, as multiscale randomly rough surfaces with self-affine PSD typically exhibit a Gaussian $P(h)$.

\begin{figure}[!h]
\centering
  \includegraphics[height=3.9cm]{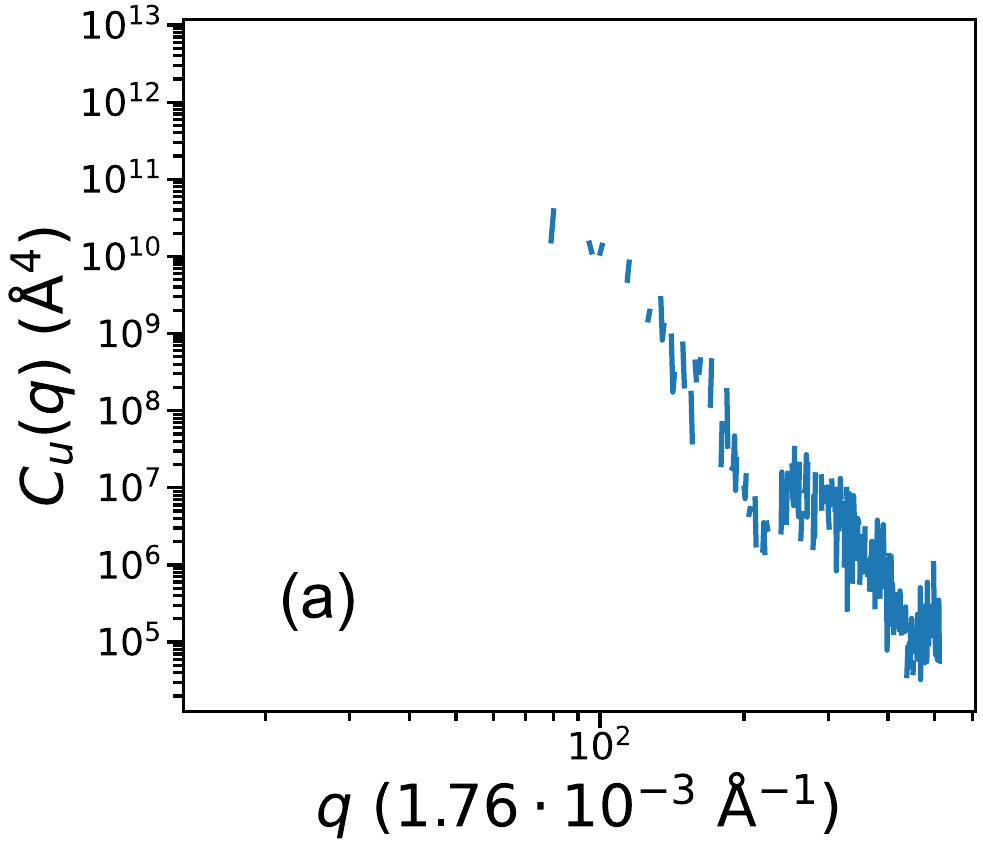}
  \includegraphics[height=3.9cm]{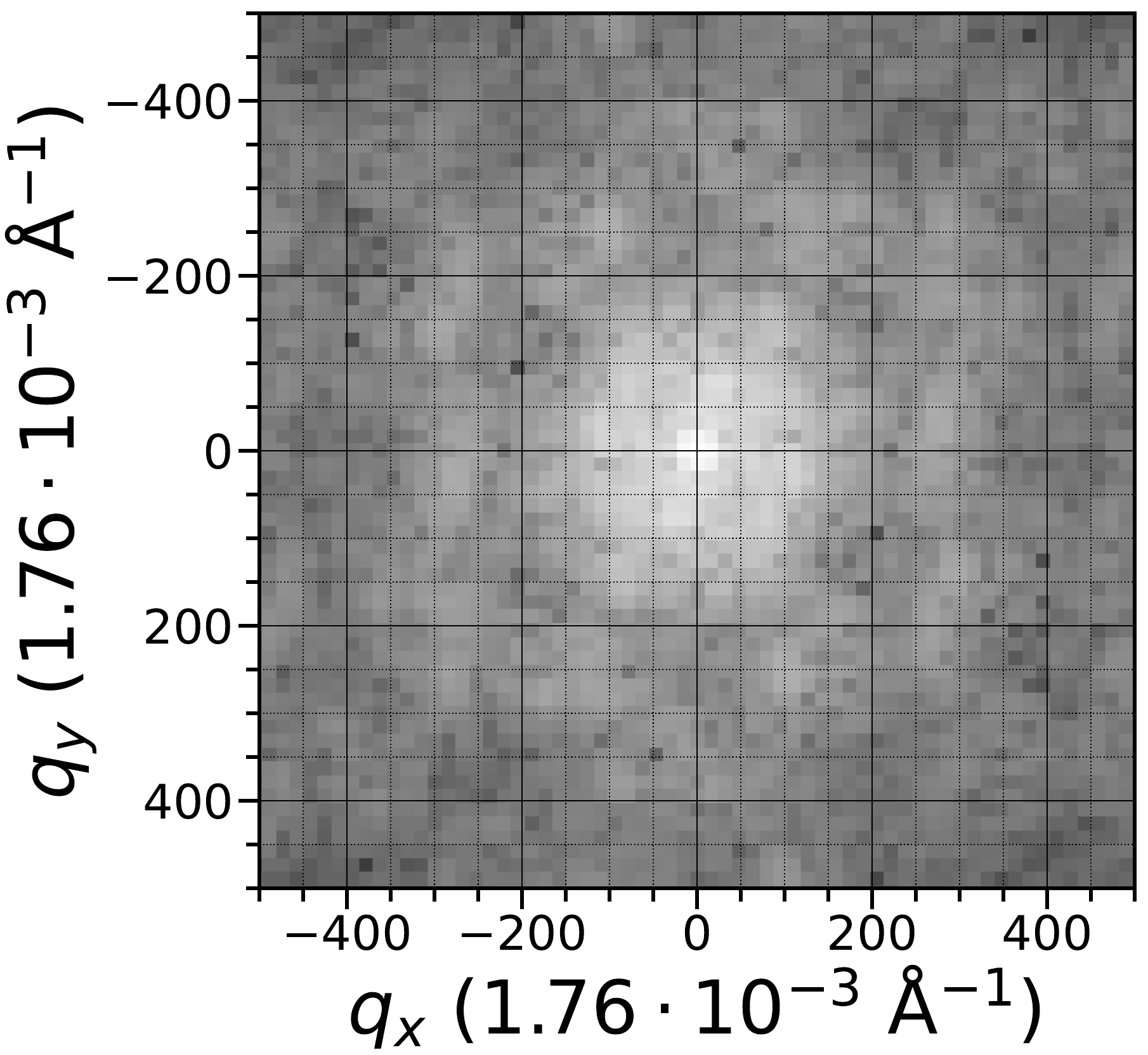}
  \includegraphics[height=3.9cm]{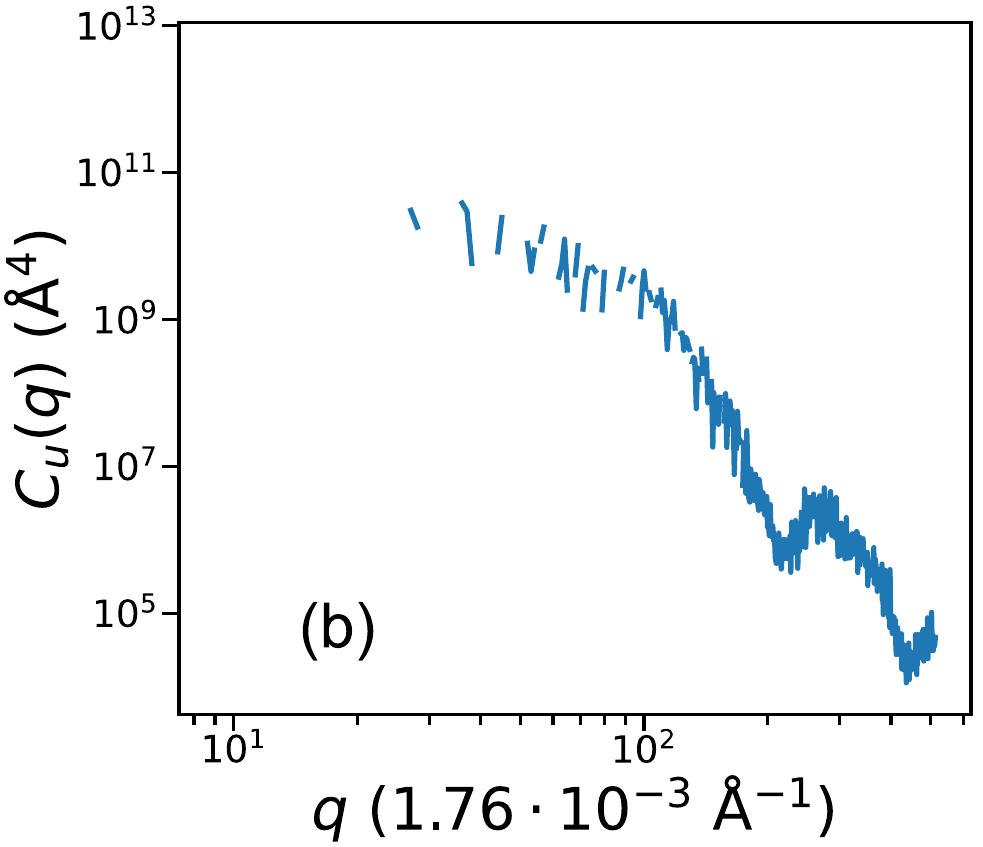}
  \includegraphics[height=3.9cm]{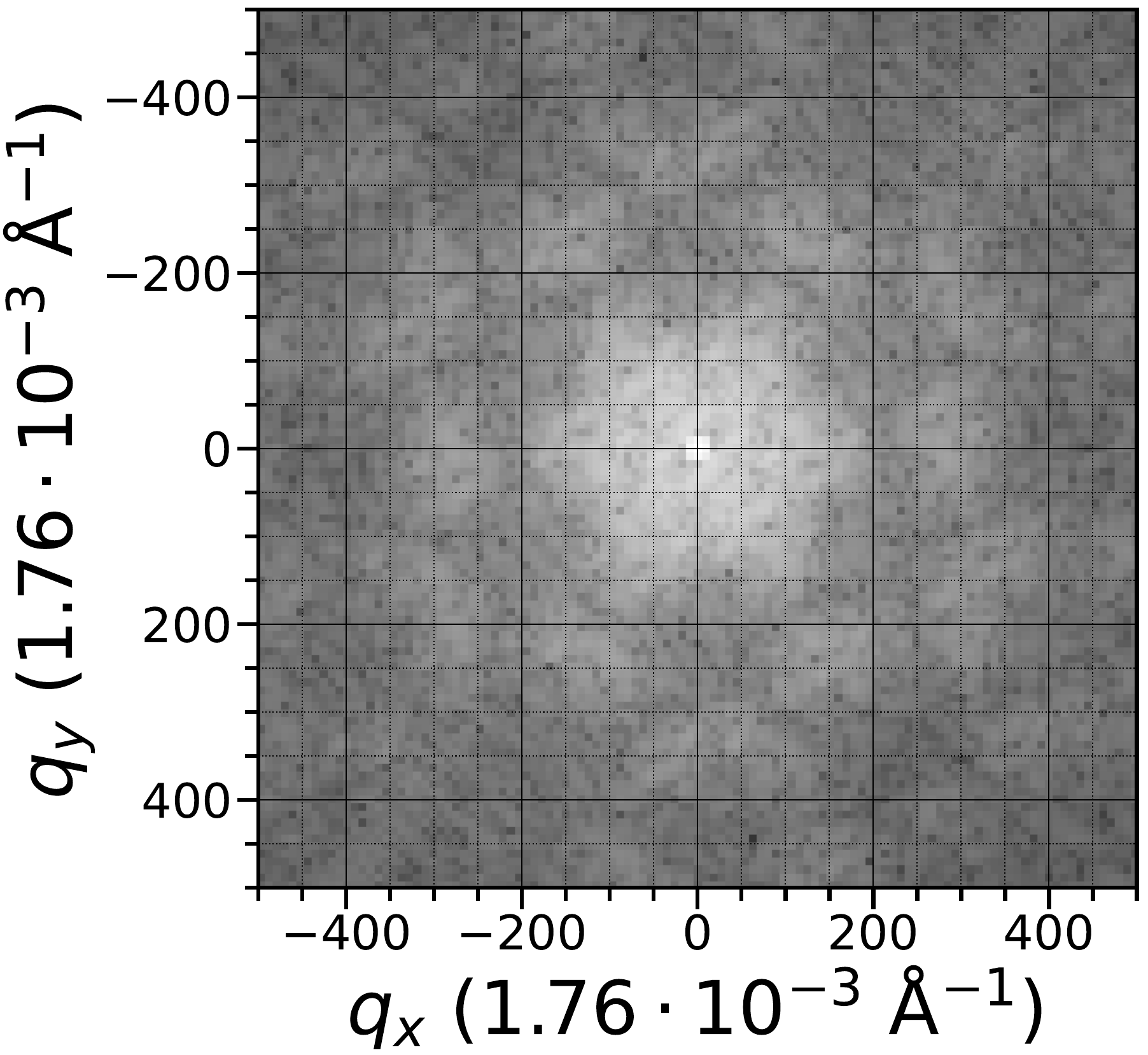}
  \includegraphics[height=3.9cm]{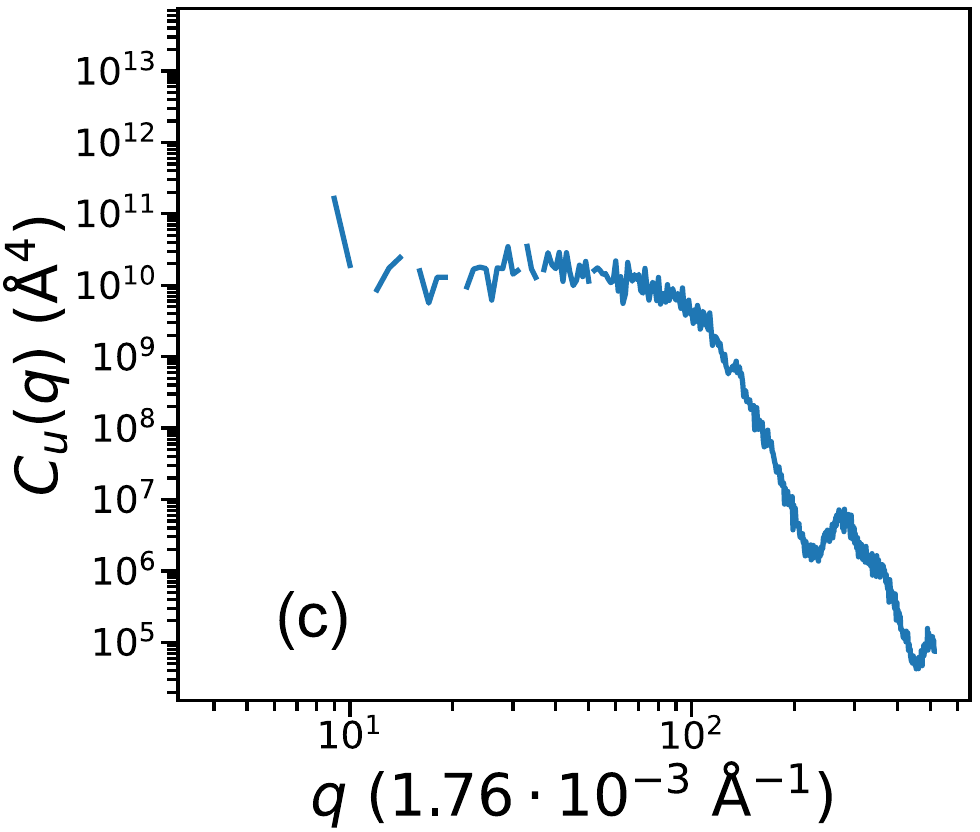}
  \includegraphics[height=3.9cm]{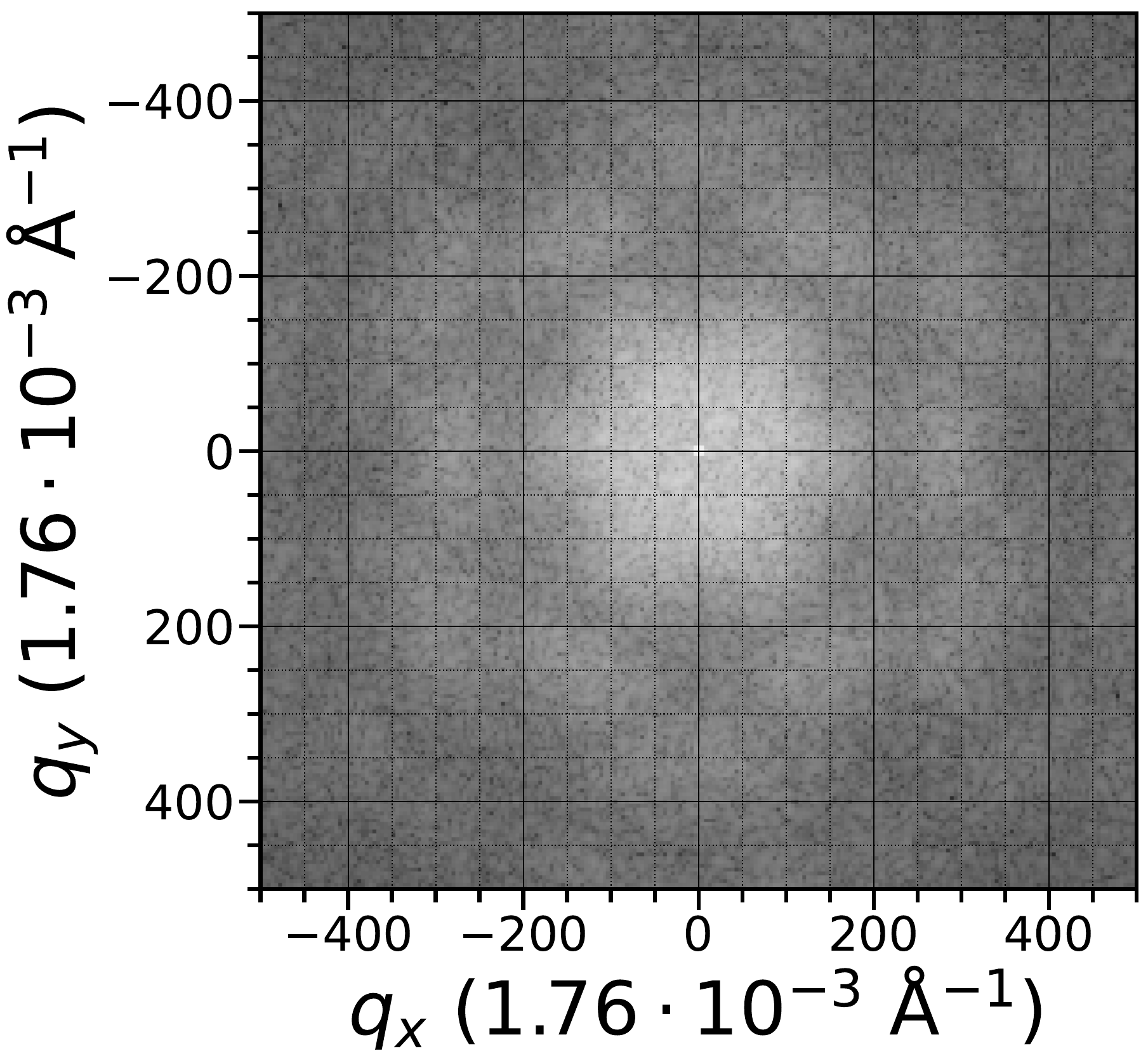}
  \includegraphics[height=3.9cm]{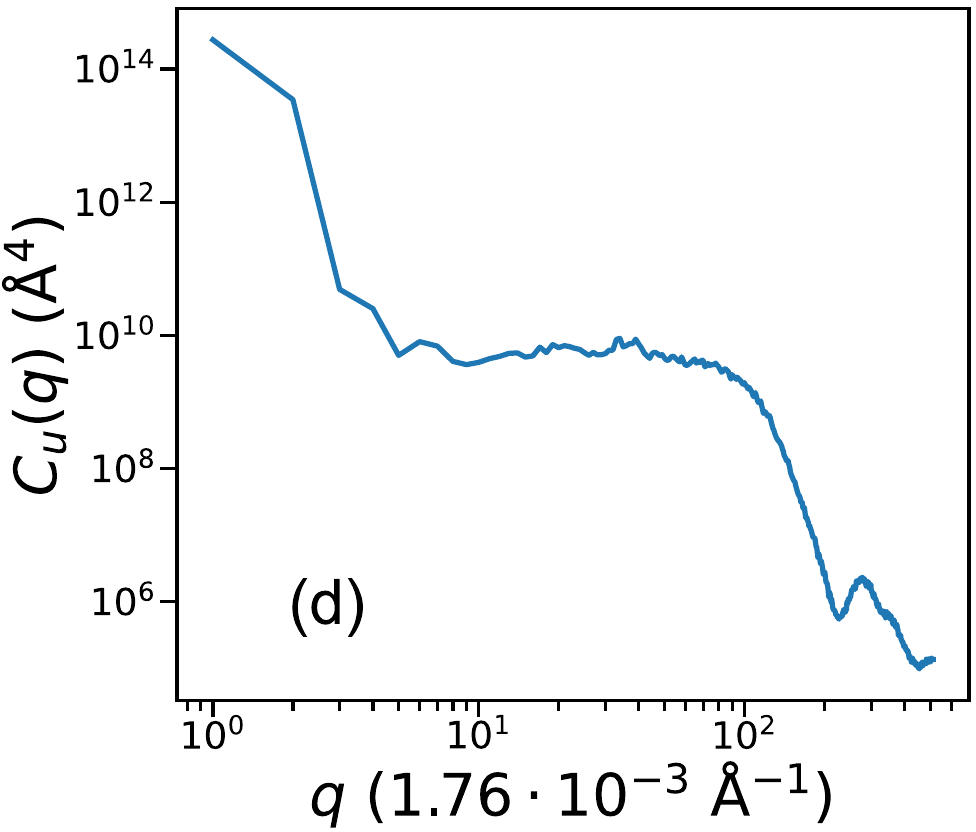}
  \includegraphics[height=3.9cm]{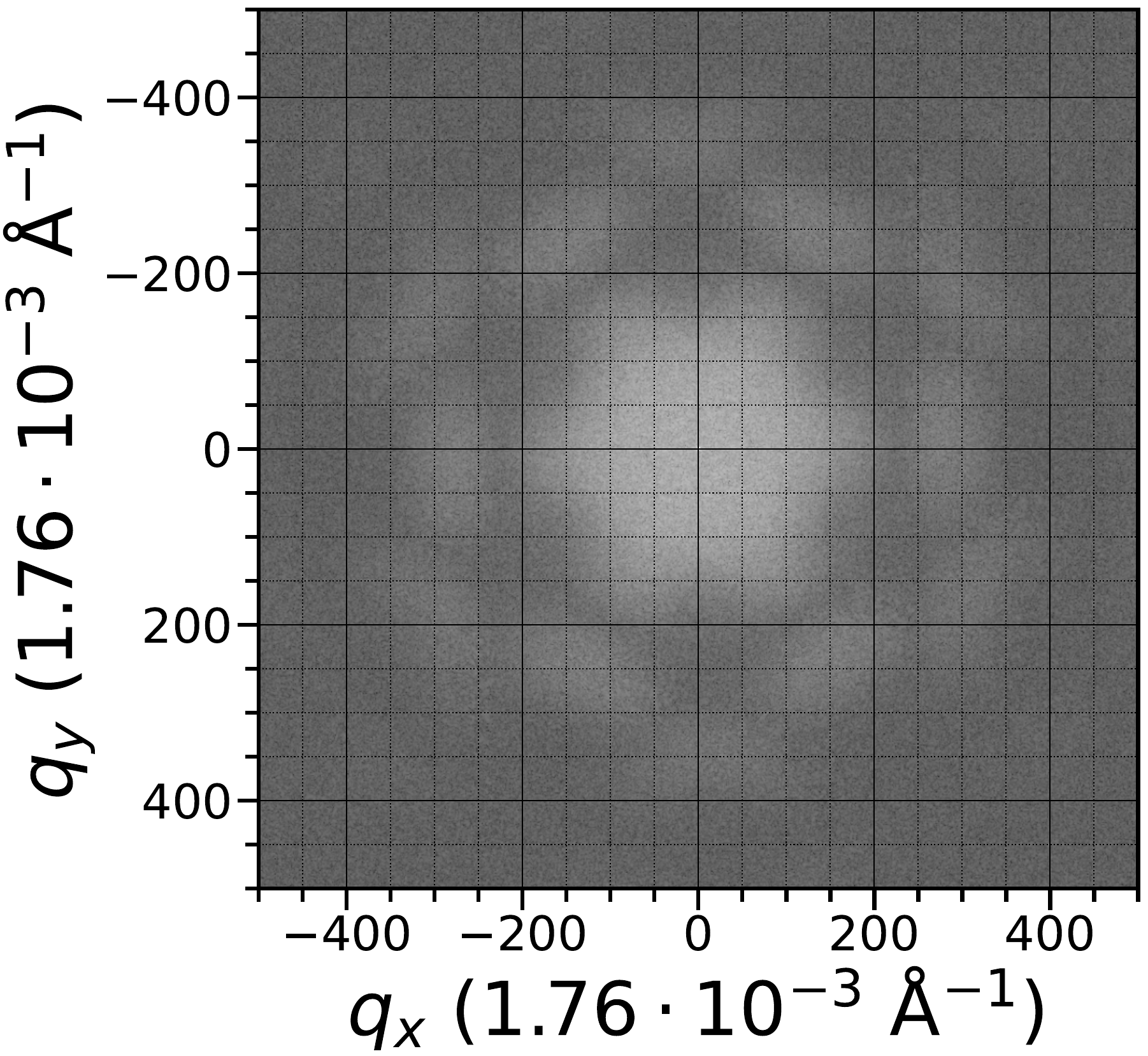}
  \caption{Isotropic $C_{u}(q)$ log-log plots and 2D PSD $C_{u}(q_{x}, q_{y})$ heatmaps for Al NPs with 800 (a), 4096 (b), 32768 (c), 1124864 (d) atoms.}
  \label{fgr:gap_bottom_psd_al}
\end{figure}

\begin{figure}[h]
\centering
  \includegraphics[height=3.9cm]{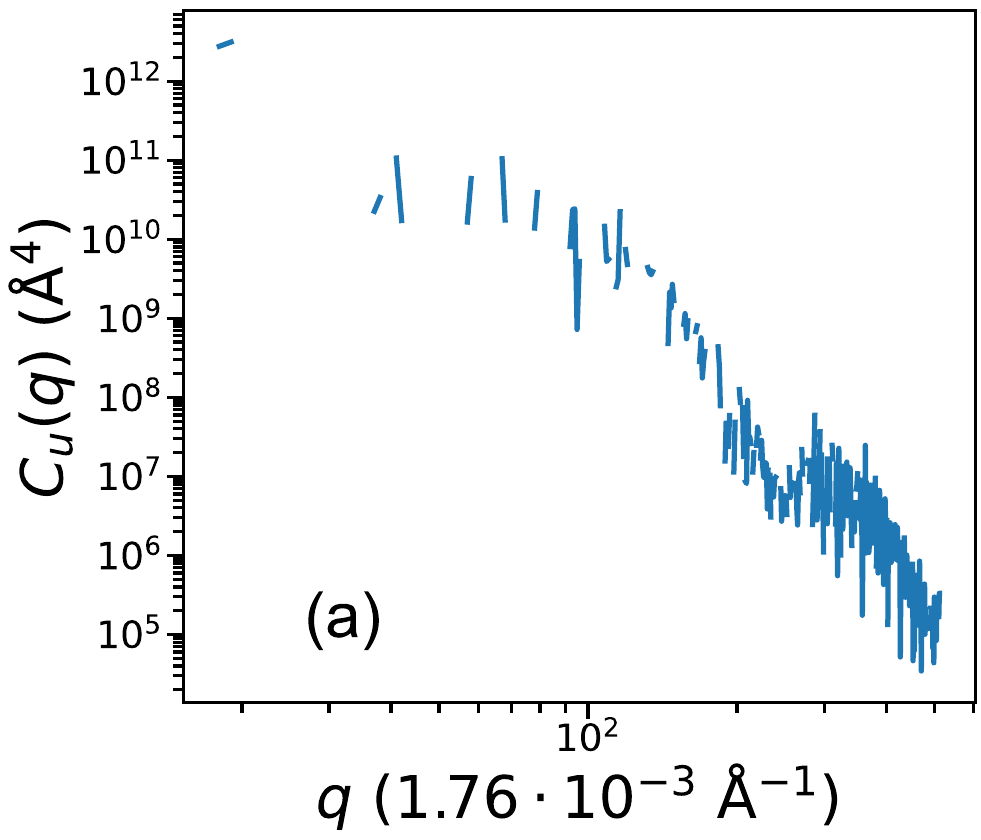}
  \includegraphics[height=3.9cm]{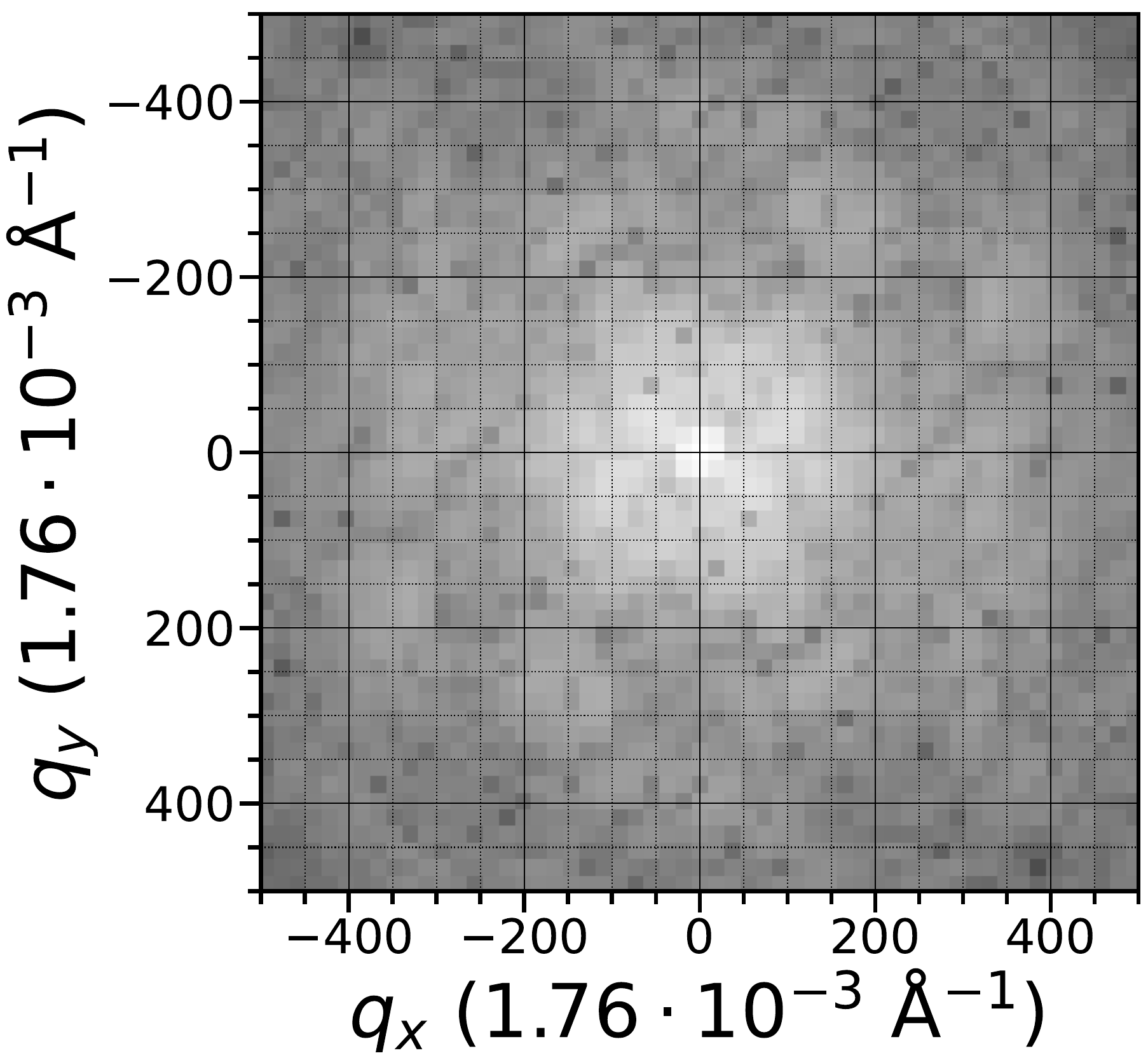}
  \includegraphics[height=3.9cm]{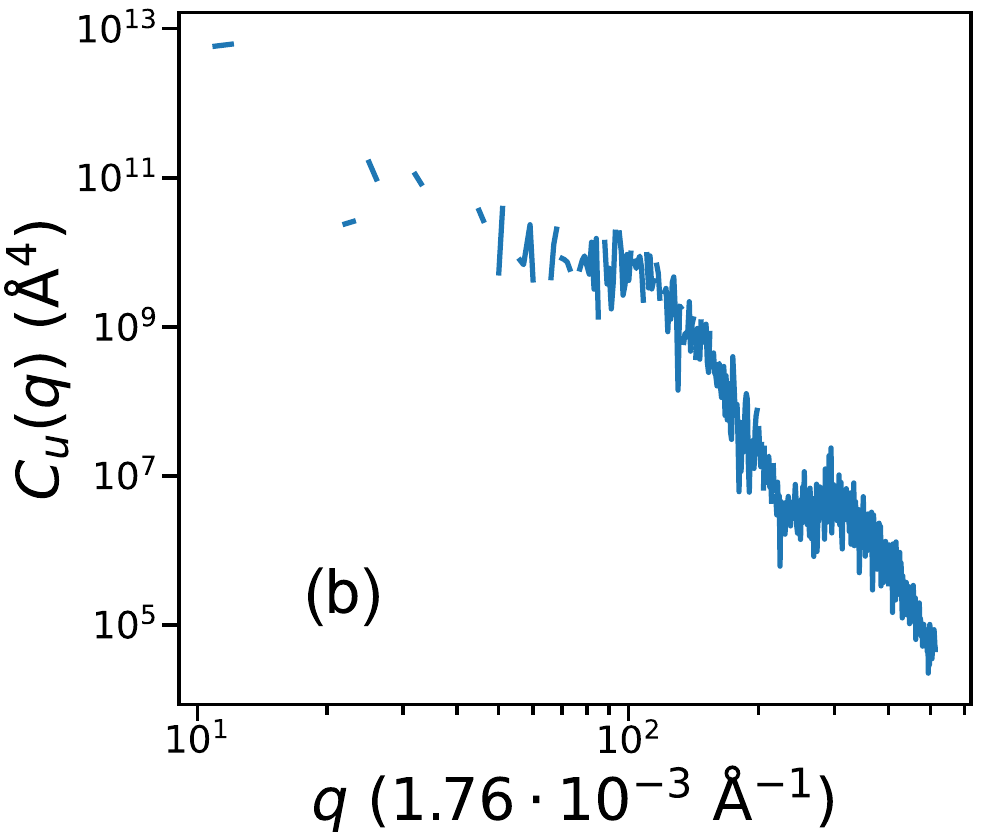}
  \includegraphics[height=3.9cm]{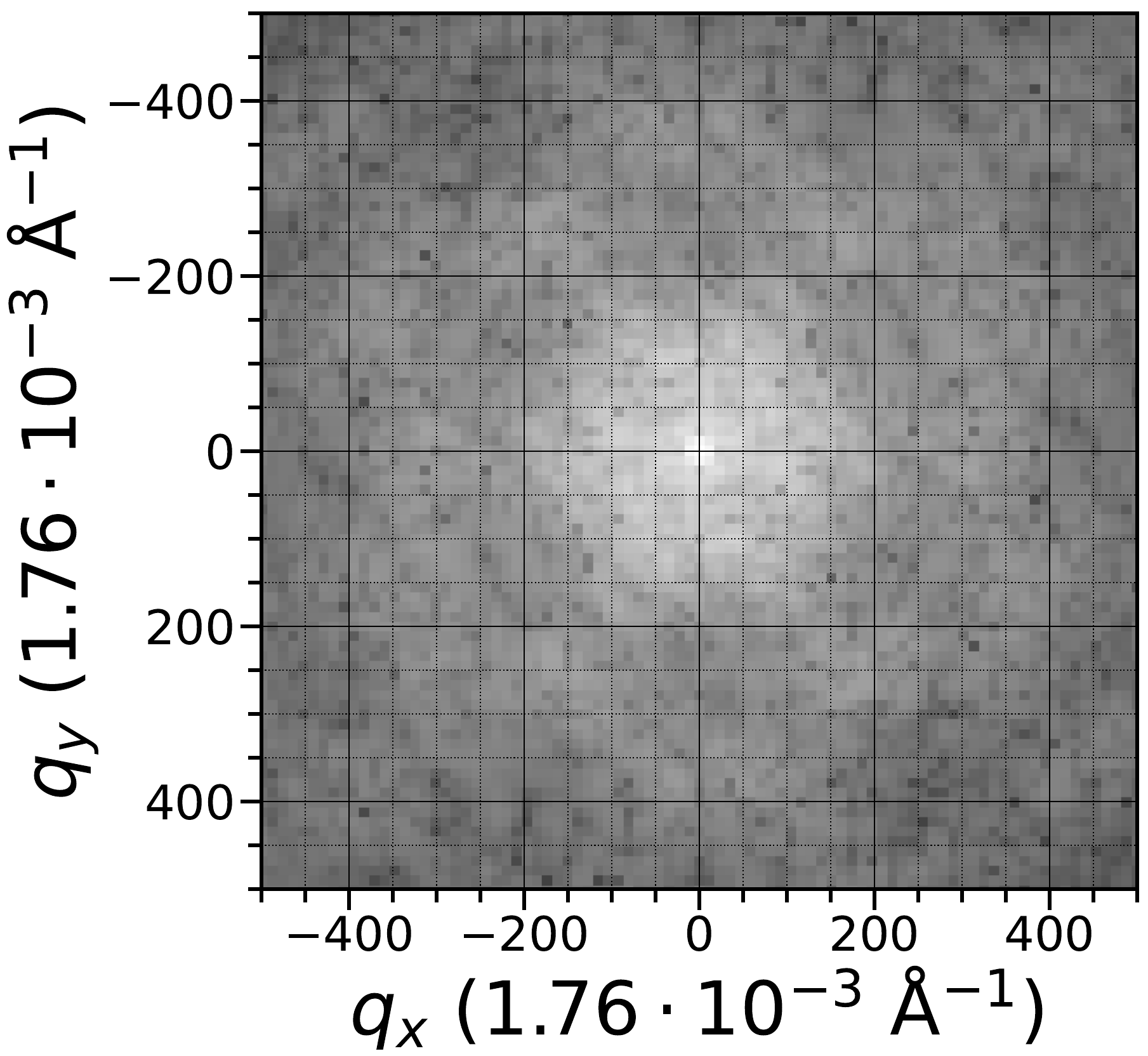}
  \includegraphics[height=3.9cm]{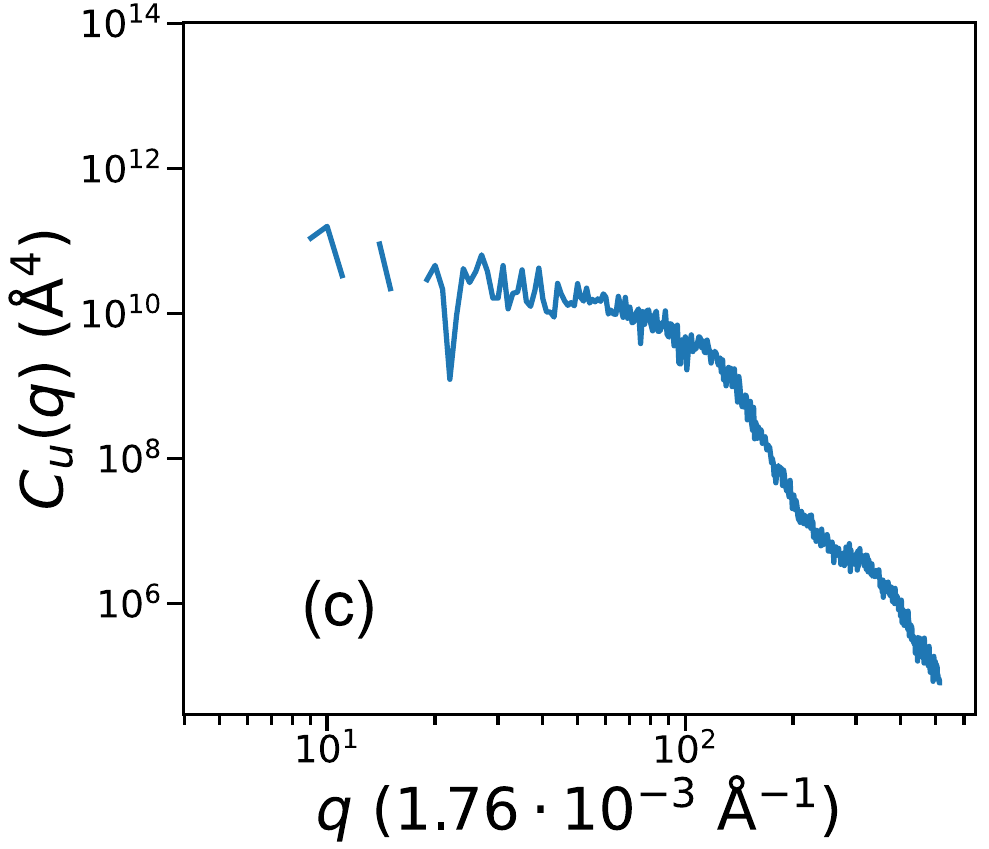}
  \includegraphics[height=3.9cm]{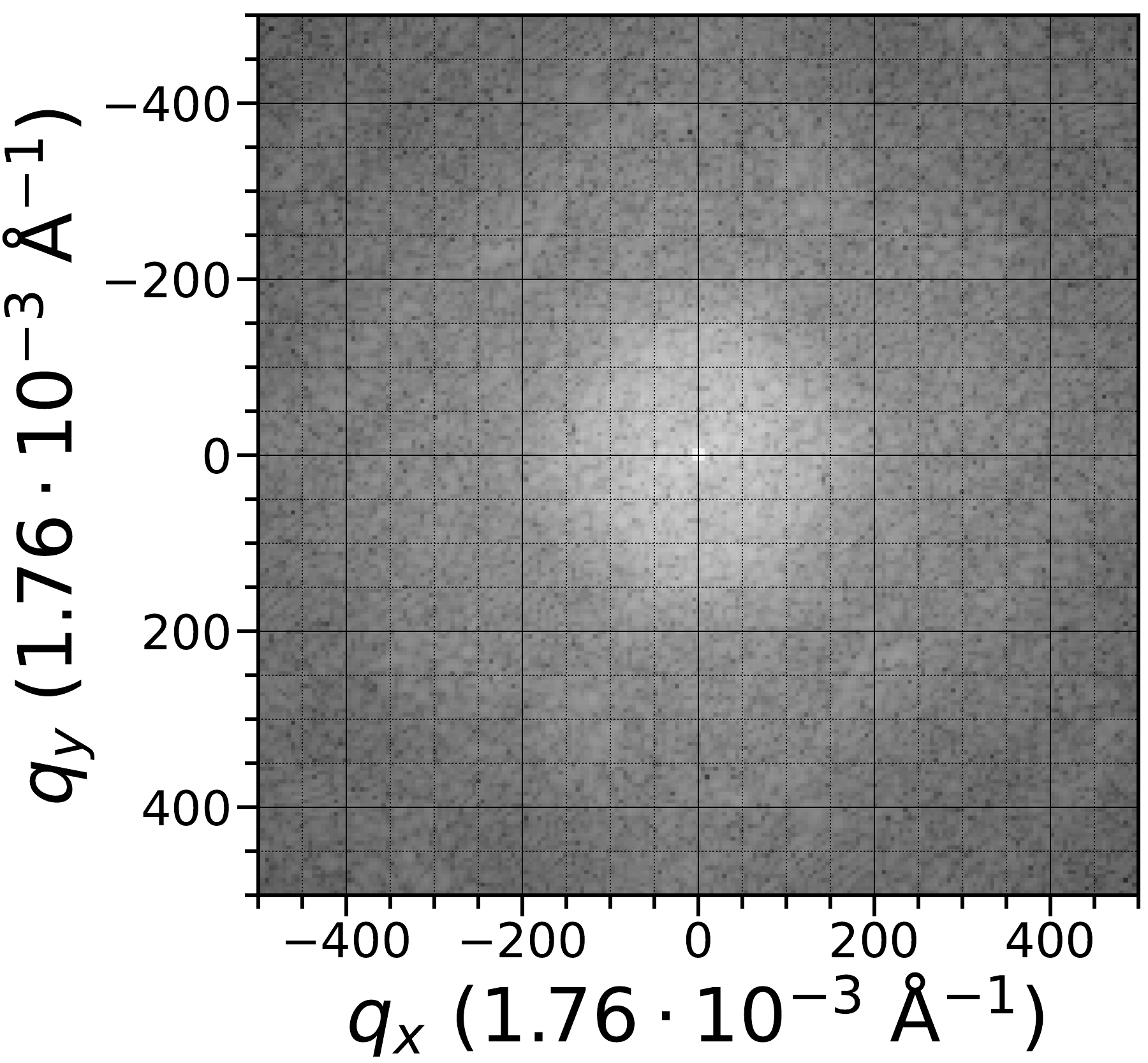}
  \includegraphics[height=3.9cm]{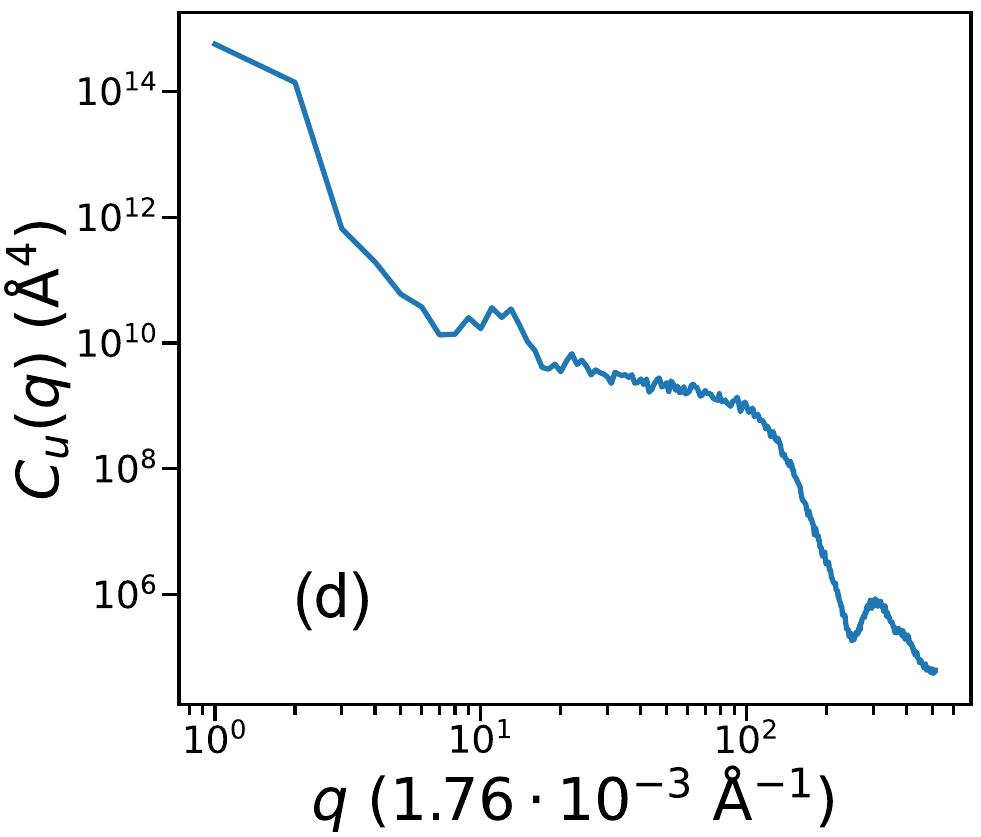}
  \includegraphics[height=3.9cm]{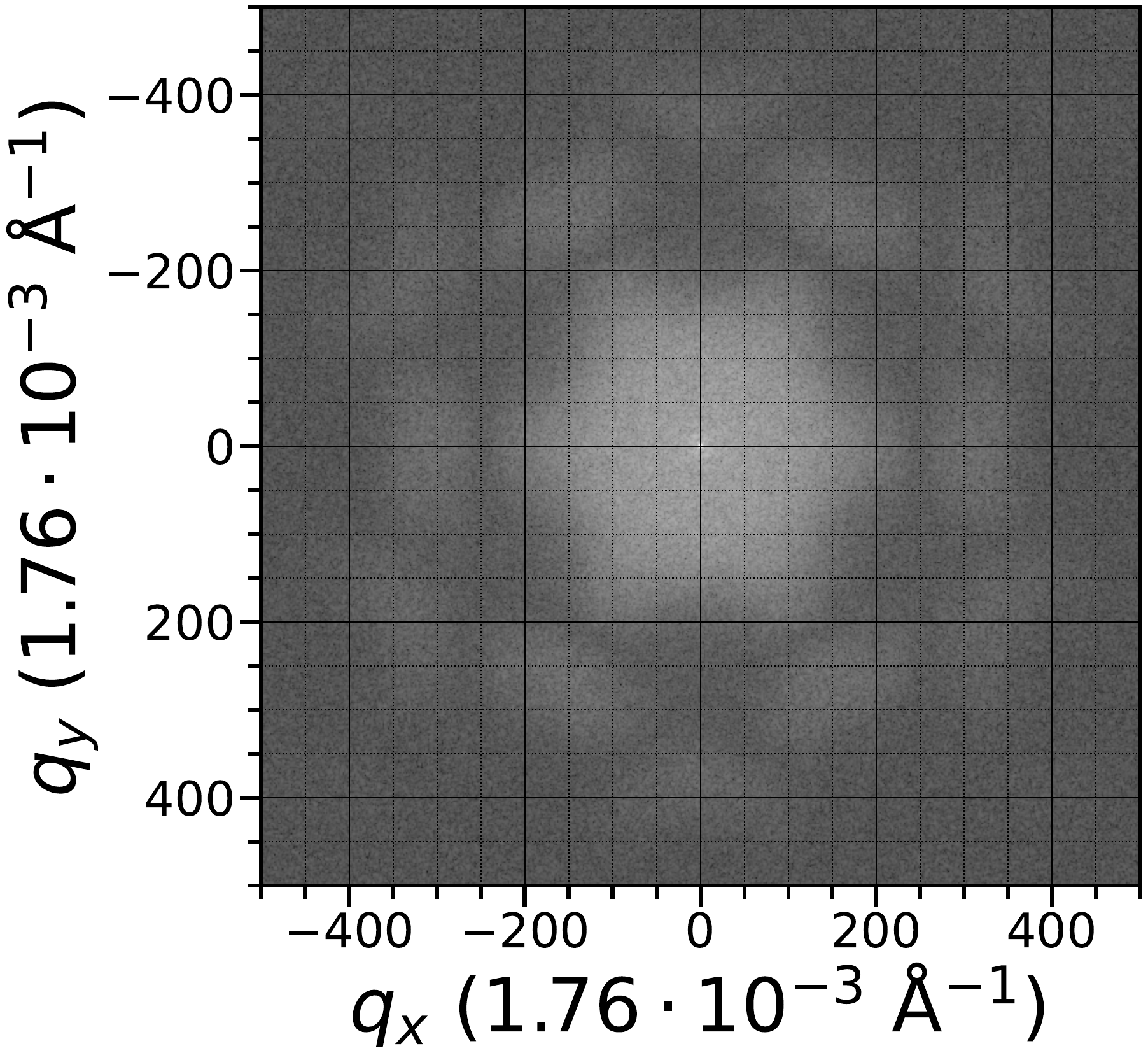}
  \caption{$C_{u}(q)$ log-log plots and $C_{u}(q_{x}, q_{y})$ heatmaps for Cu NPs with 800 (a), 4096 (b), 32768 (c), 1124864 (d) atoms.}
  \label{fgr:gap_bottom_psd_cu}
\end{figure}

$C_{h}(q)$ for both metals also has a high-frequency, quickly decaying region, starting at $\lambda \sim$~5~\AA~and ending at approximately the nearest-neighbor distance of each metal. Even though this region can at least partially be approximated by a power law, it does not exhibit self-affine scaling. It reflects the atomic-scale roughness contributions to the height power.

\subsection{Contact mechanics}
\label{sec:cm}

\subsubsection{Interfacial separation.}
\label{sec:gap}

Fig.~\ref{fgr:side_top_view_bot_gap_map} shows the $u(x, y)$ heatmap of the interface atomic layer for the largest NPs. The gap values are randomly distributed over the surface area in contrast to the surface height $h(x, y)$, which exhibits regions of gradually changing values, cf.  Figs.~\ref{fgr:side_top_view_bot_height_map},~\ref{fgr:side_top_view_bot_height_map_largest}. For Cu, some grouping of similar $u$ values can be observed; see the yellow regions in Fig.~\ref{fgr:side_top_view_bot_gap_map} (b). The gap distributions $P(u)$ in Figs.~\ref{fgr:gap_distributions_and_fits},~\ref{fgr:bot_gap_distribution} confirm the random nature of $u$: for most NPs, $P(u)$ is close to a Gaussian. Fig.~\ref{fgr:gap_distributions_and_fits} shows $P(u)$ and corresponding Gaussian fits with corresponding values of the fitting parameters for NPs containing 140608 and 1124864 atoms. The drastically different behavior of $P(u)$ compared to $P(h)$ may be due to elastic deformations of the graphene substrate, which adapt to the NP surface topography, thereby eliminating the narrow spike in $P(u)$. Also note that the width of $P(u)$ is less than 2~\AA~even for the largest NPs, which is an order of magnitude smaller than the height spread of more than 1~nm. Interestingly, $P(u)$ is consistent with the behavior of macroscopic randomly rough surfaces, where the gap distribution is Gaussian under zero external (squeezing) pressure~\cite{Almqvist2011jmps}.

The gap distributions $P_{tot}(u)$ for all atoms comprising a NP (where $u$ is the distance from an atom to the substrate) exhibit a clear layered ordering, confirming the polycrystalline structure of both Al and Cu NPs; see Fig.~\ref{fgr:interfacial_distances_all} for an example of $P_{tot}(u)$ for a NP with 681472 atoms. Other NPs have similar histograms. The bottom layer atoms correspond to the bar with the smallest $u$.

The PSDs $C_{u}(q_{x}, q_{y})$ and $C_{u}(q)$ of Al and Cu NPs are displayed in Figs.~\ref{fgr:gap_bottom_psd_al},~\ref{fgr:gap_bottom_psd_cu}, respectively. Compared to the height PSDs in Figs.~\ref{fgr:h_bottom_psd_al},~\ref{fgr:h_bottom_psd_cu}, the high-frequency components with hexagonal symmetry are present for all NP sizes, not just the smaller ones. The high frequency component wavelength $\lambda_{x} \sim$~2.1--2.5~\AA~along the $x$-direction (zigzag graphene edge). Along the $y$-direction (armchair) $\lambda_{y} \sim$~1.4--2.3~\AA. Here, 1.4~\AA~corresponds to the covalent bond length in graphene. This might indicate that these contributions to $C_{u}$ come from the substrate atoms. Additionally, $C_{u}(q)$ can have a maximum or a roll-off region starting at around 5~\AA~($q \sim 110$ in the plots). There can be additional PSD spikes, e.g., as can be seen in Fig.~\ref{fgr:gap_bottom_psd_cu} (d) for Cu at $q \sim 12$ or $\lambda = 56.8$~\AA, which may be caused by spurious roughness due to the peculiarities of the NP production. However, such local extrema are not consistent for all the NPs. Therefore, the gap PSDs clearly indicate a contribution to $C_{u}$ from the smaller wavelengths down to the graphene bond lengths, even though this contribution is orders of magnitude smaller than that of the larger wavelengths. One similarity of $C_{u}(q)$ with $C_{h}(q)$ is the presence of the quickly decaying high-frequency region within approximately the same frequency range. Another similarity is the smeared $C_{u}(q_{x}, q_{y})$ for the Cu NP having 32768 atoms in Fig.~\ref{fgr:gap_bottom_psd_cu} (c), suggesting that the contributions for $u$ are spread over a range of wavelengths, which is also confirmed by the smooth shape of the corresponding $C_{u}(q)$.

\begin{figure}[h]
\centering
  \includegraphics[height=3.8cm]{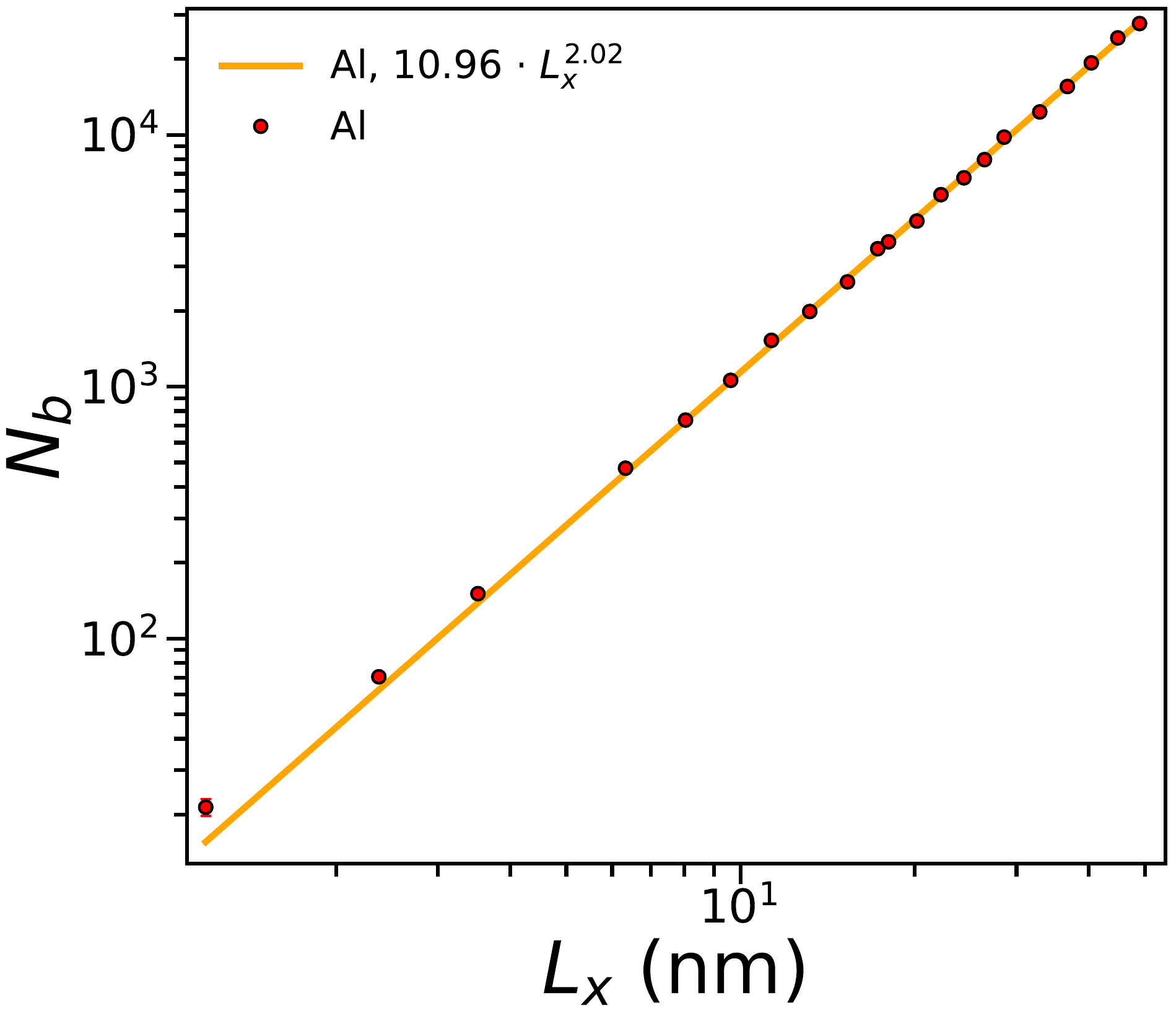}
  \includegraphics[height=3.8cm]{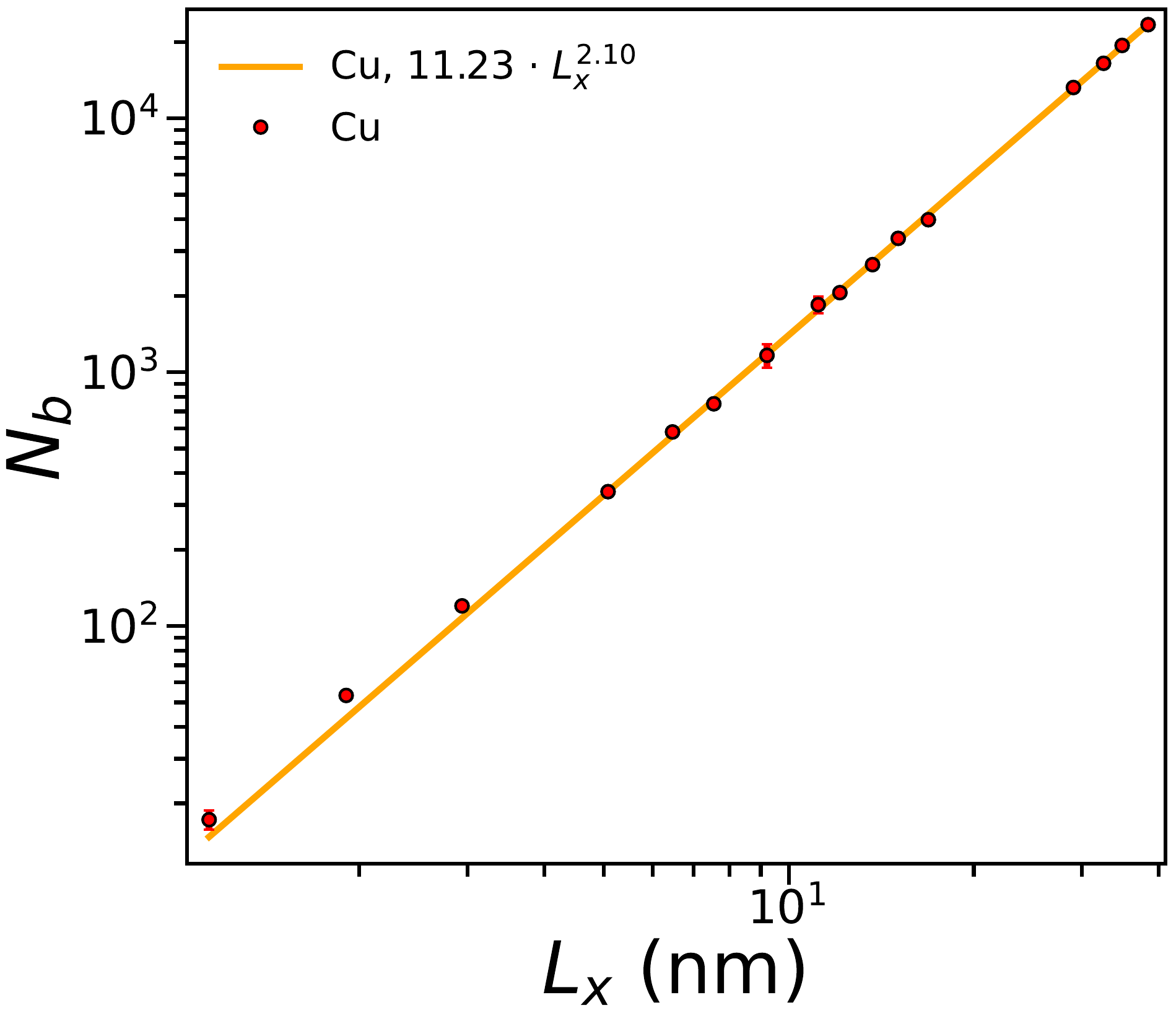}
  \caption{Number of atoms in the bottom layer vs $L_{x}$, with corresponding power-law fit.}
  \label{fgr:nbot_vs_size_x}
\end{figure}

The number of atoms $N_{b}$ in the contact layer, that directly affects the mean gap $\bar{u}$, also exhibits size-dependent scaling: it grows quadratically with $L_{x}$ for most NPs, except for the smallest ones whose size is $\lesssim$ 3--6~nm, cf. Fig.~\ref{fgr:nbot_vs_size_x}.

\begin{figure}[h]
\centering
  \includegraphics[height=3.6cm]{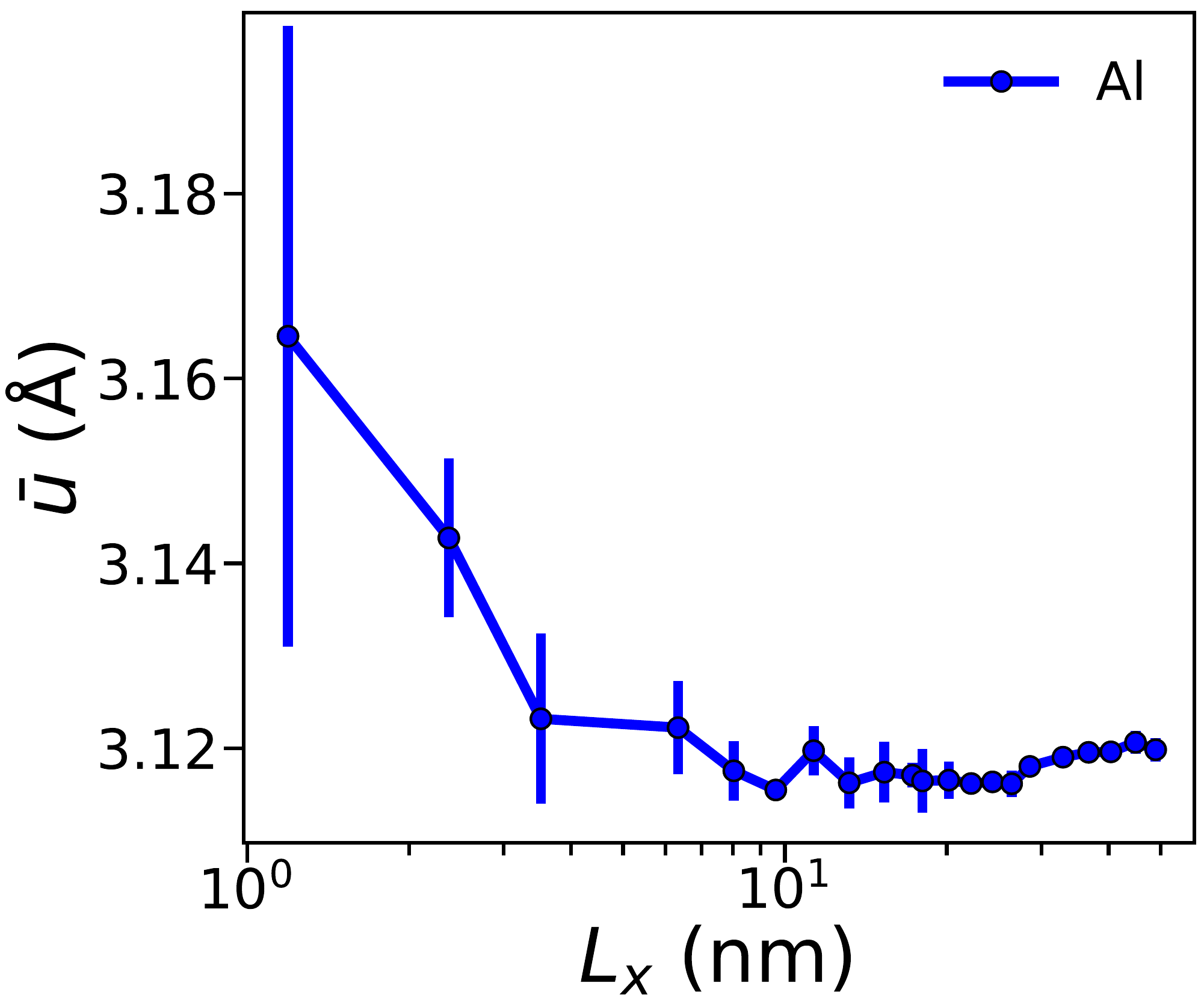}
  \includegraphics[height=3.6cm]{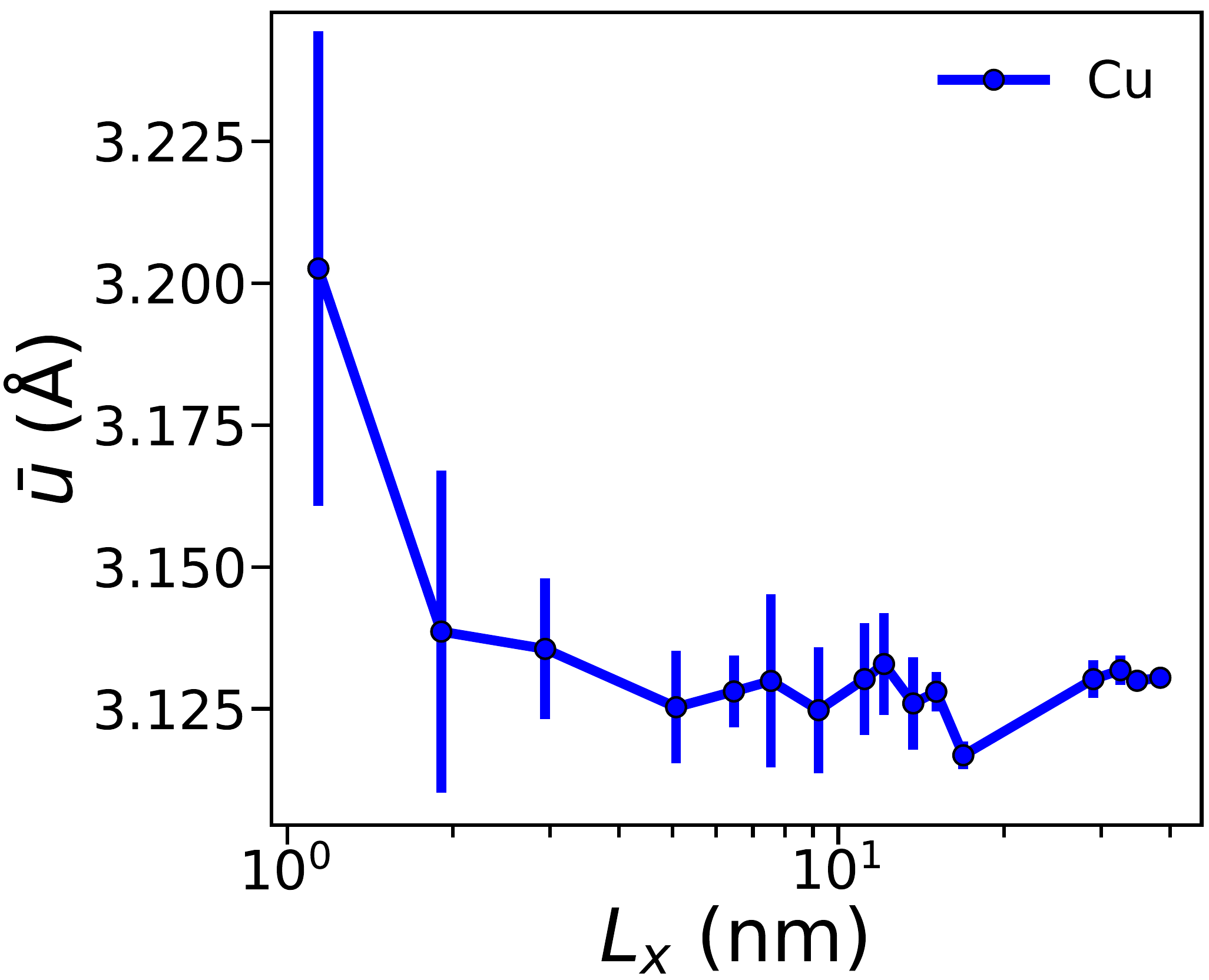}
  \caption{Mean gap $\bar{u}$ vs the lateral NP size $L_{x}$.}
  \label{fgr:mean_gap_vs_size_x}
\end{figure}

\begin{figure}[h]
\centering
  \includegraphics[height=3.65cm]{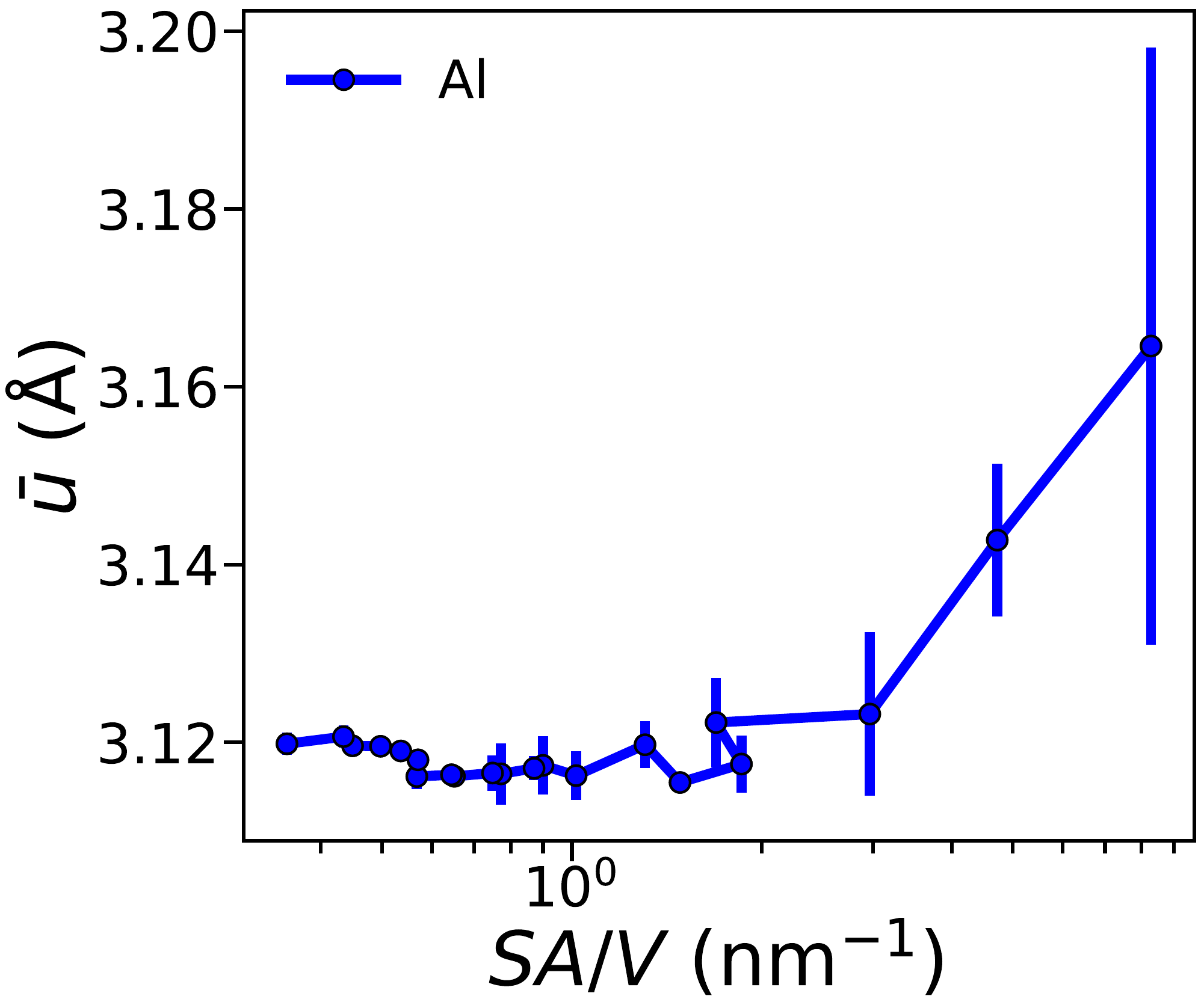}
  \includegraphics[height=3.65cm]{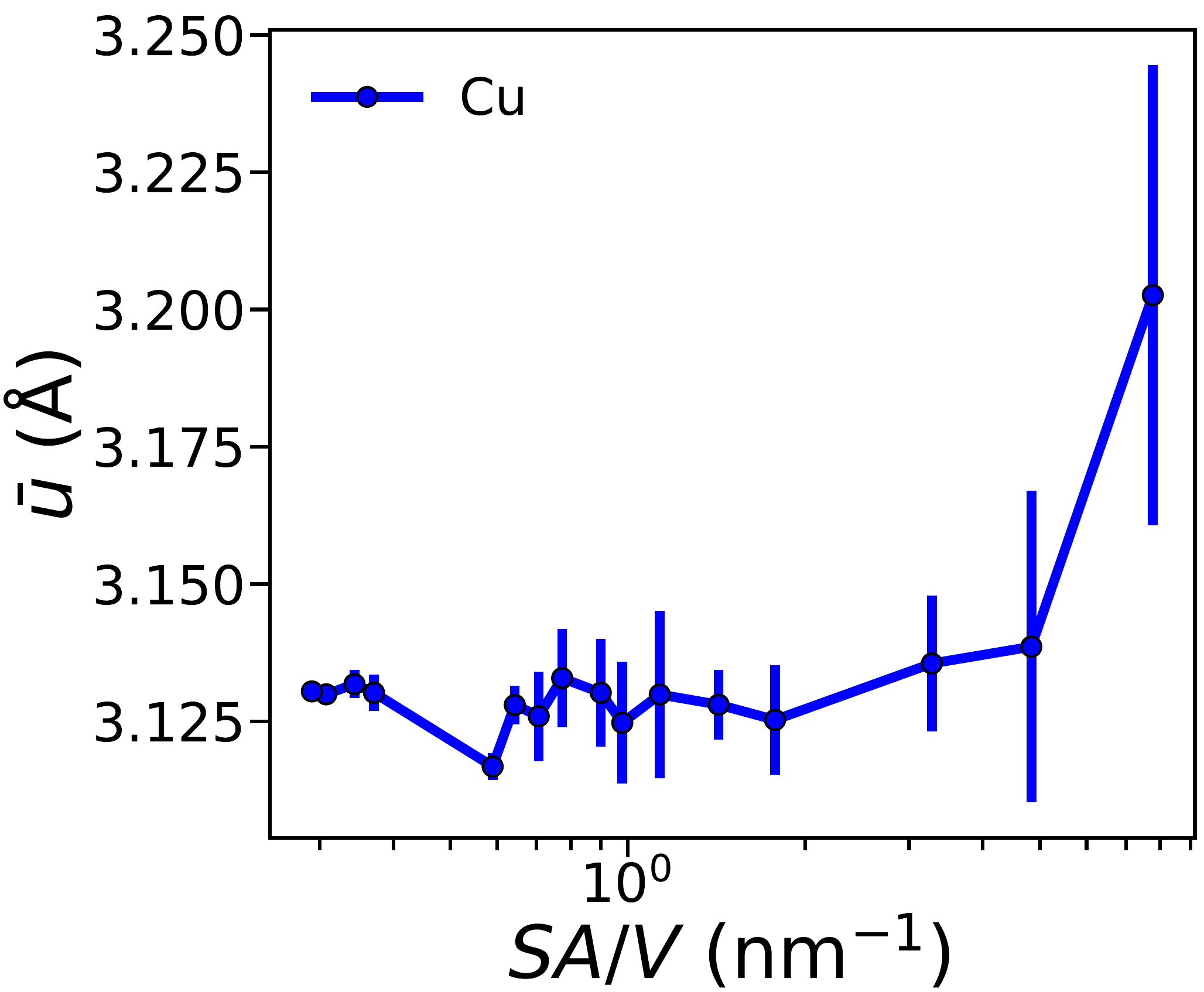}
  \caption{Mean gap $\bar{u}$ vs $SA/V$.}
  \label{fgr:mean_gap_vs_sav}
\end{figure}

\begin{figure}[!h]
\centering
  \includegraphics[height=3.6cm]{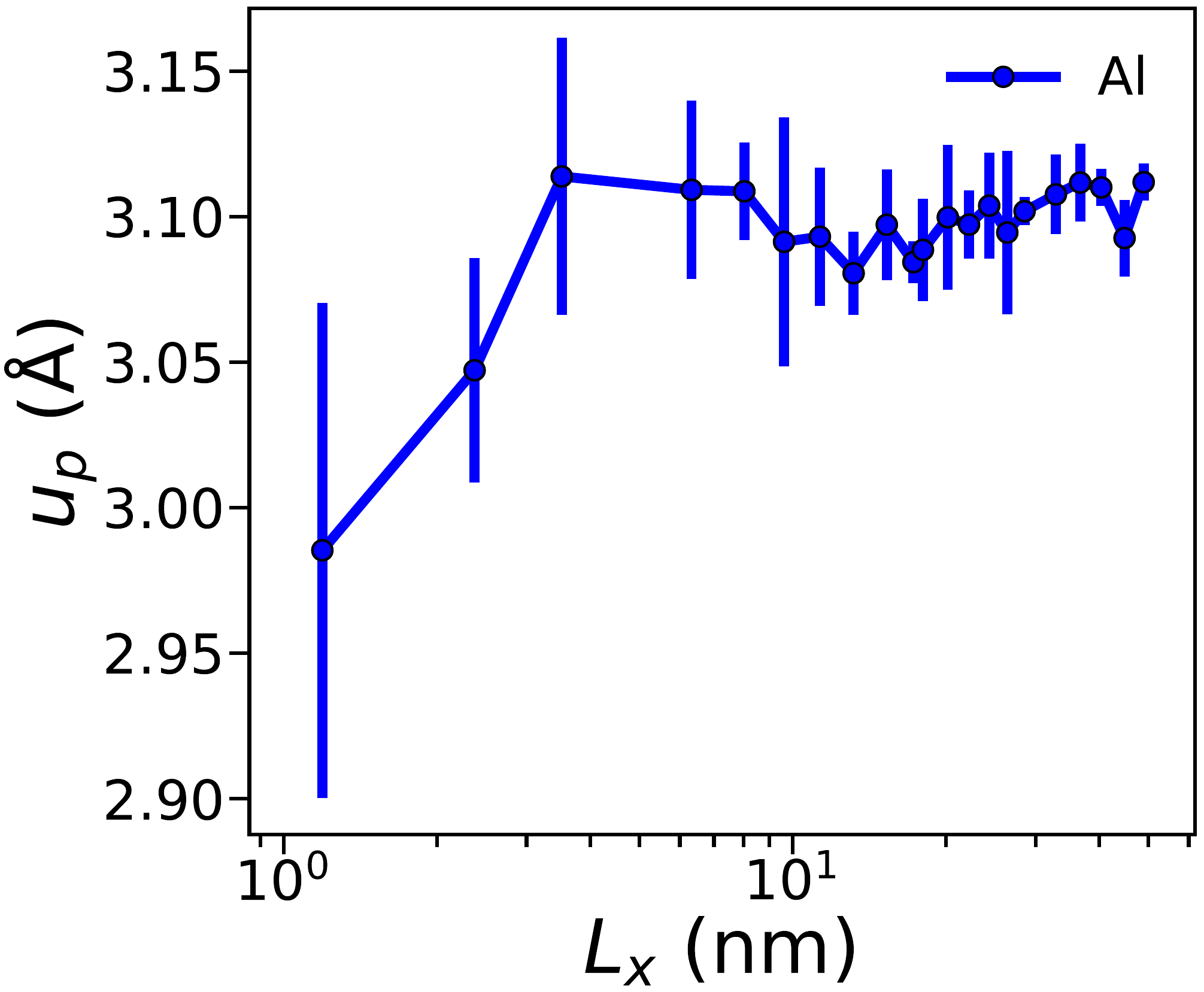}
  \includegraphics[height=3.6cm]{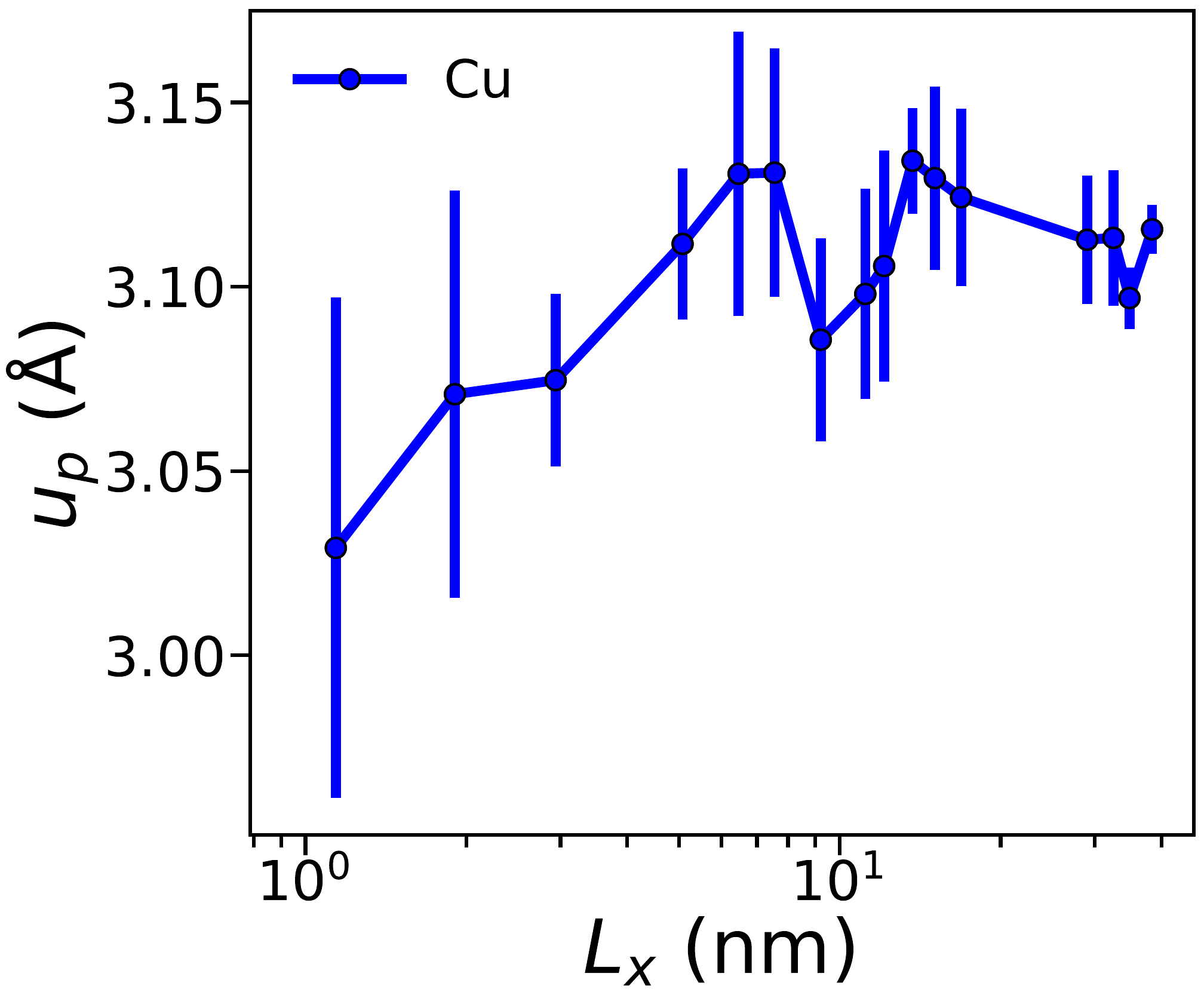}
  \caption{Most probable gap $u_{p}$ vs the lateral NP size $L_{x}$.}
  \label{fgr:probable_gap_vs_size_x}
\end{figure}

\begin{figure}[!h]
\centering
  \includegraphics[height=3.6cm]{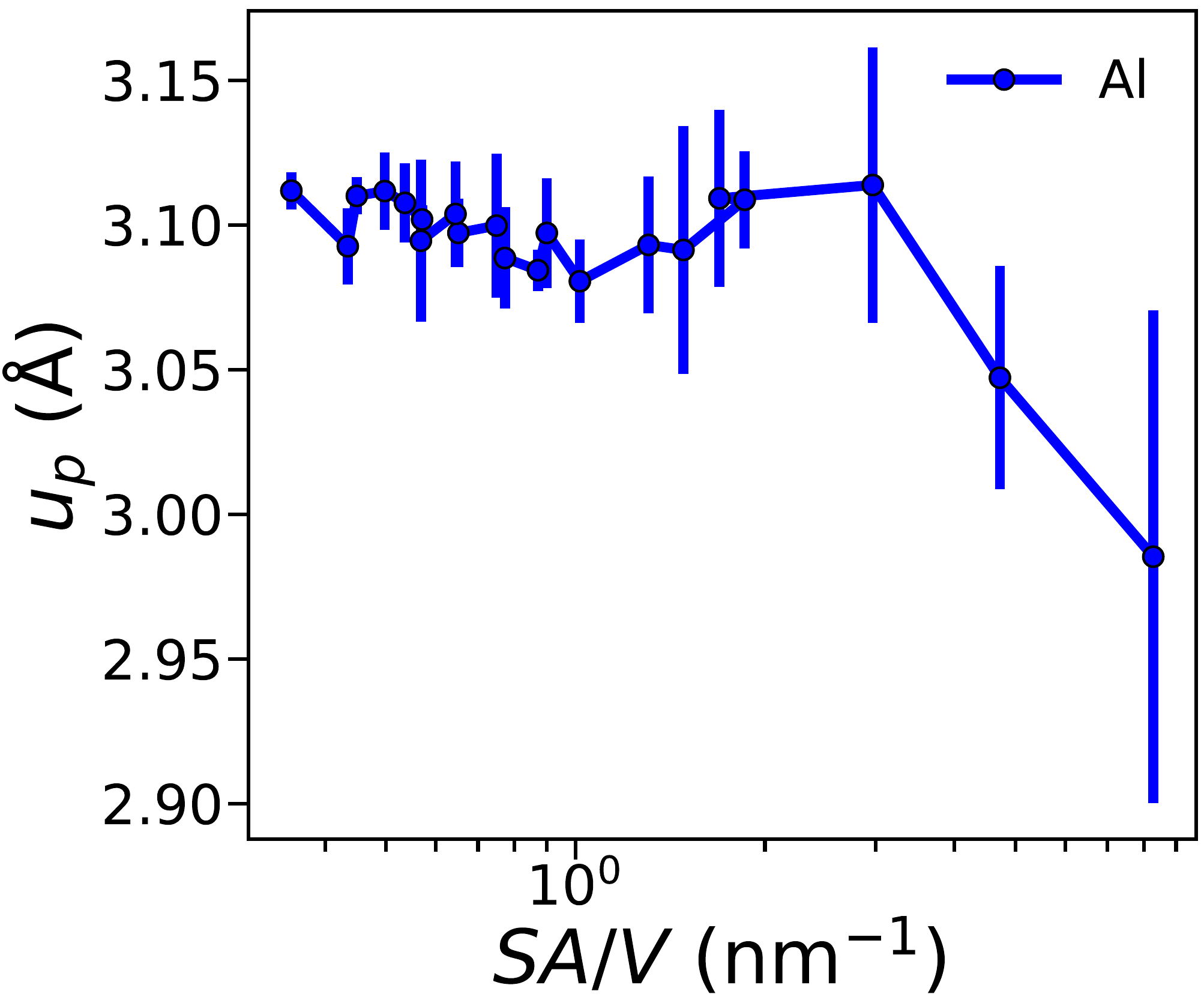}
  \includegraphics[height=3.6cm]{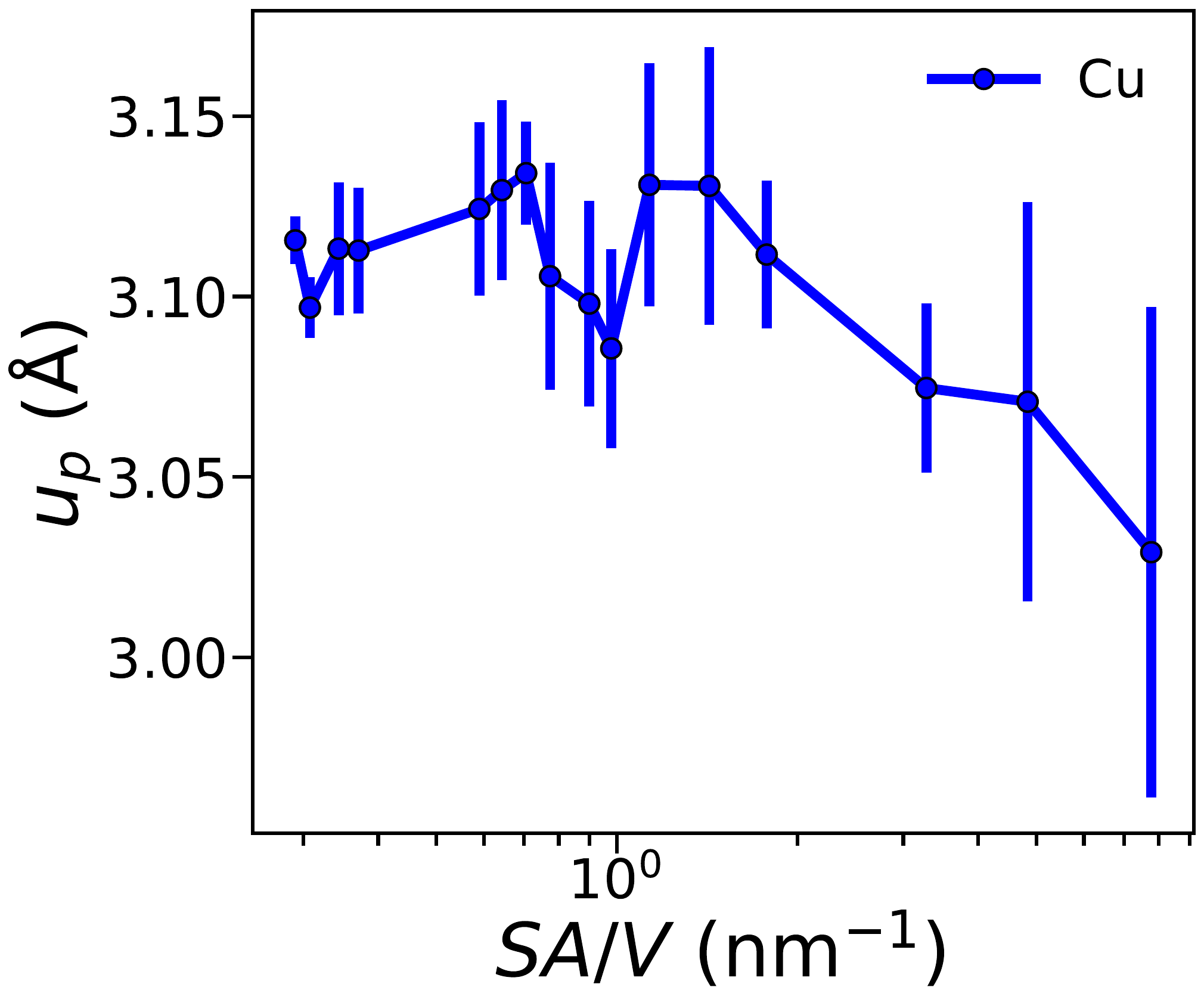}
  \caption{Most probable gap $u_{p}$ vs $SA/V$.}
  \label{fgr:probable_gap_vs_sav_x}
\end{figure}

The mean interfacial separation $\bar{u}$ was calculated from the corresponding distributions as the mean of $P(u)$. Figs.~\ref{fgr:mean_gap_vs_size_x},~\ref{fgr:mean_gap_vs_sav} suggest that Al and Cu NPs have slightly different thermodynamic values of $\bar{u}$ of around 3.12 and 3.13~\AA, respectively, even though their LJ interaction with the substrate is the same. These values are smaller than the LJ equilibrium interatomic distance ($1.122\sigma = 3.38$~\AA). The spread of $\bar{u}$ is small, around 0.1~\AA~for both metals. Within this range, $\bar{u}$ decreases with NP size: the smallest NPs ($\lesssim$~6~nm), and $SA/V$ larger than $\sim$~1.8~nm$^{-1}$ have bigger average values of $\bar{u}$ and exhibit considerable fluctuations. The larger NPs have $\bar{u}$ close to the reported thermodynamic values with negligible fluctuations in the largest NPs. The most probable gap $u_{p}$ (the peak value in $P(u)$), displays the reverse scaling in Figs.~\ref{fgr:probable_gap_vs_size_x},~\ref{fgr:probable_gap_vs_sav_x} compared to $\bar{u}$: smaller NPs have lower most probable values. The results suggest that, on average, the smallest NPs are farther from the substrate than the larger ones, but have a lower probability of larger $u$ values.

The observed behavior of $\bar{u}$ and $u_{p}$ for smaller NPs is similar to that of the melting point of Al nanoclusters~\cite{Yalamanchali2017} and may indicate a size effect. It can be attributed to the higher diffusion of the smallest NPs, their small interface size (less than 10$^3$ atoms, cf. Fig.~\ref{fgr:nbot_vs_size_x}), and the irregular structure of their contact layer, which can vary from one NP sample to another. This behavior is analogous to the finite-size effects observed for macroscopic surfaces~\cite{Pastewka2013}. When the surface size or squeezing pressure is so small that only a single surface asperity is in contact, CM properties, in particular the interfacial stiffness $K$, scale differently from higher pressures. Larger $K$ fluctuations are also observed.

\begin{figure}[!h]
\centering
  \includegraphics[height=3.8cm]{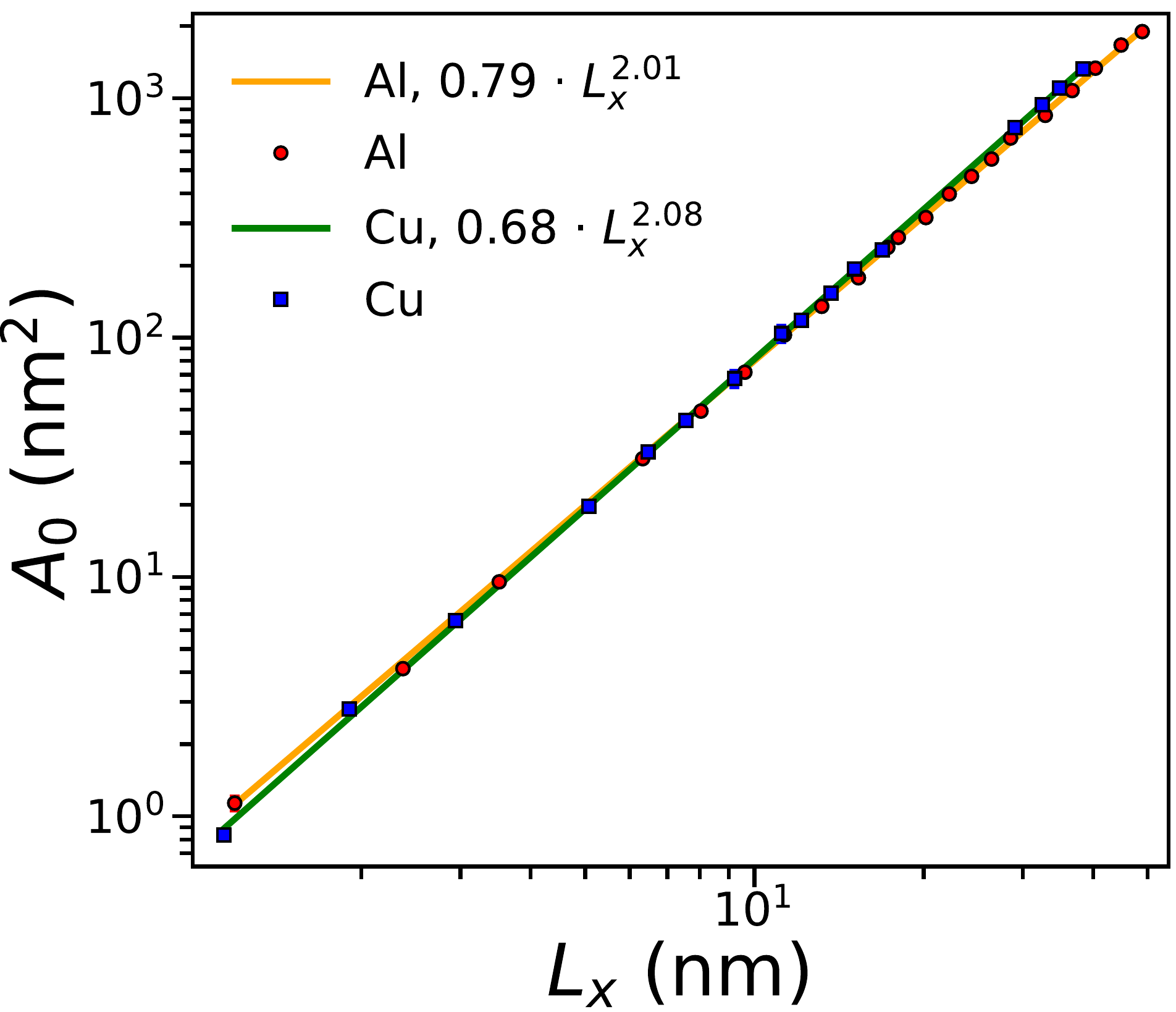}
  \includegraphics[height=3.8cm]{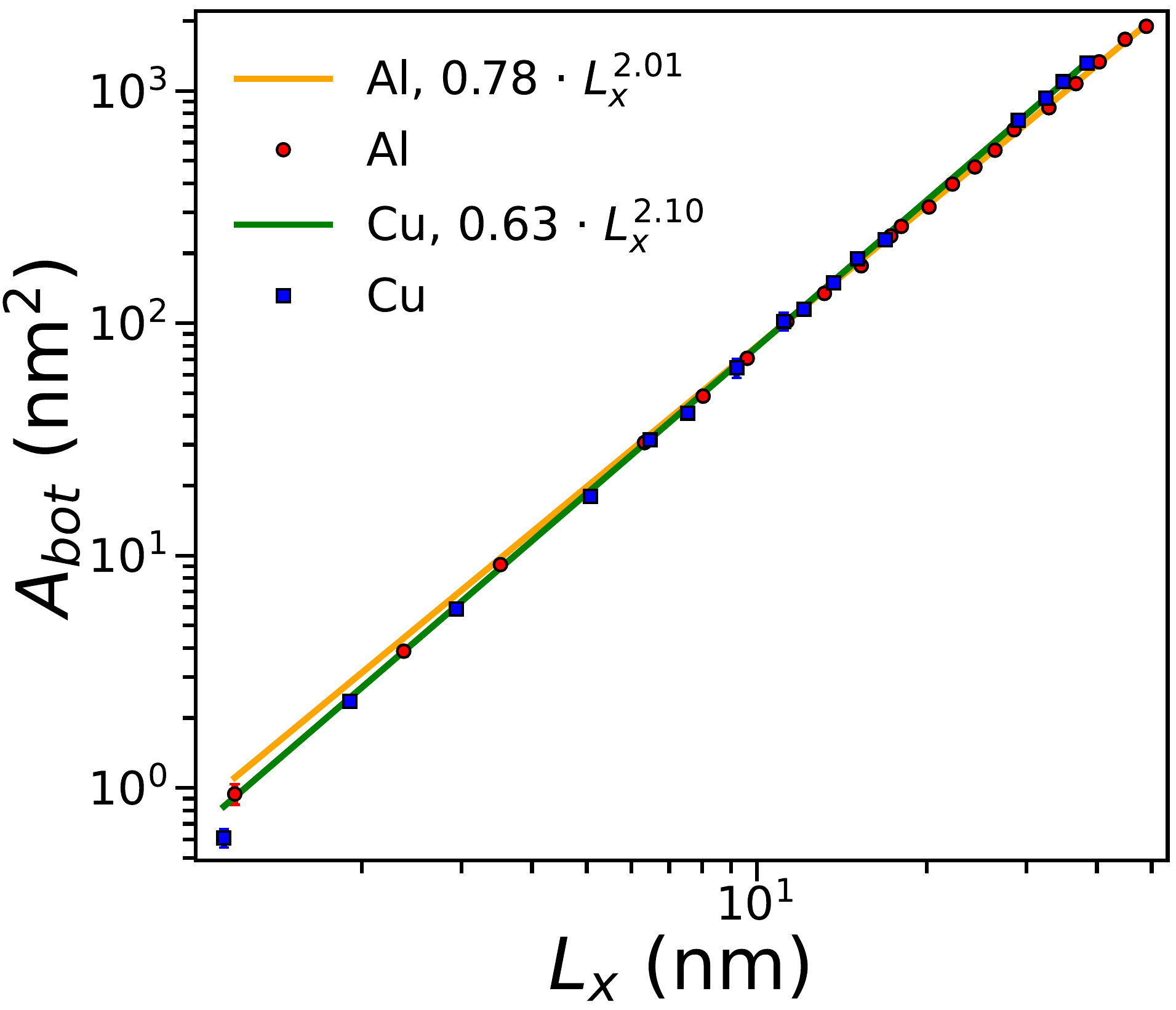}
  \caption{The apparent contact area $A_{0}$ and the polygonal fit to the bottom layer $A_{bot}$ vs $L_{x}$.}
  \label{fgr:apparent_and_bottom_areas}
\end{figure}

\begin{figure}[h]
\centering
  \includegraphics[height=3.7cm]{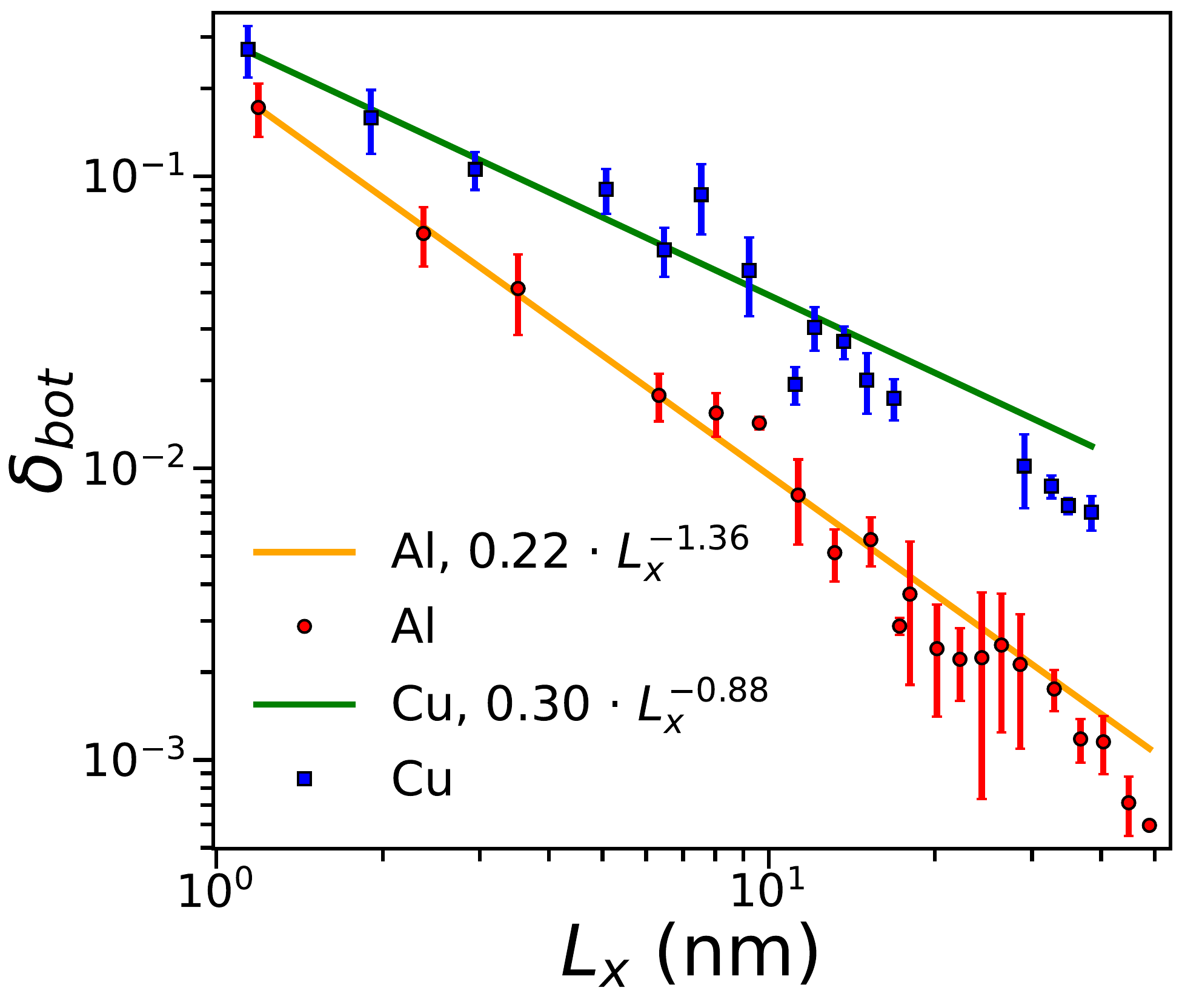}
  \includegraphics[height=3.7cm]{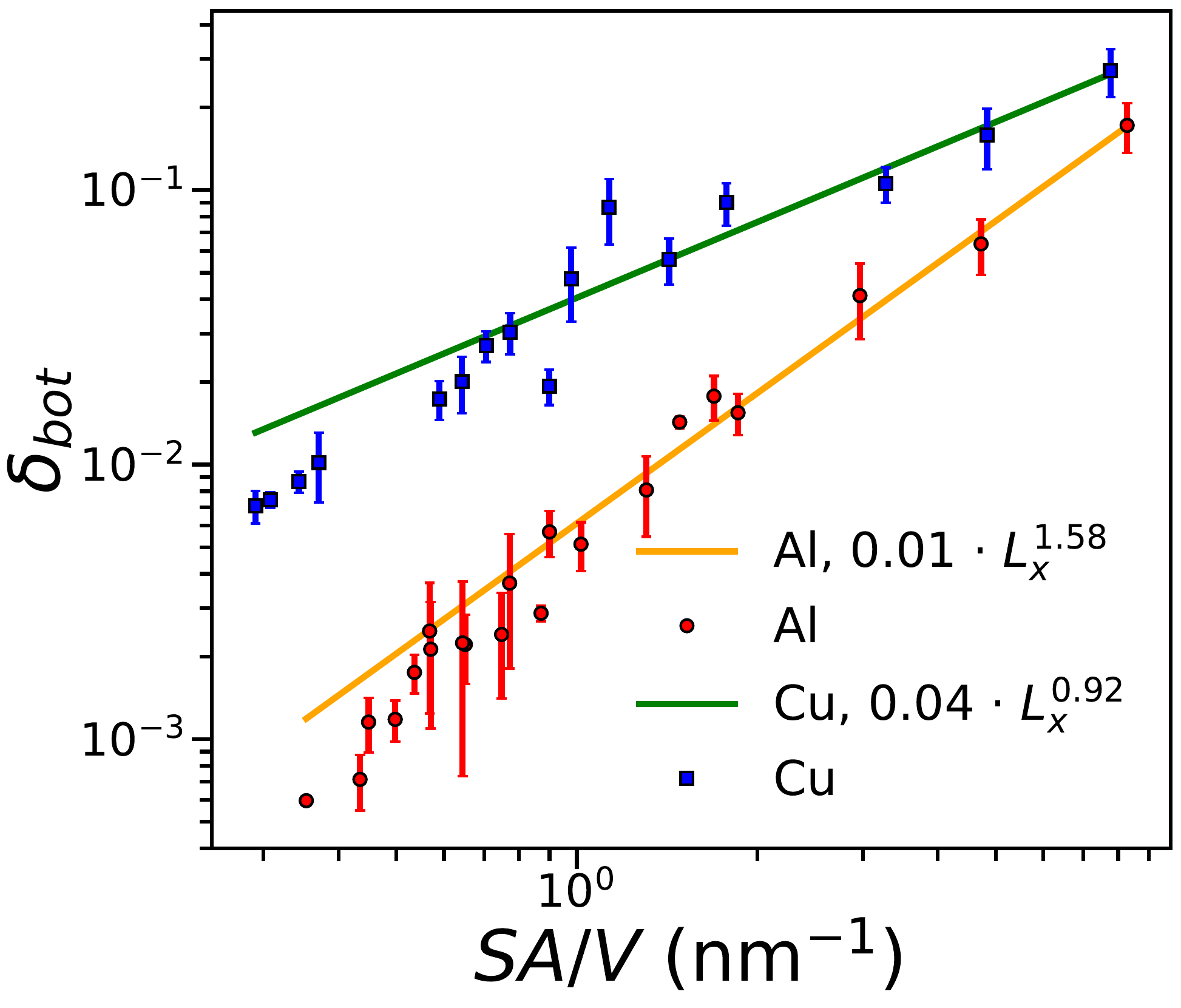}
  \caption{Relative error $\delta_{bot}$ of the bottom polygon area compared to the apparent area $A_{0}$ vs $L_{x}$ and $SA/V$.}
  \label{fgr:area_error}
\end{figure}

\begin{figure}[!h]
\centering
  \includegraphics[height=3.7cm]{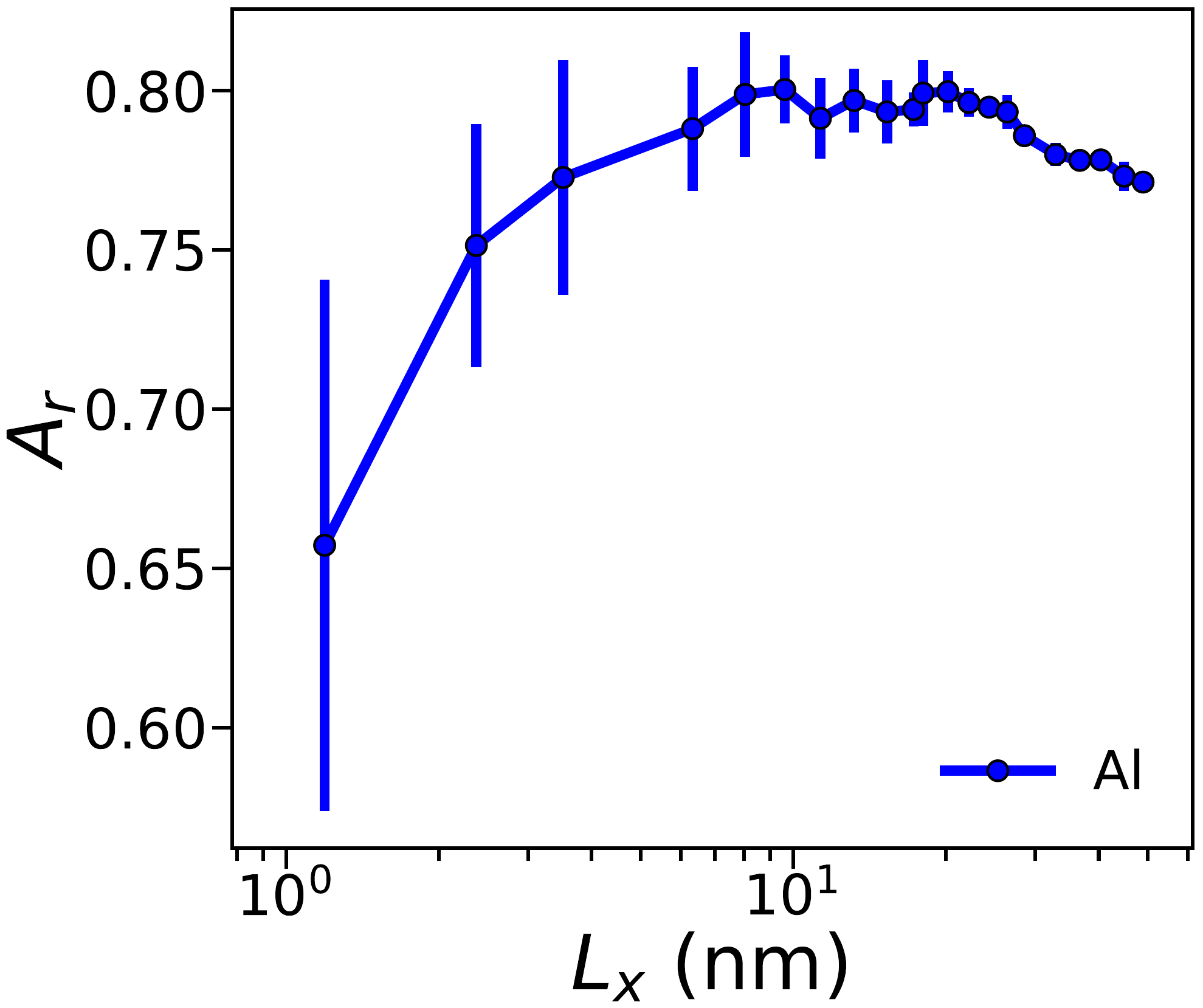}
  \includegraphics[height=3.7cm]{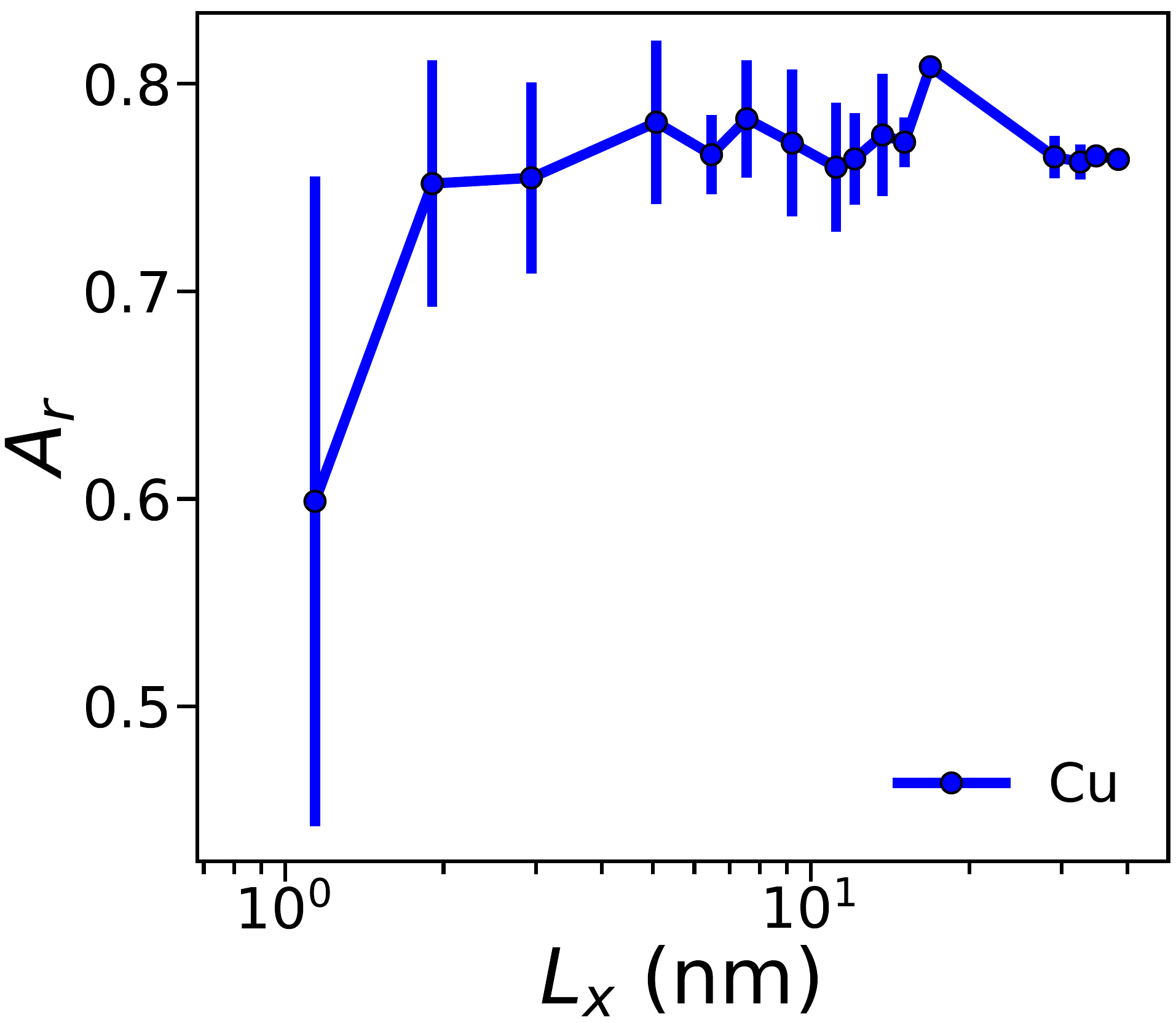}
  \caption{Relative contact area $A_{r}$ scaling with $L_{x}$.}
  \label{fgr:relative_area_vs_sizex}
\end{figure}

\begin{figure}[!h]
\centering
  \includegraphics[height=3.7cm]{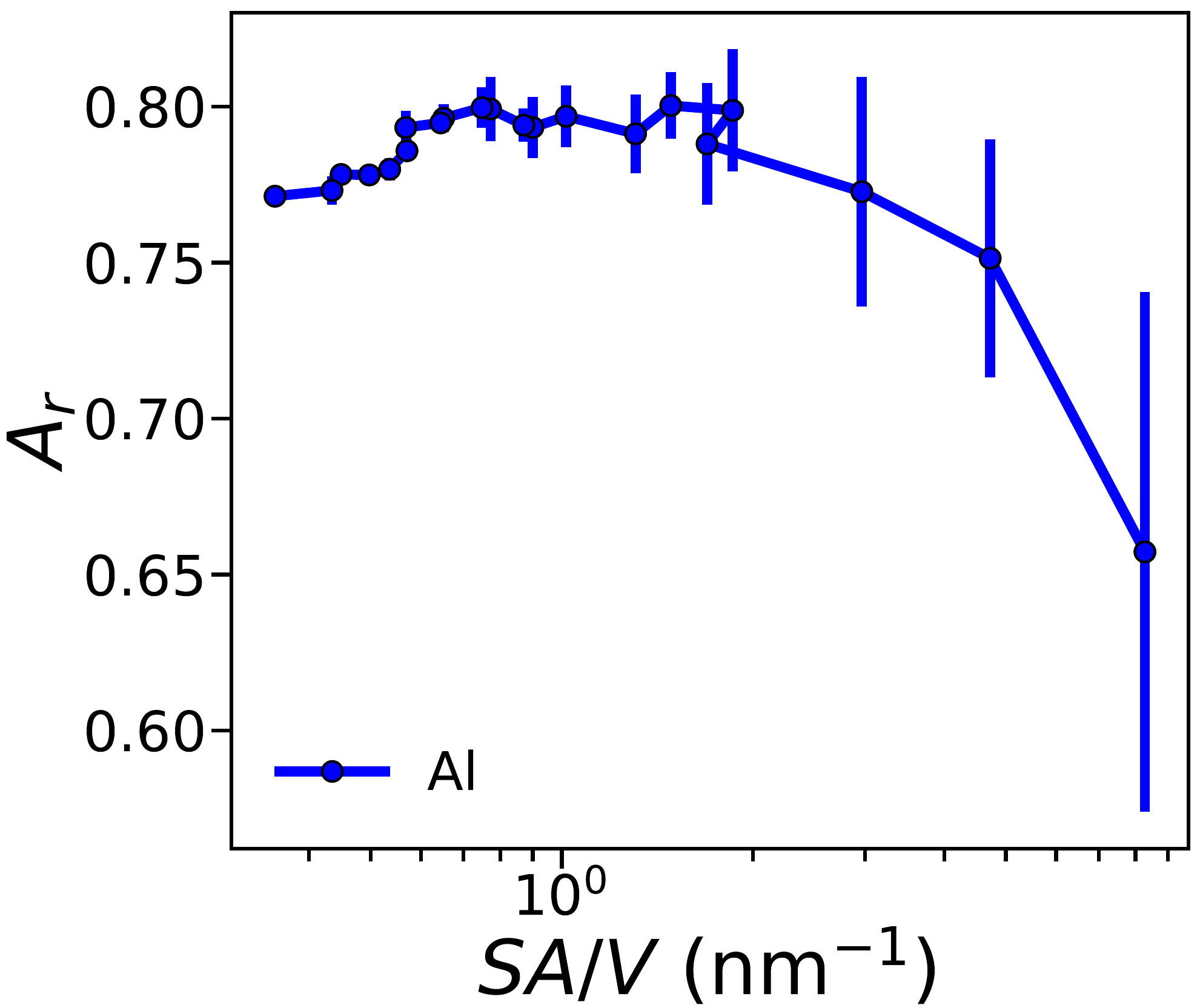}
  \includegraphics[height=3.7cm]{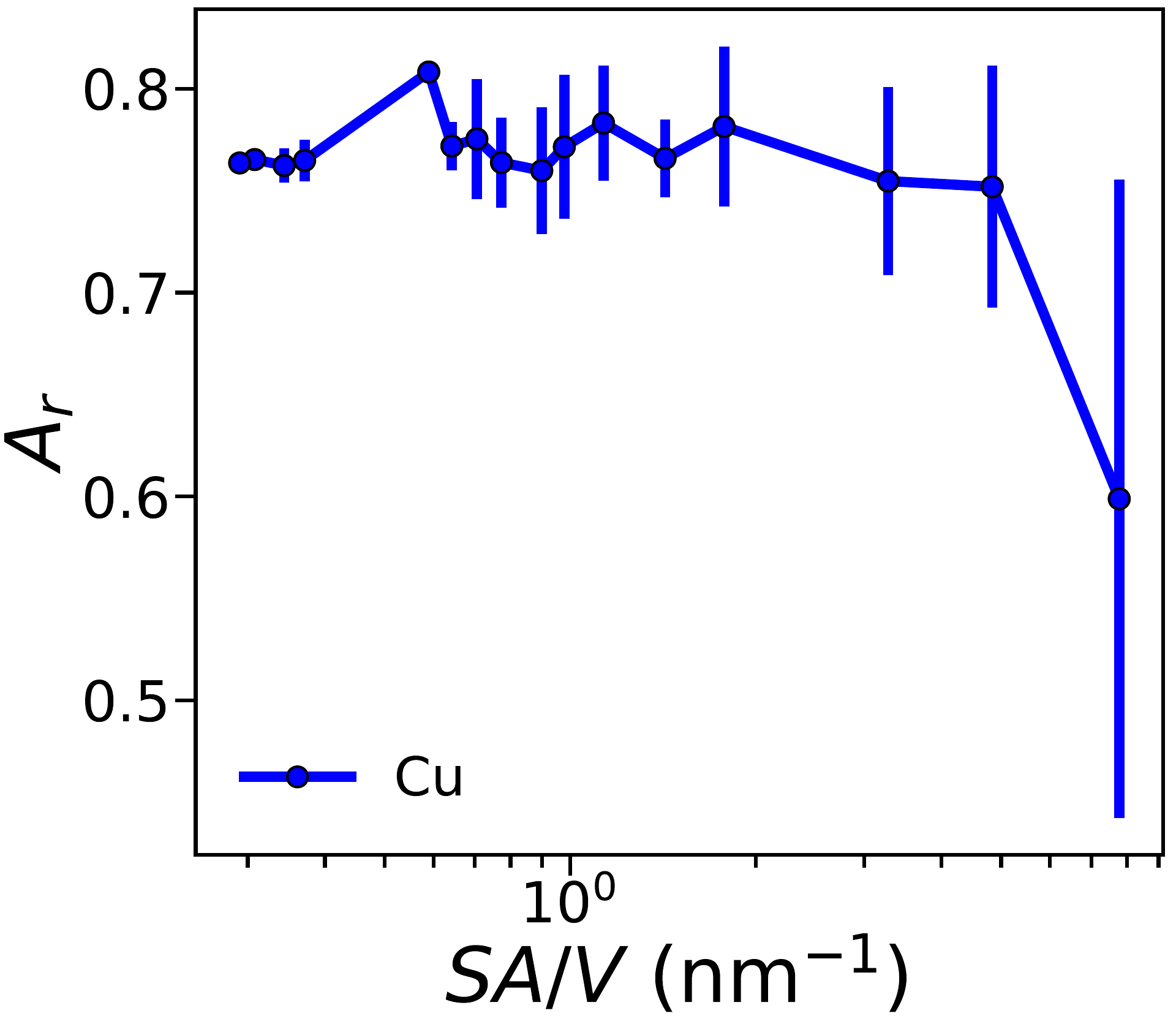}
  \caption{Relative contact area $A_{r}$ vs $SA/V$.}
  \label{fgr:relative_area_vs_sav}
\end{figure}

\subsubsection{Contact area.}
\label{sec:area}

\paragraph{Absolute contact area.}

Both the apparent contact area $A_0$ and the area $A_{bot}$ of the polygonal fit to the bottom atomic layer of a NP scale quadratically with $L_{x}$, see Fig.~\ref{fgr:apparent_and_bottom_areas}. Plots for Cu and Al almost coincide, suggesting that NPs of the same lateral size have almost the same apparent area $A_{0}$. As with most other scalings, $A_{bot}$ values are slightly off the power-law fit for the smallest NPs.

Fig.~\ref{fgr:area_error} shows the relative error $\delta_{bot} = (A_{0} - A_{bot})/A_{0}$ of the bottom polygon area compared to $A_{0}$. One can see that $\delta_{bot} < 1$~\% for NPs larger than about 10~nm and 30~nm for Al and Cu, respectively. Therefore, for most experimentally accessible NPs, the apparent area of NPs is an acceptable measure of the contact area for Al and Cu NPs adsorbed on a suspended graphene. One can argue that the flat-polygon fit area is not a good approximation for the nonflat bottom atomic layer \cite{Lyashenko2013,Khomenko_NAP2017}. However, we also considered the area of the bottom atomic-layer mesh. The results are close to those for $A_0$, but the mesh surface reconstruction required some manual, imprecise fitting, resulting in a larger error spread, which is not shown here.

\paragraph{Relative contact area.}

In macroscopic theories, the real $A_{real}$ and relative $A_{r} = A_{real}/A_{0}$ contact areas are mathematically well-defined using distance or other criteria, such as forces~\cite{Persson2005jpcm,Almqvist2011jmps}. The distance criterion states that if $u(x, y) = 0$, real contact occurs at point $(x, y)$. However, at the atomic level, $A_{real}$ is not well-defined, because, e.g., $u(x, y) \neq 0$ in all the cases. The $A_{real}$ value depends on the task at hand. For example, if liquid leakage through a contact is considered, one must account for the kinetic diameter of the smallest liquid molecules. The contact will occur in regions where the $u$-value is not big enough for molecules to pass through. For heat or electrical transfer, a simple distance criterion may not be sufficient to identify the contact. One needs to consider how the interface atoms interact to exchange heat or electrical charges. Theoretically, there can be scenarios in which the distance between the surfaces is still not sufficient for liquid molecules to leak, meaning the contact area is percolated. However, such a $u$ can already be large enough to prevent heat or electrical transfer between the surfaces.

In this work, we consider the $A_{r}$ of the NPs in the context of water leakage. Different values for the characteristic size of the kinetic diameter of a water molecule or the effective pore size for its leakage are reported in the literature~\cite{Murata2000,Yang2019}, e.g., 3--3.2~\AA. We used a value of 3.2~\AA~so that metal atoms with an interfacial gap $u < 3.2$~\AA~were considered to be in contact. The relative contact area $A_{r}$ is then the ratio of atoms in contact to the number of interface metal atoms $N_{b}$ \cite{Zakharov_NAP2018}. The size scaling of $A_{r}$ in Figs.~\ref{fgr:relative_area_vs_sizex},~\ref{fgr:relative_area_vs_sav} is consistent with the reported $\bar{u}$ dependence: smaller NPs have a lower $A_{r}$ reflecting larger $\bar{u}$. This means that, geometrically, water molecules are harder to leak through the larger NPs' contact interface. Note that for macroscopic randomly rough surfaces, the contact area percolates~\cite{Persson2012} at $A_{r} \approx 0.48$. If this holds for the NPs, water won't leak under even the smallest NPs at zero squeezing pressure because $A_{r} > 0.5$ in most cases.

\section*{Conclusions}

We investigated the size scaling of morphological and CM properties of Al and Cu NPs adsorbed on a suspended graphene sheet. The main result is that the scaling of the morphological properties, such as $S$ and $V$, as well as the CM quantities (the mean $\bar{u}$ and most probable $u_{p}$ gaps, relative contact area $A_{r}$, and the height and gap PSDs $C_h$ and $C_u$) manifest size effects. The NPs can be roughly subdivided into 2 groups. The border is around 4~K atoms or a linear size of about 3--6~nm, and $SA/V$ is around 1.8~nm$^{-1}$. Properties of the NPs smaller than the mentioned size scale differently compared to the larger ones. In particular, values of the smaller NPs strongly fluctuate and differ from the thermodynamic values of the larger NPs. PSDs of the mentioned groups differ too. PSDs are smeared for the smaller NPs, whereas the larger nanoislands exhibit a hexagonal frequency structure. Such scaling behavior can be explained by the fact that smaller NPs have higher mobility and fewer atoms in their interface layer, leading to greater fluctuations in the dependent morphological and CM values across NP samples.

The results also suggest that mimicking the thermal dewetting procedure in MD can produce NPs of different shapes and surface topographies for different metals. All the NPs had a polycrystalline structure. Typically, the interface atomic layer of the NPs obtained this way corresponded to the lowest energy (111) atomic plane, which has a hexagonal atomic ordering. PSDs confirm such a sixfold arrangement. In particular, 2D $h$ and $u$ PSDs are anisotropic and have sixfold symmetry with a central ``snowflake'' region whose dimensions are bounded by the frequencies that roughly correspond to the nearest neighbor distances of the metal atoms. PSDs can also differ across NPs due to differences in interface atomic topography.

Another prominent insight is that the polycrystalline NPs did not have atomically flat contact surfaces but rather exhibited surface-height corrugation, as evidenced by the height distributions and PSDs. In particular, the height PSD of the larger NP samples has relatively narrow (less than 1~decade in frequency) regions that scale as power laws, some of which exhibit self-affine surface roughness with Hurst exponents 0.1--0.56. Our simulations suggest that even though the graphene substrate surface is macroscopically smooth, the NPs formed on it can still exhibit random roughness with RMS value of~$\sim$\AA~and less. Additionally, we showed that the PSDs of the mean gap exhibit high-frequency contributions, even for the largest NPs, and are attributed to the atomic-level roughness of the graphene bond length scale. Our results pave the way for the quantitative description and prediction of contact mechanics and friction of NPs, providing estimates of PSD that can be used in analytical theories applied to atomistic systems.

The surface height distribution $P(h)$ has three parts: a narrow spike, a region with uniform heights, and a decaying tail. The spike and the tail of larger NPs can be fit to Gaussians. The spread of the surface height can be more than 1~nm. In contrast, the interfacial separation distribution $P(u)$ is close to a single Gaussian, similarly to macroscopic randomly rough surfaces at zero squeezing pressure, and the spread of $u$ values is only a few~\AA. This suggests that the elastic graphene substrate deforms so that the mean gap is normally distributed, in contrast to the surface height, whose topography is determined solely by the preparation methodology.

We also showed that the apparent contact area is close to the area of the polygon fit to the contact atomic layer with relative differences less than 1\% for NPs larger than about 10 and 30~nm for Al and Cu, respectively. One needs to be cautious when making assumptions about the equality of the apparent and real contact area of NPs adsorbed on a suspended graphene, as the relative difference can be more than 10\% for the smallest NPs. The relative contact area, calculated using the distance criterion in the context of water leakage, also exhibits size effects and reflects the mean gap behavior, being smaller for larger $\bar{u}$ of the smaller NPs.

Finally, the model in this work considered only two particular metals and an atomically smooth elastic graphene substrate. Future research can include other metals and substrates with different elemental compositions, elasticities, and atomic structures. The current study did not consider the behavior of the contact forces and energies. Additionally, no external pressure was applied to the metal atoms, though some tribological experiments employ a ``tip-on-top'' configuration in which the AFM tip is placed on top of an NP to move it~\cite{Oo2024}, thus, squeezing the NP. The morphology and contact mechanics of NPs under the mentioned conditions should be considered in follow-up studies.

\section*{Author contributions}

Mykola Prodanov: Conceptualization, methodology, software, running the simulations, data processing and analysis, writing -- original draft, review \& editing. Oleksii Khomenko: Project administration, resources, conceptualization, supervision, writing -- review \& editing.

\section*{Conflicts of interest}
The authors have no conflicts of interest to declare.

\section*{Data availability}
The MD code used in this study can be found at\\
https://github.com/prodk/SurfaceGrowthCrossPlatform\\
Raw data are available from the authors upon request.






\balance


\bibliography{al_cu_np_contact_mech} 
\bibliographystyle{rsc} 
\end{document}